\documentclass[12pt]{article}
\usepackage{color,amssymb,epsfig,verbatim}
\usepackage{graphicx} 
\usepackage{bm}
\usepackage{multirow}
\usepackage{amsmath}
\usepackage{lscape}
\usepackage[table,xcdraw]{xcolor}
\usepackage{natbib}
\setcitestyle{aysep={}} 
\usepackage{hyperref}
\usepackage{accents}
\hypersetup{colorlinks,linkcolor={black},citecolor={black},urlcolor={red}}

\renewcommand{\baselinestretch}{1.5}

\def\log{\hbox{log}}

\def\boxit#1{\vbox{\hrule\hbox{\vrule\kern6pt
          \vbox{\kern6pt#1\kern6pt}\kern6pt\vrule}\hrule}}

\def\bse{\begin{eqnarray*}}
\def\ese{\end{eqnarray*}}
\def\be{\begin{eqnarray}}
\def\ee{\end{eqnarray}}
\def\bq{\begin{equation}}
\def\eq{\end{equation}}
\def\bse{\begin{eqnarray*}}
\def\ese{\end{eqnarray*}}

\definecolor{emerald}{RGB}{6, 91, 82}

\usepackage{enumitem}

\usepackage{scalerel}

\usepackage{rotating} 
\usepackage[para]{threeparttable}
\usepackage{caption}

\newtheorem{theorem}{Theorem}

\usepackage{ dsfont }

\usepackage{romannum}

\newenvironment{placehere}
    {\begin{center}
    [ \it Place}  
    {About Here \rm]
    \end{center}
    }

\begin{document}
\thispagestyle{empty}


\baselineskip=28pt
\begin{center}
{\LARGE{\bf Sensitivity Analysis of Error-Contaminated Time Series Data under Autoregressive Models with Application of COVID-19 Data}}
\end{center}
\baselineskip=14pt
\vskip 2mm
\begin{center}
Qihuang Zhang and Grace Y. Yi\footnote{\baselineskip=10pt Corresponding Author: Department of Statistical and Actuarial Sciences, Department of Computer Science, University of Western Ontario, London, Ontario, Canada, N6A 5B7. Department of Statistics and Actuarial Science, University
of Waterloo. Email:  gyi5@uwo.ca}

\end{center}
\bigskip

\vspace{8mm}

\begin{center}
{\Large{\bf Abstract}}
\end{center}
\baselineskip=17pt
{ 
Autoregressive (AR) models are useful tools in time series analysis. Inferences under such models are distorted in the presence of measurement error, which is very common in practice.
In this article, we establish analytical results for quantifying the biases of the parameter estimation in AR models if the measurement error effects are neglected.  We propose two measurement error models to describe different processes of data contamination. An estimating equation approach is proposed for the estimation of the model parameters with measurement error effects accounted for. We further discuss forecasting using the proposed method. Our work is inspired by COVID-19 data, which are error-contaminated due to multiple reasons including the asymptomatic cases and varying incubation periods. We implement our proposed method by conducting sensitivity analyses and forecasting of the mortality rate of COVID-19 over time for the four most populated provinces in Canada. The results suggest that incorporating or not incorporating measurement error effects yields rather different results for parameter estimation and forecasting.

}

\vspace{8mm}

\par\vfill\noindent

\par\medskip\noindent
\underline{\bf Short title}: Time Series with Measurement Error {\color{white} COVID-19}

\clearpage\pagebreak\newpage
\pagenumbering{arabic}

\newlength{\gnat}
\setlength{\gnat}{22pt}
\baselineskip=\gnat

\clearpage



\section{Introduction} 

Time series data are common in the fields of epidemiology, economics, and engineering. Various models and methods have been developed for analyzing such data. The validity of these methods, however, hinges on the condition that time series data are precisely collected. This condition is restrictive in applications. Measurement error is often inevitable. In the study of air pollution, for example, it is difficult or even impossible to precisely obtain the true measurement of the daily air population level.

Some work on time series subject to measurement error is available in the literature. \citet{Tanaka2002} proposed a Lagrange multiplier test to assess the presence of measurement error in time series data. \citet{staudenmayer2005measurement} explored the classical measurement error model for the autoregressive process. \citet{TripodisBuonaccorsi2009} studied measurement error in forecasting using the Kalman filter. \citet{DedeckerSamson2014}  considered a non-linear AR(1) model with measurement error.  Despite available discussions of measurement error in time series, several limitations restrict the application scope of the existing work. Most available methods consider only autoregressive models without the drift and assume the simplest additive measurement error model. Furthermore, most work involves a complex formulation to adjust for the measurement error effects, which is not straightforward to implement for practitioners. In addition, to our knowledge, there is no available work addresses measurement error effects on prediction under autoregressive models.

In this article, we systematically explore analysis of error-prone time series data under autoregressive models. We propose two types of models to delineate measurement error processes: additive regression models and multiplicative models. These modeling schemes offer us great flexibility in facilitating different applications. We investigate the impact of the naive analysis which ignores the feature of measurement error in the inferential procedures, and we obtain analytical results for characterizing the biases incurred in the naive analysis. We develop an estimating equation approach to adjust for the measurement error effects on time series analysis. We establish asymptotic results for the proposed estimators, and develop the theoretical results for the forecasting of times series in the presence of measurement error. Finally, we describe a block bootstrap algorithm for computing standard errors of the proposed estimators.

Our work is partially motivated by the data of COVID-19, a wide-spread disease that has become a global health challenge and has caused over ten million infections and half million deaths as of August, 2020. Because of the special features of the disease, the COVID-19 data introduce a number of new challenges: 1) due to the asymptomatic infected cases and the patients with light symptoms who do not go to hospitals, the number of reported cases with COVID-19 is typically smaller than the true number of infected cases; 2) due to the limited test resources, many infected cases are not able to be identified instantly; and 3) the varying incubation periods lead to the delay of the identification of the infections. Consequently, the discrepancy between the reported case number and the true case number can be substantial, and ignoring these features and applying the traditional time series analysis method would no longer produce valid results.  

In this paper, we apply the developed methods to analyze the COVID-19 data. We are interested in studying how the mortality rate in a region may change over time and describing the trajectory of the death rate. While the mortality rate of a disease is defined as the death number divided by the case number, the determination of the mortality rate of COVID-19 is challenging. In contrast to the standard definition, \citet{baud2020real} estimated mortality rates by dividing the number of deaths on a given day by the number of patients with confirmed COVID-19 infections 14 days earlier, driven by the consideration of the maximum incubation time to be 14 days. Due to the unique features of COVID-19, there does not seem to be a precise way to define the mortality rate of COVID-19. In this paper, we conduct sensitivity analyses to assess the severity of the pandemic by using different definitions of the mortality rate and considering different ways of modeling measurement error in the data.

The remainder of the article is organized as follows. The notation and the setup for autoregressive time series models and the proposed measurement error models are introduced in Section~\ref{Sec:Setup}. In Section~\ref{sec:naive-est}, we present the theoretical results for characterizing the impact of measurement error on the analysis of time series data. In Section~\ref{sec:Methodology}, we develop an estimating equation approach to adjust for the biases due to measurement error. In Section~\ref{sec:Canada}, we implement the proposed method to analyze the COVID-19 data in four Canadian provinces. The article is concluded with a discussion presented in Section~\ref{sec:discussion}.

\section{Model Setup and Framework} \label{Sec:Setup}
\subsection{Time Series Model}

Consider a $T\times 1$ vector of time series,  $X^{(T)}=(X_1,X_2, \ldots, X_T)^{\rm \scalebox{0.6}{T}}$. We are interested in modeling the dependence of $X_{t}$ on it previous observations $X^{(t-1)}$ and we consider it to be postulated by an autoregressive model with lag $p$ 
\begin{equation}\label{eqn:timeseriresmodel}
    X_{t} =  \phi_0 + \sum_{j=1}^p \phi_j X_{t-j} + \epsilon_t,
\end{equation}
where $p$ is an integer smaller than $T$, $\epsilon^{(t)}=(\epsilon_1,\ldots,\epsilon_t)^{\rm\scalebox{0.5}{T}}$ is independent of $X^{(t)}=(X_1,\ldots,X_t)^{\rm\scalebox{0.5}{T}}$ with each $\epsilon_t$ having zero mean and variance $\sigma_\epsilon^2$, $\phi_0$ is a constant drift, and $\phi=(\phi_1,\ldots,\phi_p)^{\rm \scalebox{0.6}{T}}$ is the regression coefficient. 

The additive form in (\ref{eqn:timeseriresmodel}) and the zero mean assumption of $\epsilon_t$ show that $\phi_0$ and $\phi$ are constrained by 
\begin{equation}\label{eqn:constraintsphi_0}
  \phi_0=E(X_t) - \{E(\widetilde{X}_{t-1})\}^{\rm\tiny T}\phi,  
\end{equation}
where $\widetilde{X}_{t-1}=(X_{t-1},\ldots,X_{t-p})^{\rm\tiny T}$. To make the process of $X_t$ stationary, $\phi_1,\ldots,\phi_p$ are further constrained such that all the roots of the equation in $z$
\begin{equation*}
    z^p- \phi_1 z^{p-1} - \dots -\phi_p  = 0
\end{equation*}
have absolute values smaller than 1 \citep[][Sec.3.1.]{brockwell2016introduction}. For example, a stationary AR(1) process requires that $|\phi_1|<1$, and a stationary AR(2) process needs that $(\phi_1+\phi_2)<1$, $(\phi_2-\phi_1)<1$ and $|\phi_2|<1$. Here we are interested in the estimation of parameters, $\phi$ and $\phi_0$. Let $\mu$ denote the mean $E(X_t)$ of the time series, which equals $\frac{\phi_0}{1-\phi_1-\ldots-\phi_p}$ if ${X_t}$ is (weakly) stationary. When $p=1$, the stationarity of 
a time series implies $\text{Var}(X_t)=\frac{\sigma_\epsilon^2}{1-\phi_1^2}$ for $t=1,\ldots, T$.

\subsection{Estimation of Model Parameters}

The estimation of the parameters in the AR($p$) time series model (\ref{eqn:timeseriresmodel}) can be carried out by the least squares method. To see this, we first focus on estimation of $\phi=(\phi_1,\ldots,\phi_p)^{\rm\scalebox{0.5}{T}}$. Let $S(\phi)=\sum_{t=p+1}^T\{X_t-(\phi_0 + \sum_{j=1}^p \phi_j X_{t-j})\}^2$ be the sum of the squared difference between $X_t$ and its linearly combined history with lag $p$. Then applying the constraint (\ref{eqn:constraintsphi_0}) gives $S(\phi)=\sum_{t=p+1}^T\left[\{X_t- E(X_t)\} - \{\widetilde{X}_{t-1} - E(\widetilde{X}_{t-1})\}^{\rm\tiny T}\phi\right]^2$.

To minimize $S(\phi)$ with respect to $\phi$, we solve $\frac{\partial S(\phi)}{\partial \phi}=0$ for $\phi$ and obtain the solution
 \begin{align}\label{eqn:OLSphi}
  \widehat{\phi}^{\scalebox{0.5}{\rm (LS)}} &= \left(\sum_{t=p+1}^{T}\left\{\widetilde{X}_{t-1} - E(\widetilde{X}_{t-1})\right\}\left\{\widetilde{X}_{t-1} - E(\widetilde{X}_{t-1})\right\}^{\rm\tiny T}\right)^{-1}   \sum_{t=p+1}^{T}\left\{\widetilde{X}_{t-1} - E(\widetilde{X}_{t-1})\right\} \left\{X_t-E(X_t)\right\},
 \end{align}
where for $t=1,\ldots, T$, $E(X_{t})$ can be estimated by $\frac{1}{T}\sum_{t=1}^T X_t$, which is denoted as $\widehat{\mu}$. 
 
 Next, by the constraint (\ref{eqn:constraintsphi_0}), replacing $E(X_t)$ by $\widehat{\mu}$ gives an estimator of $\phi_0$:
 \begin{equation}\label{eqn:OLSphi0}
     \widehat{\phi}_0^{\scalebox{0.5}{\rm (LS)}} = \widehat{\mu} - \widehat{\mu} \sum_{j=1}^p \widehat{\phi_j}.
 \end{equation}

Re-expressing (\ref{eqn:timeseriresmodel}) as $\epsilon_t= X_t-(\phi_0 + \sum_{j=1}^p \phi_j X_{t-j}) $ and by the definition of $S(\phi)$, we may estimate  Var$(\epsilon_t)=\sigma_\epsilon^2$  by 
 \begin{align} \label{eqn:OLSsigma_epsilon}
     \widehat{\sigma}_\epsilon^{2\scalebox{0.5}{\rm (LS)}} &= 
     \frac{1}{T-p}S(\widehat{\phi}) \nonumber \\
     &= \frac{1}{T-p} \sum_{t=p+1}^T\{X_t- E(X_t)\}^2 - \frac{2}{T-p} \sum_{t=p+1}^T\{X_t- E(X_t)\}\{\widetilde{X}_{t-1} - E(\widetilde{X}_{t-1})\}^{\rm\tiny T}\widehat{\phi}  \nonumber \\
     & \qquad + \frac{1}{T-p}\sum_{t=p+1}^T \widehat{\phi}^{\rm\tiny T}  \{\widetilde{X}_{t-1} -E(\widetilde{X}_{t-1})\}\{\widetilde{X}_{t-1} -E(\widetilde{X}_{t-1})\}^{\rm\tiny T}\widehat{\phi}
 \end{align}
 with $E(X_t)$ estimated by $\widehat{\mu}$.



Estimators (\ref{eqn:OLSphi})--(\ref{eqn:OLSsigma_epsilon}) can be derived in an alternative way. First, by the stationarity of the $X_{t}$, for $k=0,\ldots,p$ and $p\le t$, $\text{Cov}(X_{t},X_{t-k})$ is time-independent and let $\gamma_k$ denote it; it is clear that $\gamma_0$ represents $\text{Var}(X_t)$ for any $t$. Let $\Gamma$ be the autocovariance matrix \begin{equation*}
    \Gamma = \begin{pmatrix}
    \gamma_0 & \cdots & \gamma_{p-1} \\
    \vdots   & \ddots & \vdots \\
    \gamma_{p-1}  & \cdots & \gamma_0
    \end{pmatrix}.
\end{equation*} 
Let $\widehat{\gamma}=(\widehat{\gamma_1},\cdots,\widehat{\gamma_p})^{\rm\scalebox{0.5}{T}}$ with $\widehat{\gamma}_k=\frac{1}{T-k}\sum_{t=k+1}^{T}(X_t-\widehat{\mu})(X_{t-k} -\widehat{\mu})$ being an estimator of $\gamma_k$ for $k=0,\ldots,p$, and let $\widehat{\Gamma}$ be the estimator of $\Gamma$ with $\gamma_k$ replaced by $\widehat{\gamma_k}$ for $k=0,\ldots,p-1$. 

Next, we examine the summation terms in (\ref{eqn:OLSphi}) and (\ref{eqn:OLSsigma_epsilon}) by using the fact that as \mbox{$T\to\infty$},  $\frac{1}{T-p} \sum_{t=p+1}^T\{X_t- E(X_t)\}^2 \xrightarrow{\,\,p\,\,} \gamma_0$, $\frac{1}{T-p} \sum_{t=p+1}^T\{X_t- E(X_t)\}\{\widetilde{X}_{t-1} - E(\widetilde{X}_{t-1})\}^{\rm\tiny T} \xrightarrow{\,\,p\,\,} \gamma$, and $\frac{1}{T-p}\sum_{t=p+1}^T  \{\widetilde{X}_{t-1} -E(\widetilde{X}_{t-1})\}\{\widetilde{X}_{t-1} -E(\widetilde{X}_{t-1})\}^{\rm\tiny T} \xrightarrow{\,\,p\,\,} \Gamma$. Then,  (\ref{eqn:OLSphi})--(\ref{eqn:OLSsigma_epsilon}) motivate an alternative method of finding estimators for $\phi$, $\phi_0$, and $\sigma_\epsilon^2$, by solving the estimating equations:
\begin{align}\label{eqn:esteqntrue}
        \phi &= \widehat{\Gamma}^{-1}\widehat{\gamma}; \nonumber\\
    \phi_0 &= \left(1-\sum_{i=1}^p \phi_i\right)\widehat{\mu};  \\
    \sigma_\epsilon^2 &= \widehat{\gamma_0} - 2\phi^{\rm\scalebox{0.5}{T}}\widehat{\gamma} + \phi^{\rm\scalebox{0.5}{T}}\widehat{\Gamma} \phi, \nonumber 
\end{align}
for $\phi$, $\phi_0$, and $\sigma_\epsilon^2$. Let $\widehat{\phi}$, $\widehat{\phi}_0$ and $\widehat{\sigma}_\epsilon^2$ denote the resultant estimators of $\phi$, $\phi_0$, and $\sigma_\epsilon^2$, respectively. These estimators are asymptotically equivalent to the least squares estimators $\widehat{\phi}^{\scalebox{0.5}{\rm (LS)}}$, $\widehat{\phi}_0^{\scalebox{0.5}{\rm (LS)}}$, and $\widehat{\sigma}_\epsilon^{2\scalebox{0.5}{\rm (LS)}}$ in a sense that $\widehat{\phi} - \widehat{\phi}^{\scalebox{0.5}{\rm (LS)}} \xrightarrow{\,\,p\,\,} 0$, $\widehat{\phi}_0 - \widehat{\phi}_0^{\scalebox{0.5}{\rm (LS)}} \xrightarrow{\,\,p\,\,} 0$ and $\widehat{\sigma}_\epsilon^2 - \widehat{\sigma}_\epsilon^{2\scalebox{0.5}{\rm (LS)}} \xrightarrow{\,\,p\,\,} 0$, as $T\to\infty$, and hence, they are consistent \citep[][Ch.7, A.7.4]{box2015time}. 

Estimating equations (\ref{eqn:esteqntrue}) offer a unified estimation framework in its connections with not only the least squares estimation but also the maximum likelihood method under the assumption of Gaussian error as well as the Yule-Walker method. Similar to the least squares method, finding estimators using one of those approaches is asymptotically equivalent to solving (\ref{eqn:esteqntrue}) for $\phi$, $\phi_0$ and $\sigma_\epsilon^2$ \citep[][Ch.7, A.7.4]{box2015time}.

\section{Measurement Error and Impact}\label{sec:naive-est}
\subsection{Measurement Error Models}\label{sec:measurementerror}
Suppose that for $t=1,\ldots,T$, the observation of $X_t$ is subject to measurement error  and the precise measurement of $X_t$ may not be observed, but its surrogate measurement $X_t^*$ is available. We consider two measurement error models.

The first measurement error model takes an additive form
\begin{equation}\label{eqn:measmodelClassic}
    X_t^* = \alpha_0+\alpha_1 X_{t} + e_t 
\end{equation}
for $t=1,\ldots, T$, where the error term $e_t$ is independent of $X_t$ with mean 0 and time-independent variance $\sigma_e^2$ and is assumed to be independent for $t=1,\ldots,T$, and $\alpha=(\alpha_0,\alpha_1)^{\rm\scalebox{0.5}{T}}$ is the parameter vector. Here, $\alpha_0$ represents the systematic error and $\alpha_1$ represents the constant inflation (or shrinkage) due to the measurement error. For instance, if $\alpha_0 = 0$, then setting $\alpha_1 < 1$ (or $\alpha_1>1$) features the scenario where $X_t^*$ tends to be smaller (or larger) than $X_t$ if the noise term is ignored. This model generalizes the classical additive model considered by \citet{staudenmayer2005measurement} who considered the case with $\alpha_0=0$ and $\alpha_1=1$. 

By the stationarity of the $X_t$, we note that model (\ref{eqn:measmodelClassic}) yields $E(X_t^*)=\alpha_0+\alpha_1\mu$ and 
\begin{align}\label{apx:gamma0star}
   \text{Var} (X_t^*)  &= \alpha_1^2 \gamma_0 + \sigma_e^2;
\end{align}
the variability of the $X_t^*$ can be greater or smaller than that of the $X_t$, depending on the value of $\alpha_1$.

The second measurement error model assumes a multiplicative form:
\begin{equation}\label{eqn:measmodelmultiplicative}
    X_{t}^* = \beta_0 u_tX_{t},  
\end{equation}
for $t=1,\ldots, T$, where $\beta_0$ is a positive scaling parameter, and the $u_t$ are the error terms which are independent of each other as well as of the $X_t$, and have mean one and time-independent variance $\sigma_u^2$.  Depending on the distribution of the error term $u_t$, (\ref{eqn:measmodelmultiplicative}) can feature different types of discrepancy between $X_t$ and $X_t^*$.          

The stationarity of the $X_t$ together with model (\ref{eqn:measmodelmultiplicative}) implies $E(X_t^*)=\beta_0\mu$, and
\begin{align}\label{apx:Thm4-gamma0star}
  \text{Var}(X_t^*) &= \beta_0^2 \left\{(\sigma_u^2+1) \gamma_0 + \sigma_u^2  \mu^2\right\},
\end{align}
where we use the independence of $X_t$ and $u_t$. 

Since $E(X_t^*)$ is time-independent for both (\ref{eqn:measmodelClassic}) and (\ref{eqn:measmodelmultiplicative}), in the following discussion, we let $\mu^*$ denote $E(X_t^*)$ for $t=1,\ldots, T$. The modeling of the measurement error process by (\ref{eqn:measmodelClassic}) or (\ref{eqn:measmodelmultiplicative}) introduces extra parameters \{$\alpha_0$, $\alpha_1$, $\sigma_e^2$\} or \{$\beta_0$,  $\sigma_u^2$\}, where the variance of the error term is bounded by the variability of $X_t^*$ together with others. Clearly, (\ref{apx:gamma0star}) shows that $\sigma_e^2 < \text{Var}(X_t^*)$ and (\ref{apx:Thm4-gamma0star}) implies that $\sigma_u^2 < \frac{\text{Var}(X_t^*)}{\beta_0^2\mu^2}$. 

\subsection{Naive Estimation and Bias for AR(1) Model}\label{sub:p=1}

Estimating equations (\ref{eqn:esteqntrue}) are useful when measruements of $X_t$ are available. However, due to the measurement error, $X_t$ is not observed so (\ref{eqn:esteqntrue}) cannot be directly used for estimation of the parameters for model (\ref{eqn:timeseriresmodel}). As the surrogate $X_t^*$ for $X_t$ is available, one may attempt to employ the naive analysis to model (\ref{eqn:timeseriresmodel}) with $X_t$ replaced by $X_t^*$. Here we study the impact of measurement error on the naive analysis disregarding the difference between $X_t$ and $X_t^*$. We start with the AR(1) model, i.e., model (\ref{eqn:timeseriresmodel}) with $p=1$.

If we naively replace $X_t$ in (\ref{eqn:timeseriresmodel}) by $X_t^*$, then the time series model
(\ref{eqn:timeseriresmodel}) becomes
\begin{equation}\label{eqn:working-p1}
    X_t^* = \phi_0^* + \phi_1^* X_{t-1}^* + \epsilon_t^*,
\end{equation}
where $(\phi_0^*,\phi_1^*)^{\rm\scalebox{0.5}{T}}$ and $\epsilon_t^*$ show possible differences from the corresponding quantity in the model (\ref{eqn:timeseriresmodel}). To estimate $\phi_0^*$ and $\phi_1^*$, we may employ the ordinary least squares (OLS) method. Specifically, we minimize $S(\phi_0^*,\phi_1^*) = \sum_{t=2}^T(X_t^*-\phi_0^*-\phi_1^*X_{t-1}^*)^2$ with respective to $\phi_0^*$ and $\phi_1^*$, yielding the OLS estimators of $\phi_1^*$ and $\phi_0^*$:
\begin{flalign}\label{eqn:estimator-OLS}
    && \widehat{\phi_1^*} &= \frac{\sum_{t=2}^{T} (X_{t-1}^* -  \bar{X}_{(-1)}^*)(X_{t}^* -  \bar{X}^*)}{\sum_{t=2}^{T}(X_{t-1}^* -  \bar{X}_{(-1)}^*)^2},  && \nonumber\\
 \text{and}   && \widehat{\phi_0^*} &= \bar{X}_{t}^* - \widehat{\phi_1^*}  \bar{X}^*, &&
\end{flalign}
where $\bar{X}_{(-1)}^* = \frac{1}{T-1}\sum_{t=2}^{T}X_{t-1}^*$ and $\bar{X}^* = \frac{1}{T-1}\sum_{t=2}^{T}X_t^*$.  


\begin{theorem}\label{thm:theoremclassic1}
Let $\omega_1= \frac{\alpha_1^2\sigma_\epsilon^2}{\alpha_1^2\sigma_\epsilon^2+\sigma_e^2(1-\phi_1^2)}$, $\phi_1^* = \phi_1 \omega_1,$ and $\phi_0^* = \left(\alpha_0 + \frac{\alpha_1\phi_0}{1-\phi_1}\right) \left(1-\phi_1 \omega_1 \right)$. Assume the stationarity of the times series. If the measurement error process satisfies (\ref{eqn:measmodelClassic}), then
\begin{itemize}
    \item[(1)] $\widehat{\phi}_1^* \xrightarrow{\,\,\,p\,\,\,}\phi_1^*$ and $\widehat{\phi}_0^* \xrightarrow{\,\,\,p\,\,\,}\phi_0^*$ as $T \to \infty$,
    \item[(2)] $\epsilon_t^*=\alpha_0(1-\phi_1^*)+\alpha_1\phi_0-\phi_0^*+\alpha_1(\phi_1-\phi_1^*)X_{t-1}+(1-\phi_1^*)e_t+\alpha_1\epsilon_t$ for $t=1,\ldots,T$, \\
    and hence  $\text{Var}(\epsilon_t^*)=\phi_1^2\alpha_1^2 (1-\omega_1)^2 \left(\frac{\sigma_\epsilon^2}{1-\phi_1^2}\right) + (1- \omega_1\phi_1)^2 \sigma_e^2 + \alpha_1^2 \sigma_\epsilon^2$.
\end{itemize}
\end{theorem}

The proof of the theorem is included in Supplementary~Appendix~A.2. This theorem essentially implies that the naive estimator under the additive form in (\ref{eqn:measmodelClassic}) is inconsistent because $\phi_1^*\ne\phi_1$ and $\phi_0^*\ne\phi_0$. The naive estimator $\widehat{\phi}_1^*$ attenuates and the attenuation factor $\omega_1$ depends on the parameters $\alpha_1$ and $\sigma_e^2$ of the measurement error model (\ref{eqn:measmodelClassic}) as well as $\phi_1$ and $\sigma_\epsilon^2$ in the time series model (\ref{eqn:timeseriresmodel}).  The coefficient $\alpha_1$ in the measurement error model (\ref{eqn:measmodelClassic}) affects the estimation of the both naive estimators $\widehat{\phi}_1^*$ and $\widehat{\phi}_0^*$, while the intercept $\alpha_0$ influences the estimation of $\phi_0^*$ only, but not $\phi_1^*$ or $\text{Var}(\epsilon^*)$.

\begin{theorem}\label{thm:theoremmulti1}
Let $\omega_2 = \{1 + \sigma_u^2 +\frac{(1+\phi_1)\sigma_u^2\phi_0^2}{(1-\phi_1)\sigma_\epsilon^2}\}^{-1}$, $\phi_1^* = \phi_1 \omega_2$, and $\phi_0^* = \frac{\beta_0\phi_0}{1-\phi_1}\left(1-\omega_2 \phi_1\right)$. If the times series is stationary and the measurement error process satisfies (\ref{eqn:measmodelmultiplicative}), then
\begin{itemize}
    \item[(1)] $\widehat{\phi}_1^* \xrightarrow{\,\,\,p\,\,\,}\phi_1^*$ and $\widehat{\phi}_0^* \xrightarrow{\,\,\,p\,\,\,}\phi_0^*$ as $T \to \infty$,
    \item[(2)] $\epsilon_t^*=\beta_0\phi_0 u_t -\phi_0^* + \beta_0X_{t-1}(\phi_1 u_t - \omega_2\phi_1u_{t-1}) + \beta_0 u_t \epsilon_t$ for $t=1,\ldots,T$, \\ and hence $\text{Var}(\epsilon_t^*)= \beta_0^2\{\sigma_u^2\phi_0^2 + (1+\sigma_u^2)\sigma_\epsilon^2\}+\beta_0^2\phi_1^2\frac{(1+\omega_2^2)}{\omega_2}\frac{\sigma_\epsilon^2}{(1-\phi_1^2)}.$
\end{itemize}
\end{theorem}

The proof of the theorem is included in Supplementary~Appendix~A.3. This theorem says the attenuation effect resulting from the measurement error on estimation of $\phi_1$. The constant scaling parameter $\beta_0$ in the measurement error model (\ref{eqn:measmodelmultiplicative}) does not influence the estimation of $\phi_1$ but affects the estimation of $\phi_0$ and $\sigma_\epsilon^2$. The attenuation factor $\omega_2$ is determined by the magnitude $\sigma_u^2$ of measurement error as well as the values of $\phi_0$, $\phi_1$, and $\sigma_\epsilon^2$ of the time series model (\ref{eqn:timeseriresmodel}).

\subsection{ Naive Estimation and Bias for AR($p$) Model with $p\ge 2$}\label{sub:generalp}

We now extend the discussion in Section~\ref{sub:p=1} to the AR($p$) model with $p\ge 2$. Replacing $X_t$ with $X_t^*$ in (\ref{eqn:timeseriresmodel}) gives the working model
\begin{equation}\label{eqn:workingmodel-p}
    X_{t}^* =  \phi_0^* + \sum_{j=1}^p \phi_j^* X_{t-j}^* + \epsilon_t^*,
\end{equation}
where $\phi^*=(\phi_1^*,\ldots,\phi_p^*)^{\scalebox{0.5}{T}}$ and $\epsilon_t^*$ may differ from the corresponding symbol in (\ref{eqn:timeseriresmodel}). If mimicking the procedure of using (\ref{eqn:esteqntrue}) with $X_t$ replaced by $X_t^*$ to estimate $\phi^*$, $\phi_0^*$ and $\sigma_\epsilon^{2*}$ in (\ref{eqn:workingmodel-p}), then we let  $\widehat{\phi}^*=(\widehat{\phi}_1^*,\ldots,\widehat{\phi}_p^*)^{\scalebox{0.5}{T}}$, $\widehat{\phi}_0^*$ and $\widehat{\sigma}_\epsilon^{*2}$ denote the resultant estimators. Similar to $\widehat{\gamma}_k$ and $\widehat{\mu}$, we define $\widehat{\mu}^* = \frac{1}{T}\sum_{t=1}^T X_t^*$  and $\widehat{\gamma}_k^*=\frac{1}{T-k}\sum_{t=1}^{T-k}(X_t^*-\widehat{\mu}^*)(X_{t+k}^* -\widehat{\mu}^*)$ for $k=1,\ldots,p$.  Let $\widehat{\gamma}^*=(\widehat{\gamma}_1^*,\ldots,\widehat{\gamma}_p^*)^{\rm\tiny T}$ and $\widehat{\gamma}_0^*=\frac{1}{T}\sum_{t=1}^{T}(X_t^*-\widehat{\mu}^*)(X_t^*-\widehat{\mu}^*)$.

We now discuss the asymptotic results of the naive estimators under different measurement error models.

\begin{theorem}\label{thm:classic-dim_p}
Let $\mathds{1}_p$ be the $p\times 1$ unit and let $I_p$ be the $p\times p$ identity matrix. Define $\gamma^*= \alpha_1^2 \gamma$, $\gamma_0^* = \alpha_1^2 \gamma_0 + \sigma_e^2$, $\phi^* =  \alpha_1^2 (\alpha_1^2\Gamma+\sigma_e^2I_{p})^{-1}\gamma$, $\phi_0^* =  \left(1-\phi^*\cdot \mathds{1}_{p}\right)\left(\alpha_0 + \alpha_1 \mu \right)$ and $\sigma_\epsilon^{2*}= \alpha_1^2\gamma_0+\sigma_e^2 - \alpha_1^4 \gamma^{\rm\scalebox{0.5}{T}}\left(\alpha_1^2\Gamma+\sigma_e^2 I_{p}\right)^{-1}\gamma$.
Under regularity conditions, if the time series is stationary and the measurement error process satisfies (\ref{eqn:measmodelClassic}), then
\begin{itemize}
    \item[(1)] $\widehat{\gamma}^* \xrightarrow{\,\,\,p\,\,\,}\gamma^*$ and $\widehat{\gamma}_0^* \xrightarrow{\,\,\,p\,\,\,} \gamma_0^*$ \quad as $T \to \infty$. 
    \item[(2)] $\widehat{\phi}^* \xrightarrow{\,\,\,p\,\,\,}\phi^*$, $\widehat{\phi}_0^* \xrightarrow{\,\,\,p\,\,\,}\phi_0^*$, and $\widehat{\sigma}_\epsilon^{2*} \xrightarrow{\,\,\,p\,\,\,}\sigma_\epsilon^{2*}$ \quad  as $T \to \infty$.
    \item[(3)]  Let $Q_1$ denote the $(p+1)\times(p+1)$ asymptotic covariance matrix of $\sqrt{T} \left\{(\widehat{\gamma}_0^* ,\widehat{\gamma}^{*\rm\tiny T})^{\rm\tiny T} -({\gamma}_0^*, {\gamma}^{*\rm\tiny T})^{\rm\tiny T} \right\}$ as $T\to\infty$.  Then the elements of $Q_1$ are given by 
    \begin{align*}
        q_{100}^* &= \alpha_1^4 q_{00} + 4 \alpha_1^2 \gamma_0 \sigma_e^2 + E(e_t^4) - \sigma_e^4;\\
        q_{10p}^* &= \alpha_1^4 q_{0p} + 4 \alpha_1^2 \gamma_p \sigma_e^2;\\
        q_{1pr}^* &= \alpha_1^4 q_{pr} + 2\alpha_1^2 \sigma_e^2 (\gamma_{|p-r|}+\gamma_{p+r}) \text{ for } r \ne 0, r\ne p;\\
        q_{1pp}^* &= \alpha_1^4 q_{pp} + 2\alpha_1^2 \sigma_e^2 (\gamma_{0}+\gamma_{2p})  + \sigma_e^4;
    \end{align*}
    for $p\ge 1$, where  $q_{jk}$ is the $(j,k)$ element of the asymptotic covariance matrix of $(\widehat{\gamma}_0,\widehat{\gamma}^{\rm\tiny T})^{\rm\tiny T}$, given by \citep[Sec. 7.3]{brockwell1991time}
    \begin{equation}\label{eqn:brockwell-7.3.13}
        q_{jk} = (\eta-3) \gamma_j\gamma_k + \sum_{i=-\infty}^{\infty} (\gamma_i\gamma_{i-j+k}+\gamma_{i+k}\gamma_{i-j})
    \end{equation}
    for $(j,k)=(0,0),(0,p),(p,p)$ and $(p,r)$ with $r\ne 0$ and $r\ne p$, with $\eta = E(\epsilon_t^4)/\sigma_{\epsilon}^4$.
\end{itemize}
\end{theorem}

The proof of Theorem~\ref{thm:classic-dim_p} is presented in Supplementary~Appendix~A.4. Similar to the results in Theorem~\ref{thm:theoremclassic1}, the intercept $\alpha_0$ only influence $\phi_0$ and does not influence $\phi$. 



\begin{theorem}\label{thm:multipli-dim_p}
Let $\gamma^* =  \beta_0^2 \gamma$, $\gamma_0^* = \beta_0^2 \left\{(\sigma_u^2+1)\gamma_0 + \sigma_u^2\mu^2\right\}$, $\phi^* =  \left\{\Gamma + \sigma_u^2(\gamma_0+\mu^2) I_{p}\right\}^{-1}\gamma$, $\phi_0^* =  \beta_0  \left(1- \phi^{*\scalebox{0.5}{\rm T}}\cdot \mathds{1}_{p}\right) \mu$, and $\sigma_\epsilon^{2*} = \beta_0^2(\sigma_u^2+1)\gamma_0 + \beta_0^2\sigma_u^2\mu^2 - \beta_0^2 \gamma^{\rm\scalebox{0.5}{T}}\left\{\Gamma + \sigma_u^2(\gamma_0+\mu^2) I_{p}\right\}^{-1}\gamma$. Under regularity conditions, if the time series are stationary and the measurement error process satisfy (\ref{eqn:measmodelmultiplicative}), then
\begin{itemize}
    \item[(1)] $\widehat{\gamma}^* \xrightarrow{\,\,\,p\,\,\,}\gamma^*$ and $\widehat{\gamma}_0^* \xrightarrow{\,\,\,p\,\,\,} \gamma_0^*$ \quad as $T \to \infty$.
    \item[(2)] $\widehat{\phi}^* \xrightarrow{\,\,\,p\,\,\,}\phi^*$, $\widehat{\phi}_0^* \xrightarrow{\,\,\,p\,\,\,}\phi_0^*$, and $\widehat{\sigma}_\epsilon^{2*} \xrightarrow{\,\,\,p\,\,\,}\sigma_\epsilon^{2*}$ \quad as $T \to \infty$.
    
    \item[(3)] Let $Q_2$ denote the $(p+1)\times(p+1)$ asymptotic covariance matrix of $\sqrt{T} \left\{(\widehat{\gamma}_0^* ,\widehat{\gamma}^{*\rm\tiny T})^{\rm\tiny T} -({\gamma}_0^*, {\gamma}^{*\rm\tiny T})^{\rm\tiny T} \right\}$ as $T\to\infty$.  Then the elements of $Q_2$ are given by 
    {\footnotesize 
    \begin{align*}
        q_{200}^* &= \beta_0^4(\sigma_u^2+1)^2q_{00} +  \beta_0^4\{E(u_t^4)-(\sigma_u^2+1)^2\}E(X_t-\mu)^4 \nonumber \\
    & \qquad + 4\mu\beta_0^4\sigma_u^2(\sigma_u^2+1) v_{0} + 4\mu\beta_0^4\{E(u_t^4) - E(u_t^3)-\sigma_u^2(\sigma_u^2+1)\} 
    E(X_t-\mu)^3 \nonumber \\
    & \qquad+ 2\mu^2\beta_0^4 \left\{ E(u_t^4)-2E(u_t^3) +1 - \sigma_u^4  \right\} \gamma_0 \nonumber \\
    & \qquad + 4\mu^2\beta_0^4 \left[\sigma_u^4 \sum_{h=-\infty}^\infty \gamma_h + \left\{ E(u_t^4)-2E(u_t^3) + \sigma_u^2 +1 - \sigma_u^4 \right\} \gamma_0\right] + \mu^4\beta_0^4 \left[E\{(u_t-1)^4\} - \sigma_u^4\right]; \\
       q_{20p}^* &=\beta_0^4 q_{p} (\sigma_u^2+1) + \beta_0^4 \left\{E(u_t^3)-(\sigma_u^2+1)\right\} \left[ E\{(X_t-\mu)^3(X_{t+p}-\mu)\} + E\{(X_t-\mu)^3(X_{t-p}-\mu)\}\right] \\
   & \qquad + 2\mu\beta_0^4\sigma_u^2 v_{0p} + \mu\beta_0^4 E\{3u_t^3 - 3u_t^2-2\sigma_u^2\}\left[E\{(X_t-\mu)^2(X_{t-p}-\mu)\}+E\{(X_t-\mu)^2(X_{t+p}-\mu)\}\right] \\
   & \qquad + 6\mu^2\beta_0^4 E(u_t-1)^3 \gamma_p + 4\mu^2\beta_0^4\sigma_u^2 \gamma_p;\\
       q_{2pr}^* &= \beta_0^4 q_{pr}  + \beta_0^4 \sigma_u^2 \left[ E\{(X_t-\mu)^2(X_{t+p}-\mu)(X_{t+r}-\mu)\} +   E\{(X_t-\mu)(X_{t+p}-\mu)^2(X_{t+p+r}-\mu)\} \right. \nonumber \\
   &\qquad \left. + E\{(X_{t-r}-\mu)(X_{t}-\mu)^2(X_{t+p}-\mu)\} +   E\{(X_t-\mu)(X_{t+p-r}-\mu)(X_{t+p}-\mu)^2\} \right] \nonumber \\ 
    &\qquad + \mu \beta_0^4 \sigma_u^2 \left[ E\{(X_t-\mu)(X_{t+p}-\mu)(X_{t+r}-\mu)\} +   E\{(X_t-\mu)(X_{t+p}-\mu)(X_{t+p+r}-\mu)\} \right. \nonumber \\
   &\qquad \left. + E\{(X_{t-r}-\mu)(X_{t}-\mu)(X_{t+p}-\mu)\} +   E\{(X_t-\mu)(X_{t+p-r}-\mu)(X_{t+p}-\mu)\} \right] \nonumber \\  
   &\qquad + 2\mu^2\beta_0^4\sigma_u^2(\gamma_{|p-r|}+\gamma_{p+r}) \text{ for }  r\ne p, r\ne 0;\\
        q_{2pp}^* &= \beta_0^4 q_{pp}  + \beta_0^4 (\sigma_u^4 + 2\sigma_u^2)  \text{Var}\{(X_t-\mu)(X_{t+p}-\mu)\} + 2\beta_0^4  E\{(X_t-\mu)(X_{t+p}-\mu)^2(X_{t+2p}-\mu)\} \\
   &\qquad + \mu \beta_0^4 \sigma_u^2 \left[ E\{(X_t-\mu)(X_{t+p}-\mu)^2\} +   2E\{(X_t-\mu)(X_{t+p}-\mu)(X_{t+2p}-\mu)\}\right.  \nonumber \\  
   &\qquad + \left.  E\{(X_t-\mu)^2(X_{t+p}-\mu)\} \right] + 2\mu^2\beta_0^4 \sigma_u^4 \gamma_p + 2\mu^2\beta_0^4 \sigma_u^2 (\gamma_0 + \gamma_{2p}) + \mu^4\beta_0^4\sigma_u^4; \nonumber 
     \end{align*}
    }
   where the $q_{jk}$ are given by  (\ref{eqn:brockwell-7.3.13}),  for $(j,k)=(0,0),(0,p),(p,p)$ and $(p,r)$ with $r\ne 0$ and $r\ne p$, and $v_p = \lim_{T\to\infty}\frac{1}{T}\sum_{t=1}^T\sum_{s=1}^TE\{(X_t-\mu)(X_{t+p}-\mu)(X_s-\mu)\}$.
\end{itemize}
\end{theorem}



The proof of the theorem is presented in Supplementary~Appendix~A.5. The multiplicative measurement error $u_t$ contributes to the biasedness of the parameter estimation for $\phi$, while the scaling parameter $\beta_0$ has no effects on the naive estimator $\widehat{\phi}^*$.

\section{Methodology of Correcting Measurement Error Effects}\label{sec:Methodology}
\subsection{Estimation of Model Parameters}

In the presence of measurement error, measurements of the $X_t$ are not always available but surrogate measurements $X_t^*$ are available. It may be tempting to conduct a naive analysis by implementing (\ref{eqn:esteqntrue}) with the $X_t$ replaced by the $X_t^*$, or equivalently with $\widehat{\mu}$ and $\widehat{\gamma}_k$ replaced by $\widehat{\mu}^*$ and the $\widehat{\gamma}_k^*$, respectively, to find estimators of $\phi$, $\phi_0$ and $\sigma_\epsilon^2$. However, by Theorems~\ref{thm:classic-dim_p}--\ref{thm:multipli-dim_p}, such a procedure typically yields biased estimators. In this section, we develop new estimators accounting for the measurement error effects described by either the additive model (\ref{eqn:measmodelClassic}) or the multiplicative model (\ref{eqn:measmodelmultiplicative}). 

Our idea is still to employ (\ref{eqn:esteqntrue}) to find consistent estimators of $\phi$, $\phi_0$ and $\sigma_\epsilon^2$, but instead of replacing $\widehat{\mu}$ and the $\widehat{\gamma}_k$ with $\widehat{\mu}^*$ and the $\widehat{\gamma}_k^*$ as in the naive analysis, we replace  $\widehat{\mu}$ and the $\widehat{\gamma}_k$ in (\ref{eqn:esteqntrue}) with new functions of the $X_t^*$, denoted as $\widetilde{\mu}$ and the $\widetilde{\gamma}_k$, which adjust for the measurement error effects. Specifically, if we can find $\widetilde{\mu}$ and the $\widetilde{\gamma}_k$ such that they resemble $\widehat{\mu}$ and the $\widehat{\gamma}_k$ in the sense that as $T \to \infty$,
\begin{flalign}\label{eqn:tilde-hat-equal-inp}
    && & \widetilde{\mu} \text{ and }\widehat{\mu}\text{ have the same limit in probability,} && \nonumber\\
\text{and } && & \widetilde{\gamma}_k \text{ and }\widehat{\gamma}_k\text{ have the same limit in probability for } k=0,\ldots,p,&& 
\end{flalign}
then substituting $\widehat{\mu}$ and the $\widehat{\gamma}_k$ with $\widetilde{\mu}$ and the $\widetilde{\gamma}_k$ in (\ref{eqn:esteqntrue}) yields consistent estimators of $\phi$, $\phi_0$ and $\sigma_\epsilon^2$.

With the availability of the $\widetilde{\gamma}_k$ satisfying (\ref{eqn:tilde-hat-equal-inp}), let $\widetilde{\Gamma}$ denote $\Gamma$ with the $\gamma_k$ replaced by the $\widetilde{\gamma}_k$. Then provided regularity conditions, consistent estimators of $\phi$, $\phi_0$ and $\sigma_\epsilon^2$ can be obtained by solving the estimating equations for $\phi$, $\phi_0$, and $\sigma_\epsilon^2$:
\begin{align}\label{eqn:correctmodelp}
    \phi &= \widetilde{\Gamma}^{-1}\widetilde{\gamma}, \nonumber\\
    \phi_0 &= \left(1-\sum_{i=1}^p \phi_i\right) \widetilde{\mu}, \\
    \sigma_\epsilon^2 &= \widetilde{\gamma}_0 - 2\phi^{\rm\scalebox{0.5}{T}}\widetilde{\gamma} + \phi^{\rm\scalebox{0.5}{T}}\widetilde{\Gamma}\phi. \nonumber
\end{align}

It is immediate to obtain the following result.

\begin{theorem}\label{thm:Propose-method}
Assume regularity conditions hold and the time series are stationary. If $\widetilde{\mu}$ and the $\widetilde{\gamma}_k$ are functions of the $X_t^*$ with $t = 1, \ldots, T$ and they satisfy (\ref{eqn:tilde-hat-equal-inp}), and let $\widetilde{\phi}$, $\widetilde{\phi}_0$, and $\widetilde{\sigma_\epsilon}^2$ denote the estimators for $\phi$, $\phi_0$ and $\sigma_\epsilon^2$, respectively, obtained by solving (\ref{eqn:correctmodelp}).  Then, as \qquad $T \to \infty$  
\begin{itemize}
    \item[(1)] $\widetilde{\phi} \xrightarrow{\,\,p\,\,} \phi$, $\widetilde{\phi_0} \xrightarrow{\,\,p\,\,} \phi_0$, and $\widetilde{\sigma_\epsilon}^2 \xrightarrow{\,\,p\,\,} \sigma_\epsilon^2$;
    \item[(2)] $             \sqrt{n} (\widetilde{\phi}-\phi) \xrightarrow{\,\,\, d\,\,\,} N(0,GQG^{\rm\tiny T}),$ \\
    where $G$ is the matrix of derivatives of $\widetilde{\phi}$ with respect to the components of $(\widehat{\gamma}_0^*,\widehat{\gamma}^{*\scalebox{0.5}{\rm T}})^{\scalebox{0.5}{\rm T}}$. Here $Q=Q_1$, the matrix in Theorem~\ref{thm:classic-dim_p}, if measurement error follows the model (\ref{eqn:measmodelClassic}); and $Q=Q_2$, the matrix in Theorem~\ref{thm:multipli-dim_p}, if measurement error follows the model (\ref{eqn:measmodelmultiplicative}).
\end{itemize}
\end{theorem}

Now we discuss explicitly how to determine $\widetilde{\mu}$ and the $\widetilde{\gamma}_k$ under the measurement error model (\ref{eqn:measmodelClassic}) or (\ref{eqn:measmodelmultiplicative}). With (\ref{eqn:measmodelClassic}), take  $\widetilde{\mu}=\frac{\widehat{\mu}^*}{\alpha_1}-\alpha_0$, $\widetilde{\gamma}_0 = \frac{1}{\alpha_1^2}(\widehat{\gamma}_0^*-\sigma_e^2) $, and $\widetilde{\gamma}_k = \frac{\widehat{\gamma}_k^*}{\alpha_1^2}$ for $k=1,\ldots,p$. 
With (\ref{eqn:measmodelmultiplicative}), take $\widetilde{\mu}=\frac{\widehat{\mu}^*}{\beta_0}$, $\widetilde{\gamma}_0=\frac{\gamma_0^*}{(1+\sigma_u^2)\beta_0^2}-\frac{\sigma_u^2\mu^2}{\sigma_u^2+1}$, and $\widetilde{\gamma}_k = \frac{\widehat{\gamma}_k^*}{\beta_0^2}$ for $k=1,\ldots,p$. 
By the results in Theorem~\ref{thm:classic-dim_p}(1) and Theorem~\ref{thm:multipli-dim_p}(1), it can be easily verified that these $\widetilde{\mu}$ and the $\widetilde{\gamma}_k$ satisfy (\ref{eqn:tilde-hat-equal-inp}).

We conclude this section with a procedure of estimating the asymptotic covariance matrix for the estimator $\widetilde{\phi}$. While Theorem~\ref{thm:Propose-method} presents the sandwich form of the asymptotic covariance matrix of $\widetilde{\phi}$, its evaluation involves lengthy calculations. We may alternatively employ the block bootstrap algorithm \citep{lahiri1999theoretical} to obtain variance estimates for $\widetilde{\phi}$ using the following steps. Firstly, we set a positive integer, say $N$, as the number for the bootstrap sampling; $N$ can be set as a large number such as 1000. Next, we repeat through the following five steps:
\begin{itemize}[leftmargin=1in]
    \item [Step 1:] At iteration $n \in \{1,\ldots,N\}$, we initialize a null time series $X^{(n,0)}$ of dimension~0 and specify a block length, say $b$, which is an integer between 0 and $T$. Initialize $m$=1.
    \item [Step 2:] Sample an index, say $i$, from $\{0,\ldots,T-b\}$, and then define $X_{\rm\tiny add}^{(m-1)}=\{X_{i+1},\ldots,X_{i+b}\}$.
    \item [Step 3:]  Update the previous time series $X^{(n,m-1)}$ by appending $X_{\rm\tiny add}^{(m-1)}$ to it, and let $X^{(n,m)}$ denote the new time series.
    \item [Step 4:] If the dimension $X^{(n,m)}$ is smaller than $T$ then return to Steps~2~and~3; otherwise drop the elements in the time series with the index greater than $T$ to ensure the dimension of $X^{(n,m)}$ is identical to $T$ and then go to Step 5.
    \item [Step 5:] Obtain an estimate $\widetilde{\phi}^{(n)}$ of parameter $\phi$ by applying the times series $X^{(n,m)}$ to (\ref{eqn:correctmodelp}). If $n<N$, then set $n$ to be $n+1$ and go back to Step 1 to repeat; otherwise stop.
\end{itemize}

Let $\bar{\widetilde{\phi}}^{(n)}=\frac{1}{N}\sum_{n=1}^N\bar{\widetilde{\phi}}^{(n)}$ be the sample mean. The bootstrap variance of $\widetilde{\phi}$ is then given by,
\begin{equation*}
    \text{Var}_{\rm boot}(\widetilde{\phi}) = \frac{1}{N}\sum_{n=1}^N(\widetilde{\phi}^{(n)} - \bar{\widetilde{\phi}}^{(n)})^2.
\end{equation*}

\subsection{Forecasting and Prediction Error}\label{sub:Forecasting}

Forecasting is an important application of the autoregressive models.  Specifically, in forecasting based on the observed time series $X_{(T)}=\{x_1,\ldots,x_{T}\}$, we are interested in the predictions of $\{X_{T+1},\ldots,X_{T+H}\}$ for a positive integer $H$, which is done one by one starting from the nearest time point $T+1$ to the farthest time point $T+H$. To this end, let $h=1,\ldots,H$, the $h$-step forecasting of $X_{T+h}$ is based on its history of lag-$p$, $\{X_{T+h-1},\dots,X_{T+h-p}\}$, by using the conditional expectation $E(X_{T+h}|x_{T+h-1},\ldots,x_{T+h-p})$, denoted $\widehat{X}_{T+h}$, where for $j=T+h-1,\ldots,T+h-p$, $x_j$ is the observe value of $X_{j}$ if $j\leq T$; and $x_j$ is the predicted value of $X_{j}$, $\widehat{X}_{j}$, if $j > T$. This prediction minimizes the squared prediction error $E(\widehat{X}_{T+h}-X_{T+h})^2$ \citep[e.g.,][p.131]{box2015time}. 

If no measurement error is involved, due to the zero mean of the random error term $\epsilon_t$ in the AR$(p)$ model (\ref{eqn:timeseriresmodel}), for $h=1,\ldots, H$, the conditional expectation can be calculated by
\begin{equation}\label{eqn:predictionequation}
  \widehat{X}_{T+h}   = \phi_0+ \phi_1 x_{T+h-1} + \ldots +\phi_p x_{T+h-p}.
\end{equation}

When measurement error appears, the observe values $x_{j}$ for $j=T,\ldots,T-p+1$ in (\ref{eqn:predictionequation}) are no longer available but their surrogates $X_{j}^*$ are available. We now provide a sensible estimate of $X_j$ by using the measurement error model for characterizing the relationship of $X_j$ and $X_j^*$. If measurement error follows (\ref{eqn:measmodelClassic}), we ``estimate" ${X}_j$ by
\begin{equation}\label{eqn:pred1}
    \widehat{X}_j = \frac{1}{\alpha_1}(X_j^*-\alpha_0) \qquad \text{ for } j=t,\ldots,t-p+1;
\end{equation}
if the measurement error follows (\ref{eqn:measmodelmultiplicative}), then $\widehat{X}_j$ is ``estimated"  by
\begin{equation}\label{eqn:pred2}
    \widehat{X}_j = \frac{X_j^*}{\beta_0} \qquad \text{ for } j=t, \ldots, t-p+1.
\end{equation}
These ``estimates" are unbiased in the sense that $E(\widehat{X}_j)=X_j$ for $j=t,\ldots,t-p+1$. Consequently, for $h=1,\ldots,H$, $X_{T+h}$ is predicted as 
\begin{equation}\label{eqn:equationpredpred}
    \widehat{X}_{T+h}=\phi_0+\phi_1 \widehat{X}_{T+h-1} + \cdots + \phi_p \widehat{X}_{T+h-p}. 
\end{equation}

In contrast to the observed values $\{x_{T},\ldots,x_{T-p+1}\}$, also referred to as the initial values of the forecasting of $X_{T+1},\ldots,X_{T+H}$, the estimates determined by  (\ref{eqn:pred1}) or (\ref{eqn:pred2}) introduce additional prediction error which should be characterized. Without the loss of generality, we consider $p=1$ to illustrate the recursive calculation of the prediction error; the prediction error with higher orders of the autoregressive process can be derived recursively in a similar way but with more complex expressions.

If the measurement error follows (\ref{eqn:measmodelClassic}), the mean squared prediction error of the 1-step prediction is given by
\begin{align*}
 {\rm P}_{\rm e}^{(1)} &=  E(\widehat{X}_{T+1}-X_{T+1})^2 \\
  & = E\{(\phi_0 + \phi_1 \widehat{X}_T) - (\phi_0 +\phi_1 X_{T} + \epsilon_{T+1})\}^2 \\
  & = E\left\{\phi_1\left(X_t+\frac{e_T}{\alpha_1}\right) - \phi_1 X_{T} - \epsilon_{T+1}\right\}^2 \\
  & = \frac{\phi_1^2\sigma_e^2}{\alpha_1^2} + \sigma_\epsilon^2, 
\end{align*}
where the last step is due to the independence between $e_t$ and $\epsilon_{t+1}$, as well as $E(e_t^2)=\sigma_e^2$ and $E(\epsilon_{t}^2)=\sigma_\epsilon^2$.

Then, the $h$-step prediction error is given  by 
\begin{align}\label{eqn:MSPE_add}
     {\rm P}_{\rm e}^{(h)} &=  E(\widehat{X}_{T+h}-X_{T+h})^2 \nonumber\\
  & = E\left\{\phi_1\left(\widehat{X}_{T+h-1}-X_{T+h-1}\right) - \epsilon_{T+1}\right\}^2 \nonumber\\
  & = \phi_1^2 {\rm P}_{\rm e}^{(h-1)}  + \sigma_\epsilon^2 \nonumber\\
  & = \frac{\phi_1^{2h}\sigma_e^2}{\alpha_1^2} + \sum_{i=0}^{h-1}\phi_1^{2i}\sigma_\epsilon^2,
\end{align}
where the last step comes from the recursive evaluation of $P_e^{(h-1)}$.

Similarly, if the measurement error follows (\ref{eqn:measmodelmultiplicative}), the mean squared prediction error is given by 
\begin{align*}
 {\rm P}_{\rm e}^{(1)} &=  E(\widehat{X}_{T+1}-X_{T+1})^2 \\
  & = \phi_1^2 \left\{\frac{\sigma_\epsilon^2}{1-\phi_1^2}+\mu^2\right\} \sigma_u^2 + \sigma_\epsilon^2, 
\end{align*}
where we use the independence of $\epsilon_{t+1}$, $u_t$ and $X_t$, $E(u_t)=1$, and  $\text{Var}(X_t)=\frac{\sigma_\epsilon^2}{1-\phi_1^2}$ due to the stationary AR(1) process. Hence, 
\begin{align}\label{eqn:MSPE_multi}
     {\rm P}_{\rm e}^{(h)} &=  E(\widehat{X}_{T+h}-X_{T+h})^2 \nonumber\\
  & = E\left\{\phi_1\left(\widehat{X}_{T+h-1}-X_{T+h-1}\right) - \epsilon_{T+1}\right\}^2 \nonumber\\
  & = \phi_1^2 {\rm P}_{\rm e}^{(h-1)}  + \sigma_\epsilon^2 \nonumber\\
  & = \phi_1^{2h-2} {\rm P}_{\rm e}^{(1)} + \sum_{i=0}^{h-2}\phi_1^{2i}\sigma_\epsilon^2  \nonumber\\
  & = \phi_1^{2h} \left\{\frac{\sigma_\epsilon^2}{1-\phi_1^2}+\mu^2\right\} \sigma_u^2 + \sum_{i=0}^{h-1}\phi_1^{2i}\sigma_\epsilon^2.
\end{align}

The evaluation of the mean squared prediction error ${\rm P}_{\rm e}^{(h)}$ is carried out by replacing the parameters with their estimators. We comment that the common second term in (\ref{eqn:MSPE_add}) and (\ref{eqn:MSPE_multi}), $\sum_{i=0}^{h-1}\phi_1^{2i}\sigma_\epsilon^2$, is the mean squared prediction error for the AR(1) model for error-free settings \citep[e.g.][p.152]{box2015time}, which equals  $\frac{1-\phi_1^{2h}}{1-\phi_1^2}\sigma_\epsilon^2$.

For an $\alpha$ with $0<\alpha<1$, then $h$-step $(1-\alpha)$-prediction interval is constructed as 
\begin{equation*}
    \left[\widehat{X}_{T+h} - q_{\frac{\alpha}{2}} {\rm P}_{\rm e}^{(h)} , \widehat{X}_{T+h} + q_{\frac{\alpha}{2}} {\rm P}_{\rm e}^{(h)}\right],
\end{equation*}
where $q_{\frac{\alpha}{2}}$ the $\alpha$-level quantile of the distribution of $\widehat{X}_{T+h} - X_{T+h}$. In practice, under normal assumption of $\epsilon_t$ and $e_t$, one can take $q_{\frac{\alpha}{2}}$ to be the $\alpha$-level quantile of the standard normal distribution \citep[][p.108]{brockwell2016introduction}.


\section{Analysis of COVID-19 Death Rates} \label{sec:Canada}

\subsection{Study Objective}
Using Canadian provincial COVID-19 data containing the daily confirmed cases and deaths from April 3, 2020 to May 4, 2020, we compare the times series of death rates for British Columbia, Ontario, Quebec, and Alberta, the four provinces in Canada which experience severe situations. The daily confirmed cases and fatalities are taken from ``1Point3Acres.com" (\url{https://coronavirus.1point3acres.com/}). 

In epidemiology, the mortality rate, defined as the proportion of cumulative deaths of the disease in the total number of people diagnosed with the disease \citep{kanchan2015mortality}, is often used to measure the severeness of an infectious disease. For COVID-19, determining the mortality rate is not trivial due to the difficulty in precisely determining the number of infected cases. Due to the limited test capacity, individuals with light symptoms are not being tested. Asymptomatic infections and the incubation period make it difficult to acquire an accurate number of infections. To circumvent this,  we explore different definitions of death rates. {\it Definition 1} is from \citet{baud2020real} who estimated mortality rates by dividing the number of deaths on a given day by the number of patients with confirmed COVID-19 infection 14 days before, with the consideration of the maximum incubation time to be 14 days. On the other hand, the median time from symptom onset to intensive care unit admission is about 10 days \citep[\text{[3]} in][]{baud2020real},  so we consider {\it Definition 2} 
which is the number of deaths of COVID-19 on day $t$ divided by the number of confirmed cases at day $(t-10)$. In comparison, we also consider {\it Definition 3} by calculating the death rate on day $t$ as the ratio of the number of deaths on day $t$ to the number of confirmed cases on day $t$.

While the first two ways may help more reasonably estimate mortality rates than the third definition, these calculated rates still differ from the true mortality rates because of under-reported cases which are primarily due to limited test capacity and undetected asymptomatic infections. To reflect the discrepancy between the {\it reported} and the {\it true} mortality rates for each province, for each definition of the mortality rate, we let $X_{1,t}$, $X_{2,t}$, $X_{3,t}$, and $X_{4,t}$, represent the {\it true} mortality rate on day $t$ for British Columbia, Ontario, Quebec and Alberta, respectively; and let $X_{1,t}^*$, $X_{2,t}^*$, $X_{3,t}^*$ and $X_{4,t}^*$ denote the {\it reported} mortality rate on day $t$ in British Columbia, Ontario, Quebec and Alberta, respectively. The objective is to use the reported mortality rates $\{X_{it}^*:t=1,\ldots,31\}$ to infer the true mortality rates $X_{i,t}$ which are modeled by (\ref{eqn:timeseriresmodel}) separately for $i=1,\ldots,4$. In addition, we want to forecast the true mortality rate of COVID-19 for a future time period. Due to the undetected asymptomatic cases and untested cases for light symptoms, the reported mortality rates $X_{i,t}^*$ are typically overestimated (i.e., $X_{i,t}^*\ge X_{i,t}$) for $i=1,\ldots,4$. As there is no exact information to guide us how to characterize the relationship between $X_{it}^*$ and $X_{it}$, here we conduct sensitivity studies by considering measurement error model (\ref{eqn:measmodelClassic}) or (\ref{eqn:measmodelmultiplicative}). We use the observed data $X_{i,t}^*$ from April 3, 2020 to May 4, 2020, i.e., $\{X_{i,t}^*:t=1,...,T_i\}$ with $T_1=T_2=31$, to estimate the model parameters in (\ref{eqn:timeseriresmodel}) with measurement error effects accounted for, and then forecast the mortality rate of COVID-19, from May 5, 2020 to May 9, 2020, in British Columbia, Ontario, Quebec and Alberta, Canada.

\subsection{Models Building}\label{sub:dataAnalysis}

\begin{figure}[!p]
\centering
\includegraphics[width=1\textwidth]{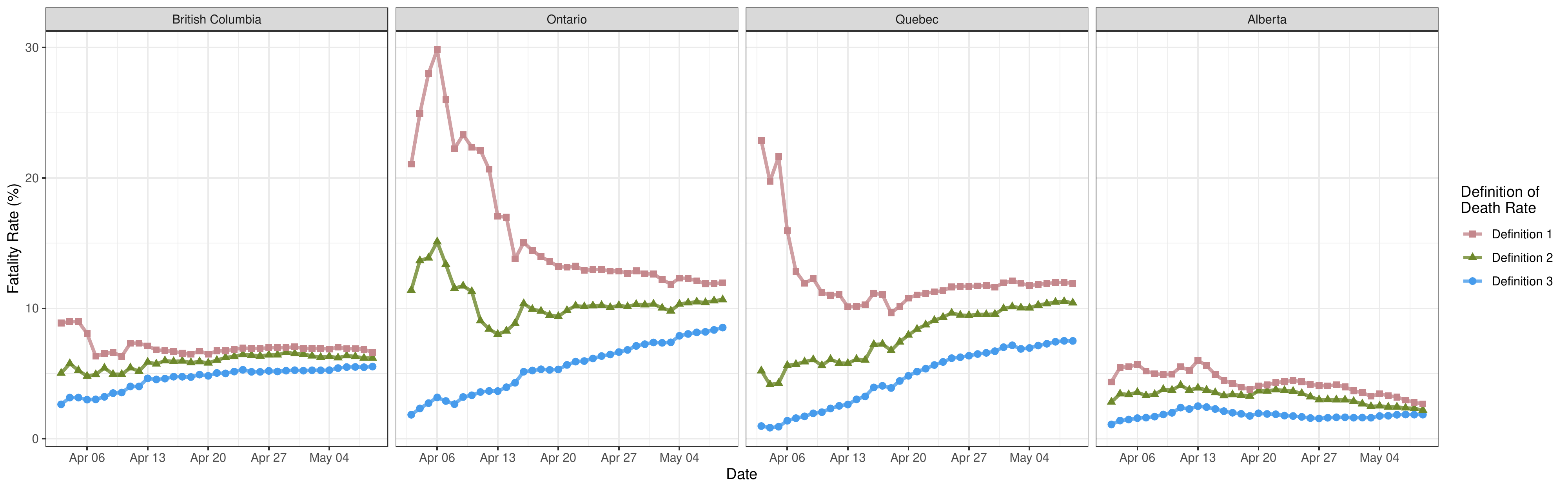}
\caption{ The time series plots of the death rate with different definitions}\label{fig:all-reportdeath}
\end{figure}

Figure~\ref{fig:all-reportdeath} displays the trajectory of the mortality rates of COVID-19 in the four provinces that are obtained from the three definitions. To assess the stationarity of the $X_{it}^*$, we conduct the augmented Dickey–Fuller (ADF) tests \citep{cheung1995lag} to times series $\{X_{i,t}^*:t=1,\ldots, T\}$, or its differencing transformation $\{X_{i,(t+1)}^*-X_{i,t}^*:t=1,\ldots,T\}$ for $i=1,\ldots, 4$ in each definition. 
Supplementary~Table~\ref{tab:ADFtest3} presents the test statistics and $p$-value of the ADF test for each time series, where ``TSV" represents a test statistics value.

\begin{placehere}
Figure~\ref{fig:all-reportdeath}
\end{placehere}


To determine the lag value $p$ for the autoregression model (\ref{eqn:timeseriresmodel}) used for the time series $\{X_{i,t}:t=1,...,T_i\}$ with $T_1=T_2=31$ for $i=1,\ldots,4$,  we fit the naive model (\ref{eqn:workingmodel-p}) with $\epsilon_t^*$ assumed to follow a normal distribution $N(0,\sigma_\epsilon^{*2})$, and use the AIC criterion by minimizing 
\begin{equation}\label{eqn:AIC}
 -2 \sum_{t=p}^T \log f(x_t^*|x_{t-1}^*,\ldots,x_{t-p}^*) + 2p,
\end{equation}
where $f(x_t^*|x_{t-1}^*,\ldots,x_{t-p}^*)$ is the conditional probability of $X_t^*$ given $X_{t-1}^*,\ldots,X_{t-p}^*$. The results are summarized in Supplementary~Table~\ref{tab:deathdef}, where no-differencing or 1-differencing is applied, the entries with ``-" indicate that the corresponding model is not applicable due to the ADF test results. 

We take those lag values for  an AR($p$) model to feature the true mortality rate $X_{i,t}$ for each definition and $i=1,\ldots,4$. 
To be specific, for the British Columbia data, with Definition~1 we consider two models: AR(1) model for the time series with 1-order differencing and AR(2) model for the time series with no-differencing; with Definitions~2~and~3, we consider AR(2) and AR(1) models, respectively, for the time series with 1-order differencing. For the Ontario data, we consider AR(1) and AR(4) for the time series with 1-order differencing in Definitions~1~and~3, respectively, and AR(2) for Definition 2 with no transformation. For the Quebec data, we consider AR(1) and AR(2) models for the times series with 1-order differencing in Definitions~1~and~2, respectively. For Alberta data, we consider an AR(1) model for the times series with 1-order differencing for both Definitions~1~and~2.


\subsection{Sensitivity Analyses}\label{sub:sensianalysis}

As there are no additional data available for estimating the parameters for the model (\ref{eqn:measmodelClassic}) or (\ref{eqn:measmodelmultiplicative}), we conduct sensitivity analyses using the findings in the literature. Different studies showed different estimates of the asymptomatic infection rates, changing from 17.9\% to 78.3\% \citep{kimball2020asymptomatic,day2020covid}. To accommodate the heterogeneity of different studies, \cite{he2020estimation} carried out a meta-analysis and obtained an estimate of the asymptomatic infection rate to be 46\%. If under-reported confirmed cases are only caused from undetected asymptomatic cases, then $X_t=(1-\tau_{A})X_t^*$, or equivalently, 
\begin{equation}\label{eqn:measCanadadata}
X_t^*=\frac{1}{1-\tau_{A}}X_t,    
\end{equation}
where $\tau_{A}$ represents the rate of asymptomatic infections. 

Now we use (\ref{eqn:measCanadadata}) as a starting point to conduct sensitivity analyses. In the multiplicative model (\ref{eqn:measmodelmultiplicative}), we take $\beta_0u_t=\frac{1}{1-\tau_{A}}$. With $E(u_t)=1$, we set $\beta_0=\frac{1}{1-\tau_{A}}$ by setting $\tau_{A}=46\%$, the value from the meta-analysis of \citet{he2020estimation}. To see different degrees of error, we consider $\sigma_u^2$ to take a small value, say $\sigma_{u1}^2$, and a large value, say, $\sigma_{u2}^2$, which is alternatively reflected by the change of the coefficient of variation, $CV=\frac{\sigma_u}{E(u_t)}$, of the error term $u_t$ from $\sigma_{u1}\times100\%$ to $\sigma_{u2}\times100\%$.

When using the additive model (\ref{eqn:measmodelClassic}) to characterize the measurement error process, motivated by (\ref{eqn:measCanadadata}), we set $\alpha_0=0$ and $\alpha_1=\frac{1}{1-46\%}$, and let $\sigma_e^2$ take a small value, say $\sigma_{e1}^2$, and a large value, say $\sigma_{e2}^2$, to feature an increasing degree of measurement error. Due to the constraints for the parameters discussed for (\ref{apx:gamma0star}) and (\ref{apx:Thm4-gamma0star}), we set the values for $\sigma_{u1}$, $\sigma_{u2}$, $\sigma_{e1}$, and $\sigma_{e2}$ case by case for each definition and for each province, which are recorded in Table~\ref{tab:par-sigmas}.


The model fitting results are reported in Tables~\ref{tab:par-Def-1}--\ref{tab:par-Def-2} and Supplementary~Table~\ref{tab:par-Def-3} for the three definitions of mortality rates, where the point estimates (EST), the associated standard errors (SE), and the p-values for the model parameters are included. Table~\ref{tab:par-Def-1} shows that with Definition~1, the estimates of $\phi_0$ in the absolute value from the proposed method are smaller than those of naive method, while the estimates of $\phi_1$ produced from the proposed and naive methods exhibit an opposite direction. As expected, the standard errors for the proposed method are generally larger than those of the naive method. However, both methods find no evidence to support that $\phi_0$ and $\phi_1$ are different from zero for the data of British Columbia and Ontario, suggesting that the mortality rates of these two provinces remain statistically unchanged. At the significance level 0.1, the naive method and the proposed method show different evidence for the data of Quebec and Alberta. The naive method suggests a likely downward trend with p-value 0.071 and 0.061 for testing of $\phi_0$ for Quebec and Alberta, respectively. The proposed method, on the other hand, show that $\phi_0$ is insignificant for these two provinces.


Table~\ref{tab:par-Def-2} displays the results for Definition~2. For the British Columbia data, the estimates of the three parameters $\phi_1$, $\phi_2$ and $\phi_3$ produced from the proposed method are smaller than those yielded from the naive method, whereas the standard errors output from the proposed method are larger than those from the naive method. However, at the significance level 0.05, both methods find no evidence to show the significance of $\phi_0$, $\phi_1$ and $\phi_2$, suggesting that the mortality rate of British Columbia remain unchanged with time. Similar findings are revealed for the Alberta data except that the parameter estimates output from the proposed method are larger than those produced from the naive method. For the Ontario and Quebec data, the revealings from the two methods are quite different. For Ontario, both methods show that $\phi_0$ is insignificant and $\phi_1$ is significant. The evidence of $\phi_2$, however, depends on the nature of measurement error. On the contrary, the findings for Quebec do not tend to show a definite direction, and they vary with the model form or degree of the measurement error process.


Table~\ref{tab:par-Def-3} shows the results for Definition~3. For the British Columbia data, the estimates produced by the proposed method are smaller than those yielded from the naive method. The standard errors output from the proposed methods inflate as the degree of measurement error increases. The naive and proposed methods reveal different evidence for the significance of $\phi_0$ and $\phi_1$, and the degree of measurement error affects the findings too. For the Ontario data, both methods uncover the same type of evidence for all the parameters at the significance level 0.05, except for the case with the large error under the multiplicative model. 

\begin{placehere}
Tables~\ref{tab:par-Def-1}--\ref{tab:par-Def-2}
\end{placehere}


\subsection{Forecasting}

With the fitted model for each time series in Section~\ref{sub:sensianalysis}, we forecast the true mortality rate for the subsequent five days (May 5 -- May 9) using the method described in Section~\ref{sub:Forecasting}.  Specifically, since the true mortality rates are not observable, we ``estimate" them using (\ref{eqn:pred1}) and (\ref{eqn:pred2}), respectively, for the measurement error models (\ref{eqn:measmodelClassic}) and (\ref{eqn:measmodelmultiplicative}), and then we forecast the values of $X_{i,32}$, $X_{i,33}$, $X_{i,34}$, $X_{i,35}$, and $X_{i,36}$ using (\ref{eqn:equationpredpred}). 

To quantify the forecasting performance, we calculate ${\rm P}_{\rm e}^{(h)}$ for $h=1,\ldots,H$ for each specified model of the mortality rates $X_{i,t}$, and we report the results, together with the total $\sum_{h=1}^H {\rm P}_{\rm e}^{(h)}$ in Tables~\ref{Tab:Table-Peh-Def-1}--\ref{Tab:Table-Peh-Def-3}, where $H$ is set as 5. For $h=1,\ldots, H$, we report the observed prediction error $({X}_{T+h}-\widehat{X}_{T+h})^2$, and the expected prediction error defined in (\ref{eqn:MSPE_add}) and (\ref{eqn:MSPE_multi}).   

\begin{landscape}
\begin{figure}[!p]
\centering
\includegraphics[width=1.2\textwidth]{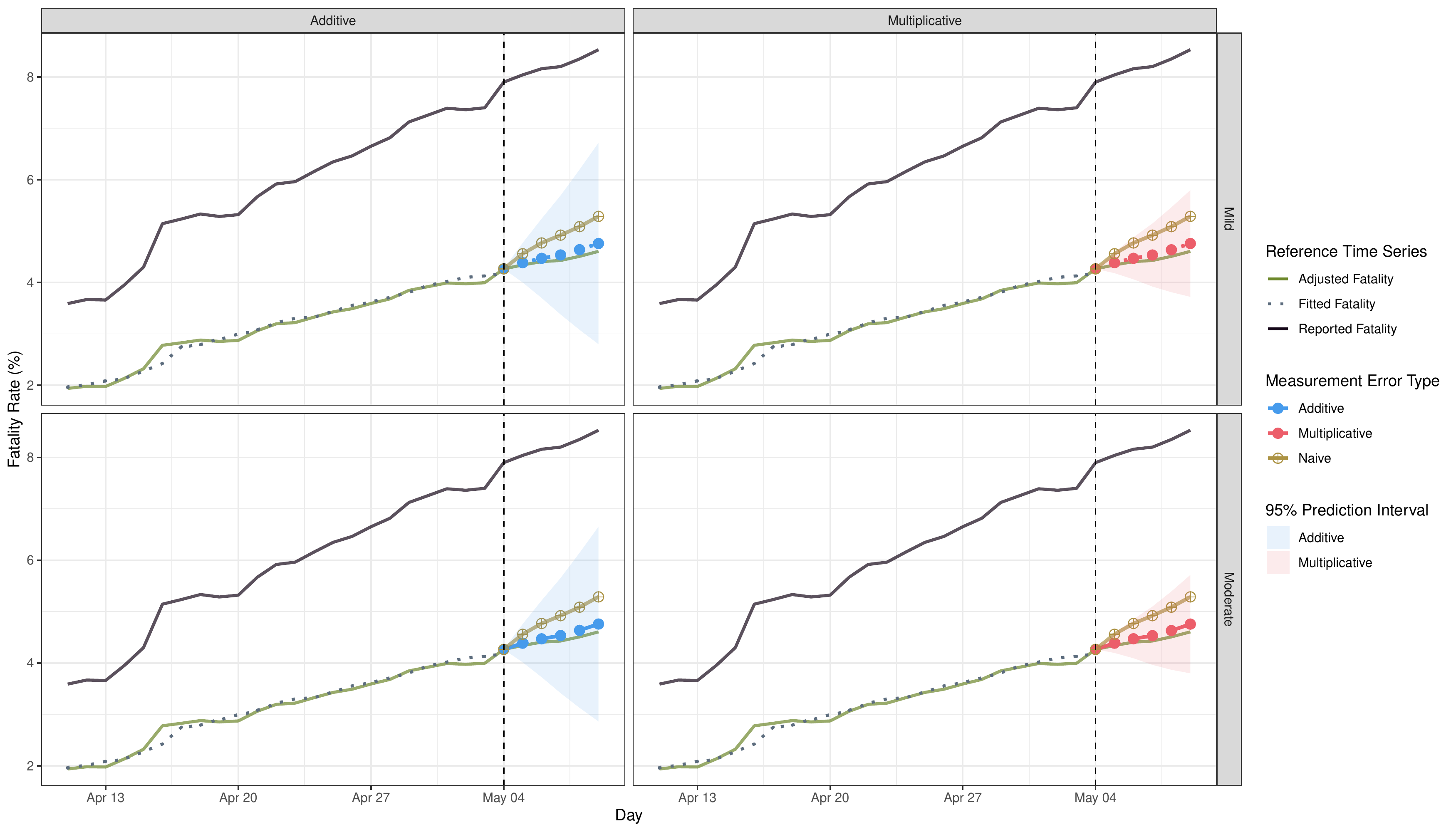}
\caption{ Ontario by Definition 3 (AR(4), order-1 differencing): A 5-day forecasting of the true mortality rate (May 5 - May 9) based on the additive (in blue) or multiplicative (in red) versus the naive model (in dark yellow); the reported mortality rates  (in black) and the adjusted true mortality rate accounting for the asymptomatic cases (in green).}\label{fig:ONfata2}
\end{figure}
\end{landscape}

\begin{landscape}
\begin{figure}[!p] 
\centering
\includegraphics[width=1.2\textwidth]{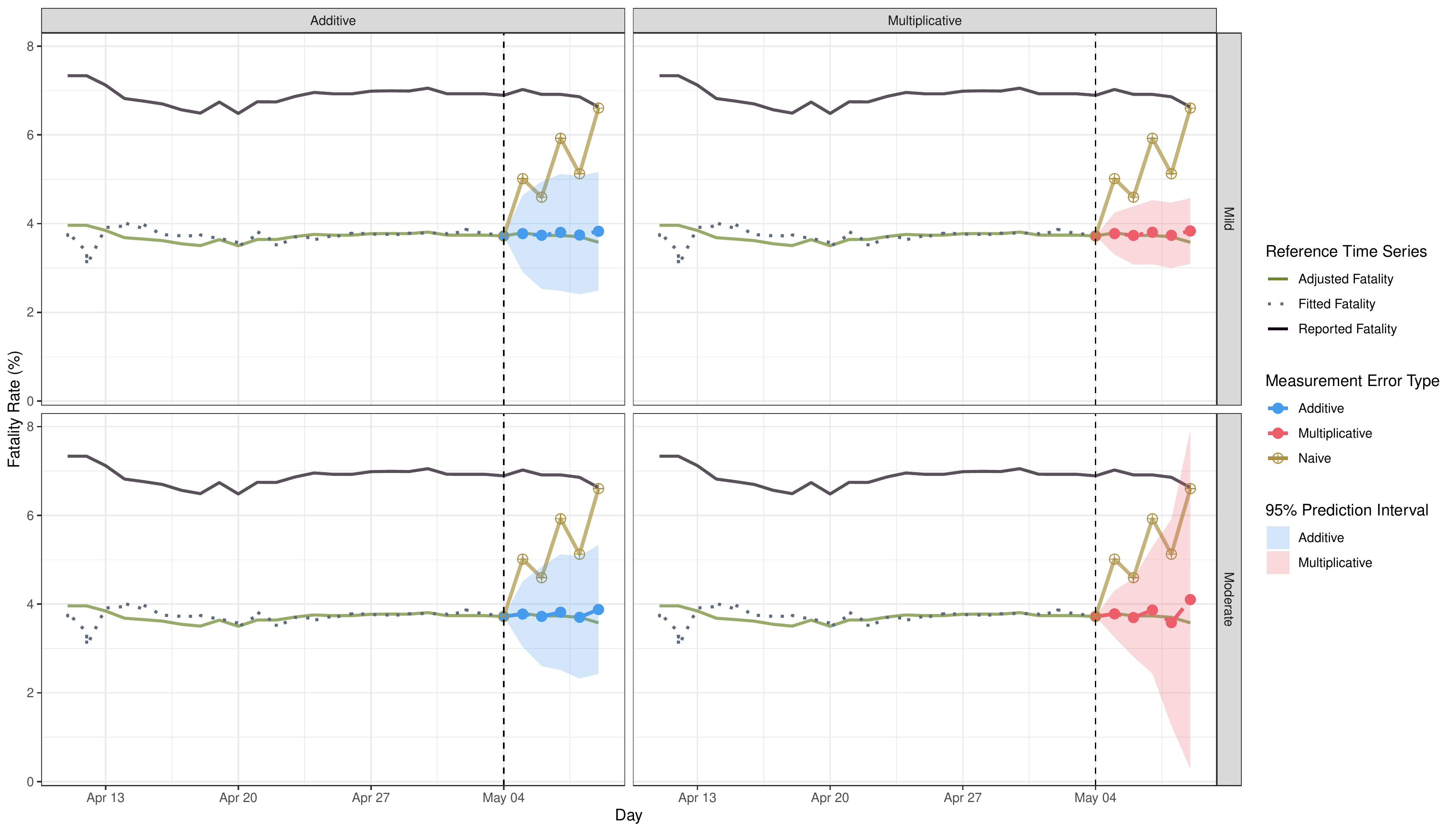}
\caption{ British Columbia by Definition 1 (AR(2), no differencing): A 5-day forecasting of the true mortality rate (May 5 - May 9) based on the additive (in blue) or multiplicative (in red) versus the naive model (in dark yellow); the reported mortality rates  (in black) and the adjusted true mortality rate accounting for the asymptomatic cases (in green).}\label{fig:BCfata3}
\end{figure}

\end{landscape}

Forecasting results based on the three definitions of mortality rates are reported in Figures~\ref{fig:BCfata3}--\ref{fig:ONfata2} for the four provinces, where the prediction results after May 4 are marked in blue and red for the measurement error models (\ref{eqn:measmodelClassic}) and (\ref{eqn:measmodelmultiplicative}), respectively, together with prediction areas marked in shaded parts, as well as the prediction results obtained from the naive method by using (\ref{eqn:equationpredpred}) with naive estimates of $\phi$ (marked in dark yellow). In comparison, we display the reported mortality rate (in black) from Apr 3, 2020 to May 9, 2020 as well as the adjusted mortality rates obtained from (\ref{eqn:measCanadadata}) (in green); in addition, we report the fitted values using (\ref{eqn:predictionequation}) in blue points. To compare the forecasting results in the presence of different degrees of measurement error. We report the results derived from a mild degree of measurement error in top subfigures and place those obtained from a large degree of measurement error in bottom subfigures.

\begin{landscape}
\begin{figure}[!p] 
\centering
\includegraphics[width=1.2\textwidth]{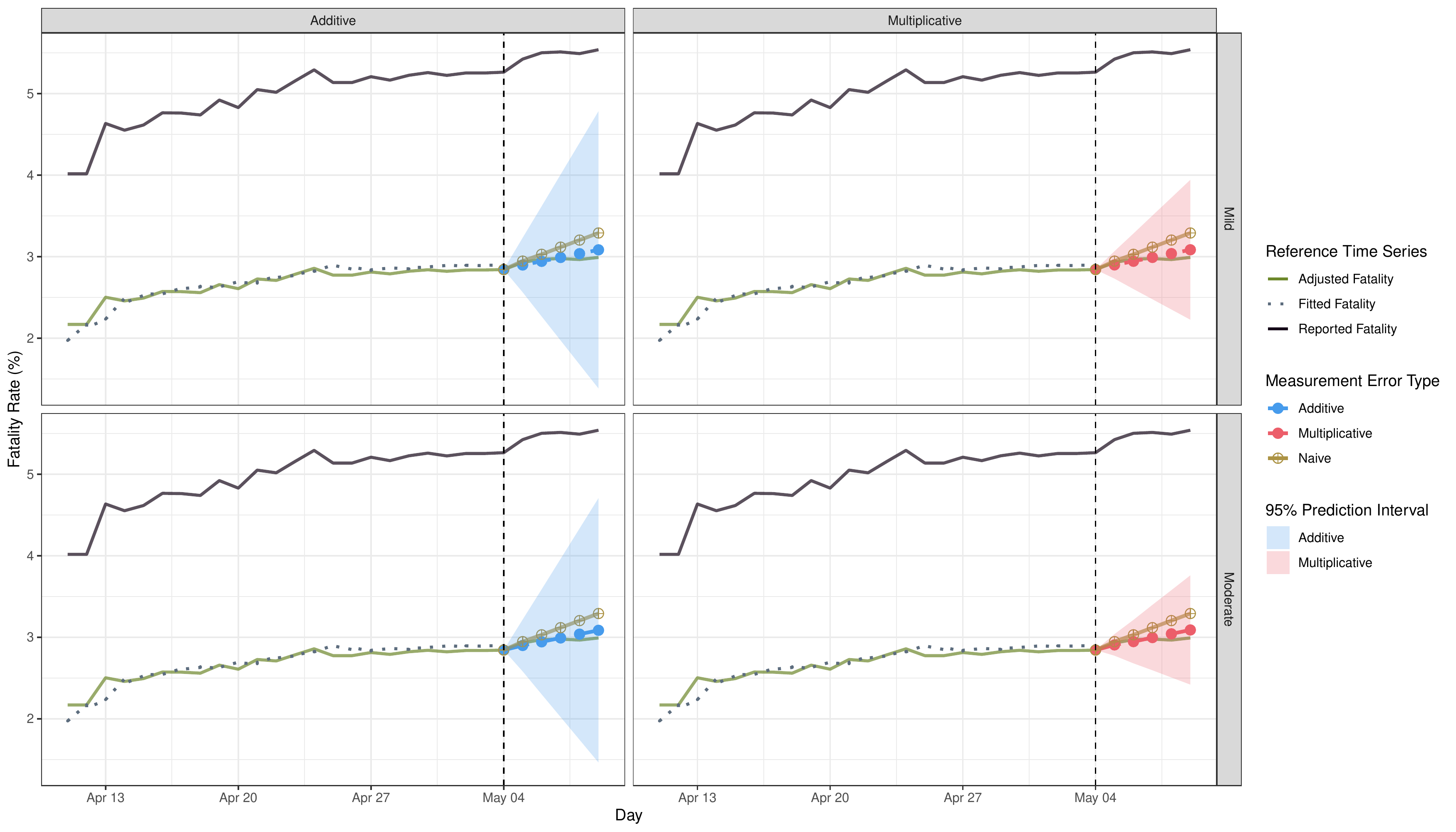}
\caption{ British Columbia by Definition 3 (AR(1), order-1 differencing): A 5-day forecasting of the true mortality rate (May 5 - May 9) based on the additive (in blue) or multiplicative (in red) versus the naive model (in dark yellow); the reported mortality rates  (in black) and the adjusted true mortality rate accounting for the asymptomatic cases (in green).}\label{fig:BCfata2}
\end{figure}

\end{landscape}

\begin{placehere}
Figure~\ref{fig:BCfata2}
\end{placehere}

\begin{landscape}
\begin{figure}[!p] 
\centering
\includegraphics[width=1.2\textwidth]{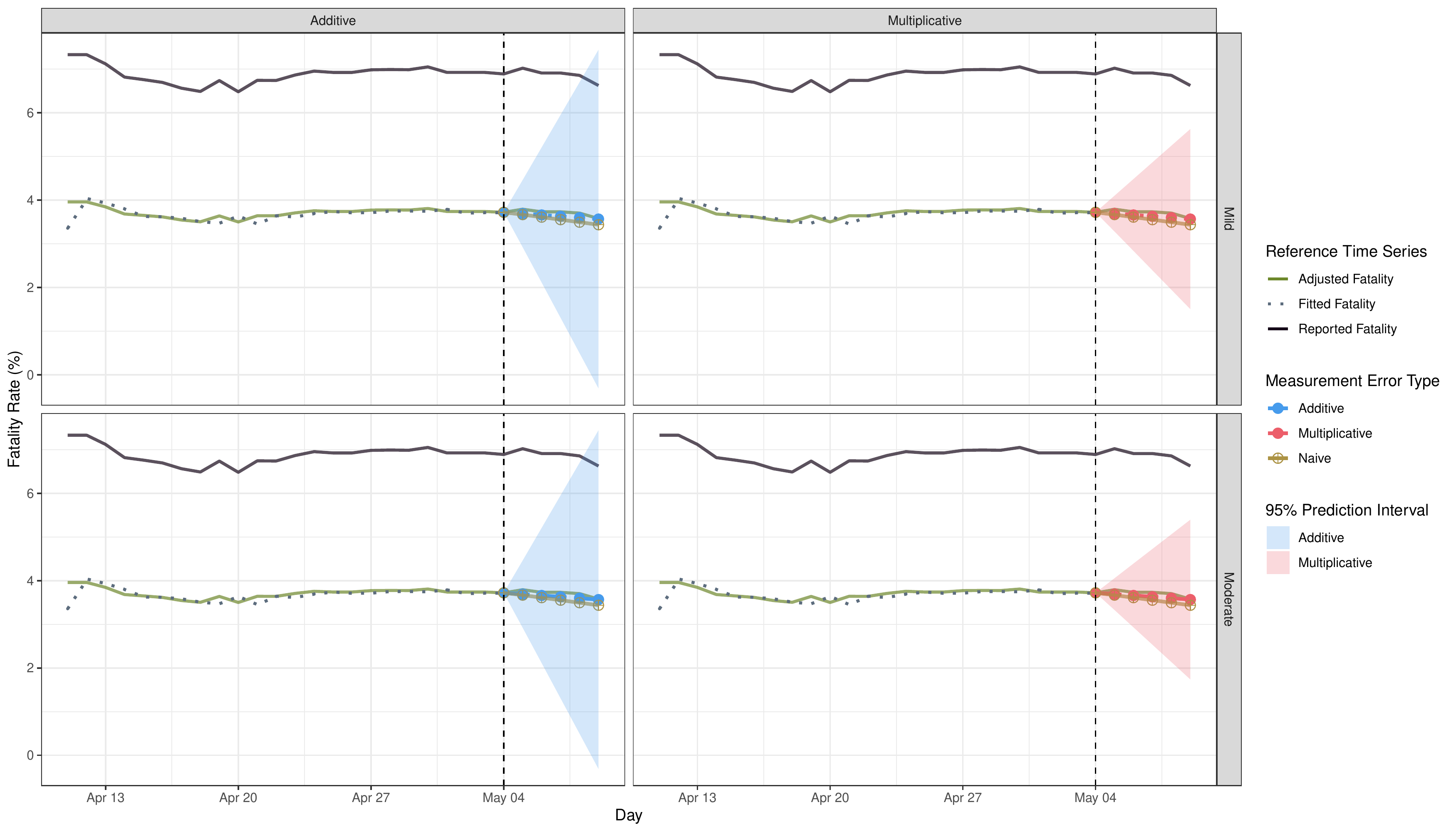}
\caption{ British Columbia by Definition 1 (AR(1), order-1 differencing): A 5-day forecasting of the true mortality rate (May 5 - May 9) based on the additive (in blue) or multiplicative (in red) versus the naive model (in dark yellow); the reported mortality rates  (in black) and the adjusted true mortality rate accounting for the asymptomatic cases (in green).}\label{fig:BCfata3.2}
\end{figure}

\end{landscape}

\begin{landscape}
\begin{figure}[!p] 
\centering
\includegraphics[width=1.2\textwidth]{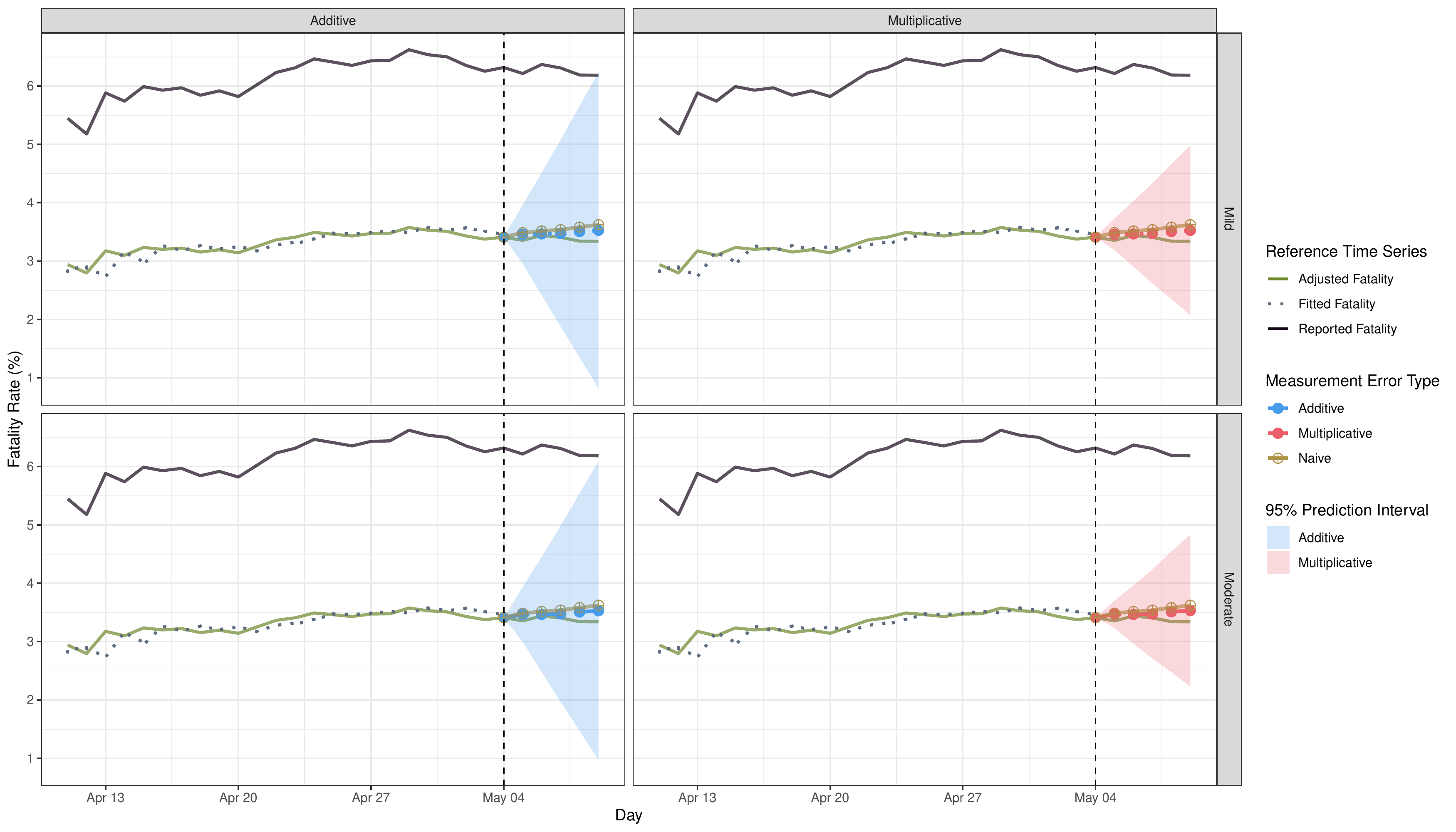}
\caption{ British Columbia by Definition 2 (AR(3), order-1 differencing): A 5-day forecasting of the true mortality rate (May 5 - May 9) based on the additive (in blue) or multiplicative (in red) versus the naive model (in dark yellow); the reported mortality rates  (in black) and the adjusted true mortality rate accounting for the asymptomatic cases (in green).}\label{fig:BCfata4}
\end{figure}
\end{landscape}

The results for British Columbia are presented in Figure~\ref{fig:BCfata2} and Web Figures~\ref{fig:BCfata3}--\ref{fig:BCfata4}. With Definition~1, the methods with measurement error effects accommodated suggest that the mortality rate in the past and its forecasting values are around 4\%, whereas the results obtained from the method without accounting for measurement error effects indicate that the mortality rates over time are higher than 6\%. With Definition~2, the methods with or without accounting for measurement error effects reveal that the mortality rates over time are, respectively, below 3.5\% and above 5\%.  With Definition~3,  the methods with or without accounting for measurement error effects indicate that the mortality rates over time are, around 3\% and above 4\%, respectively. 

\begin{placehere}
Figure~\ref{fig:ONfata2}
\end{placehere}

\begin{landscape}
\begin{figure}[!p]
\centering
\includegraphics[width=1.2\textwidth]{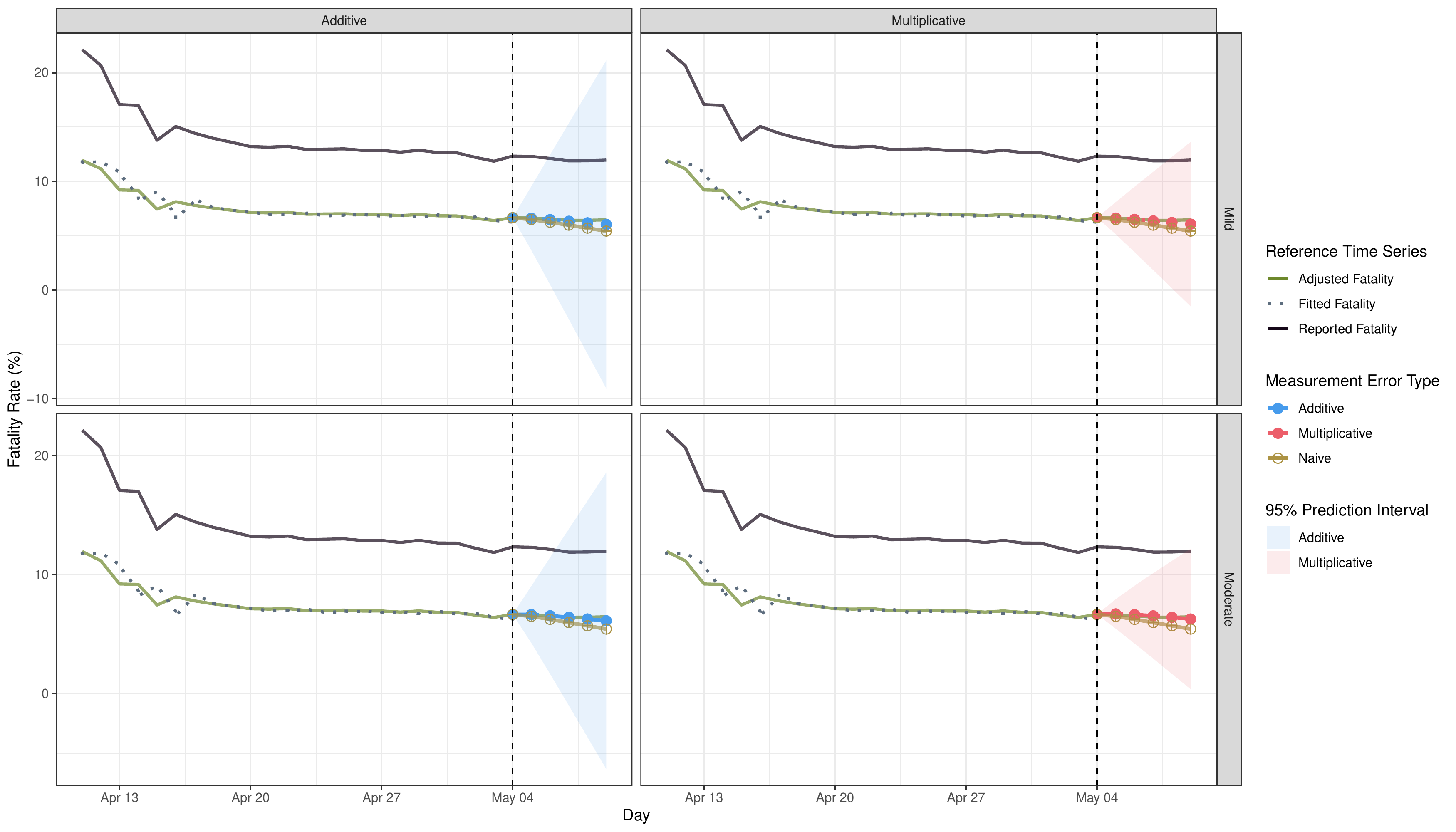}
\caption{ Ontario by Definition 1 (AR(1), order-1 differencing): A 5-day forecasting of the true mortality rate (May 5 - May 9) based on the additive (in blue) or multiplicative (in red) versus the naive model (in dark yellow); the reported mortality rates  (in black) and the adjusted true mortality rate accounting for the asymptomatic cases (in green).}\label{fig:ONfata3}
\end{figure}
\end{landscape}

\begin{landscape}
\begin{figure}[!p]
\centering
\includegraphics[width=1.2\textwidth]{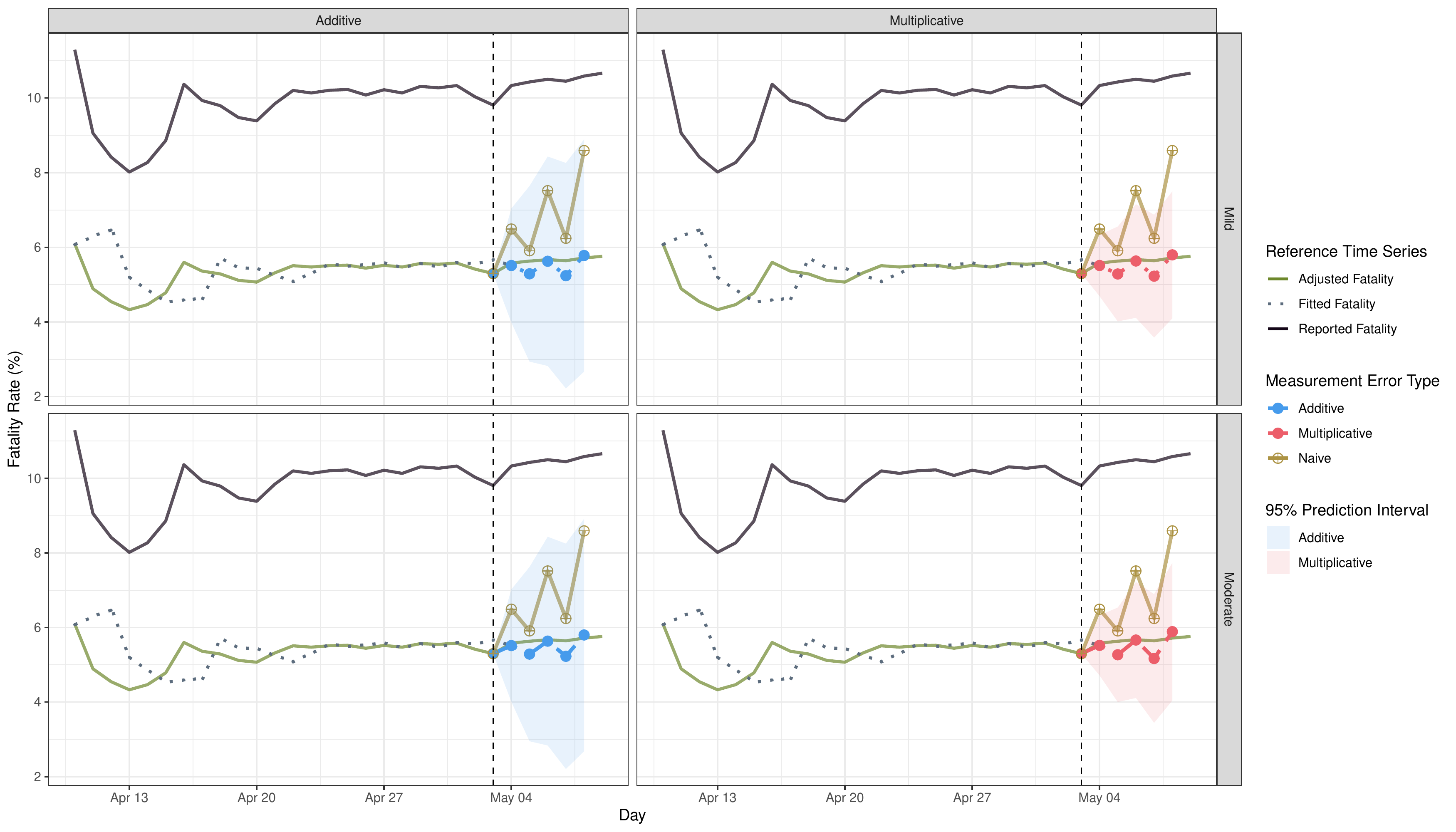}
\caption{ Ontario by Definition 2 (AR(1), no differencing): A 5-day forecasting of the true mortality rate (May 5 - May 9) based on the additive (in blue) or multiplicative (in red) versus the naive model (in dark yellow); the reported mortality rates  (in black) and the adjusted true mortality rate accounting for the asymptomatic cases (in green).}\label{fig:ONfata4}
\end{figure}
\end{landscape}

The results for Ontario are presented in Figure~\ref{fig:ONfata2} and Supplementary~Figures~\ref{fig:ONfata3}--\ref{fig:ONfata4}. With Definition~1, the methods with measurement error effects accommodated suggest that the mortality rate over time is around 7\% over time, while the reported mortality rate over time is about 12.5\%.  With Definition~2, the methods with and without incorporating the feature of measurement error indicate the mortality rate in the past and its forecasting values are, respectively, below 6\% and around 10\%. With Definition 3, the mortality rate increases over time substantially. The methods with measurement error effects accommodated suggest that the mortality rate increases from 2\% to above 4\% whereas the reported mortality rate shows that rate increases from below 4\% to above 8\%. 


The results for Quebec are presented in Supplementary~Figures~\ref{fig:QCfata3}--\ref{fig:QCfata4}. With Definition~1 the methods with measurement error effects accommodated show that the mortality rate is around 6.5\% over time, whereas the method without considering measurement error indicates the mortality rate is over 10\%. With Definition~2, the methods with or without addressing the measurement error effects show that the mortality rates over time are, respectively, below 6\% and above 7.5\%. 


The results for Alberta are presented in Supplementary~Figures~\ref{fig:ABfata3}--\ref{fig:ABfata4}. With Definition~1 the methods with and without measurement error accommodated suggest that the mortality rates are, respectively, around 2\% and 4\% over time. With Definition~2, the methods with or without addressing the measurement error effects show that the historical mortality rate and its predictions are, respectively, below 2\% and above 2\%.


\subsection{Model Assessment}

The specification of lag $p$ for model (\ref{eqn:timeseriresmodel}) of the true mortality rates $\{X_{i,t}:t=1,\ldots, T\}$ is based on (\ref{eqn:AIC}) which is derived from the reported mortality rates $\{X_{i,t}^*:t=1,\ldots, T\}$, but not from $\{X_{i,t}: t=1,\ldots, T\}$ itself. This discrepancy introduces the possibility of model misspecification when featuring the series $X_{i,t}$ using (\ref{eqn:timeseriresmodel}). To investigate this, we conduct a sensitivity analysis by considering the AR($p$) with a different value of $p$ for the $X_{i,t}$ from Definition~1.  As Table~\ref{tab:deathdef} indicates the feasibility of using AR(1) for all four provinces, here we further employ the AR(2) model to do forecasting for the period from May~5 to May~9. 

In Table~\ref{Tab:Table-Peh-AR}, we report the observed and expected prediction errors of the forecasting using AR(2) models in comparison with AR(1) models. Comparing different lag orders of the autoregressive models, we find that in terms of the observed prediction error, the selected AR(1) models have better performance than the AR(2) models for the data of Ontario and Alberta, and the results for British Columbia and Quebec are fairly similar. It is noticed that both the observed prediction error and the expected prediction error associated with the proposed method tend to become small when the degree of measurement error increases for British Columbia, Ontario, and Quebec.

\begin{placehere}
Table~\ref{Tab:Table-Peh-AR}
\end{placehere}

\section{Discussion} \label{sec:discussion}

In this article, we investigate the impact of measurement error on time series analysis under autoregressive models and establish analytic results under the additive and multiplicative measurement error models. We propose an estimating equation method to correct for the biases induced from the naive analysis which disregards the differences between the true measurements and their surrogate measurements. We rigorously establish the theoretical results for the proposed method. As a genuine application, we apply to the proposed method to analyze the mortality rates of COVID-19 data in four provinces, British Columbia, Ontario, Quebec, and Alberta, which have the most severe virus outbreaks in Canada. The real data analysis clearly demonstrates that incorporating measurement error in the analysis can uncover various different results.

Our method has the flexibility or robustness in that distribution assumptions are required to describe the measurement error process as well as the time series autoregressive process. While our research is motivated by the faulty nature of COVID-19 data, the proposed method can be applied to handle other problems related to error-contaminated time series. Our development here is directed to using autoregressive models to delineate time series data. The same principles can be applied to other model forms such as moving average models or autoregressive moving average models which may be used to handle error-prone time series data, where technical details can be more notationally involved. 

When checking the stationarity of time series, we apply the ADF test to the observed time series $X_{t}^*$, which is mainly driven by the unavailability of the true values of $X_t$, as well as the fact that the weakly stationarity of observed time series implies the weakly stationarity of the true time series if measurement error is featured with (\ref{eqn:measmodelClassic}) or (\ref{eqn:measmodelmultiplicative}). It is interesting to rigorously develop a formal test similar to the ADF test to handle time series subject to measurement error.

\section*{Acknowledgements}

This research is partially supported by the Natural Sciences and Engineering Research
Council of Canada (NSERC) as well as the Rapid Response Program – COVID-19 of the Canadian
Statistical Sciences Institute (CANSSI). Yi is Canada Research Chair in Data Science
(Tier 1). Her research was undertaken, in part, thanks to funding from the Canada Research
Chairs Program. \\


\bibliographystyle{apa}
\phantomsection  
\renewcommand*{\bibname}{References}

\addcontentsline{toc}{chapter}{\textbf{References}}

\bibliography{COVID}

\begin{thebibliography}{}

\bibitem[\protect\astroncite{Baud et~al.}{2020}]{baud2020real}
Baud, D., Qi, X., Nielsen-Saines, K., Musso, D., Pomar, L., and Favre, G.
  (2020).
\newblock Real estimates of mortality following {COVID}-19 infection.
\newblock {\em The Lancet Infectious Diseases}.

\bibitem[\protect\astroncite{Box et~al.}{2015}]{box2015time}
Box, G.~E., Jenkins, G.~M., Reinsel, G.~C., and Ljung, G.~M. (2015).
\newblock {\em Time Series Analysis: Forecasting and Control}.
\newblock New Jersey, NJ: John Wiley \& Sons.

\bibitem[\protect\astroncite{Brockwell and
  Davis}{2002}]{brockwell2016introduction}
Brockwell, P.~J. and Davis, R.~A. (2002).
\newblock {\em Introduction to Time Series and Forecasting}.
\newblock New York, NY: Springer-Verlag.

\bibitem[\protect\astroncite{Brockwell et~al.}{1991}]{brockwell1991time}
Brockwell, P.~J., Davis, R.~A., and Fienberg, S.~E. (1991).
\newblock {\em Time Series: Theory and Methods}.
\newblock New York, NY: Springer Science \& Business Media.

\bibitem[\protect\astroncite{Cheung and Lai}{1995}]{cheung1995lag}
Cheung, Y.-W. and Lai, K.~S. (1995).
\newblock Lag order and critical values of the augmented dickey-fuller test.
\newblock {\em Journal of Business \& Economic Statistics}, 13(3):277--280.

\bibitem[\protect\astroncite{Day}{2020}]{day2020covid}
Day, M. (2020).
\newblock {COVID-19}: four fifths of cases are asymptomatic, {China} figures
  indicate.
\newblock {\em The BMJ}, 369.

\bibitem[\protect\astroncite{Dedecker et~al.}{2014}]{DedeckerSamson2014}
Dedecker, J., Samson, A., and Taupin, M.-L. (2014).
\newblock Estimation in autoregressive model with measurement error.
\newblock {\em ESAIM: Probability and Statistics}, 18:277--307.

\bibitem[\protect\astroncite{He et~al.}{2020}]{he2020estimation}
He, W., Yi, G.~Y., and Zhu, Y. (2020).
\newblock Estimation of the basic reproduction number, average incubation time,
  asymptomatic infection rate, and case fatality rate for {COVID-19}:
  Meta-analysis and sensitivity analysis.
\newblock {\em Journal of Medical Virology}.

\bibitem[\protect\astroncite{Kanchan et~al.}{2015}]{kanchan2015mortality}
Kanchan, T., Kumar, N., and Unnikrishnan, B. (2015).
\newblock Mortality: Statistics.
\newblock In Payne-James, J. and Byard, R.~W., editors, {\em Encyclopedia of
  Forensic and Legal Medicine: Second Edition}, pages 572--577. Oxford,
  OX:Elsevier.

\bibitem[\protect\astroncite{Kimball}{2020}]{kimball2020asymptomatic}
Kimball, A. (2020).
\newblock Asymptomatic and presymptomatic {SARS-CoV-2} infections in residents
  of a long-term care skilled nursing facility—{King County}, {Washington},
  {March} 2020.
\newblock {\em Morbidity and Mortality Weekly Report}, 69:377--381.

\bibitem[\protect\astroncite{Lahiri}{1999}]{lahiri1999theoretical}
Lahiri, S.~N. (1999).
\newblock Theoretical comparisons of block bootstrap methods.
\newblock {\em The Annals of Statistics}, 27(1):386--404.

\bibitem[\protect\astroncite{Staudenmayer and
  Buonaccorsi}{2005}]{staudenmayer2005measurement}
Staudenmayer, J. and Buonaccorsi, J.~P. (2005).
\newblock Measurement error in linear autoregressive models.
\newblock {\em Journal of the American Statistical Association},
  100(471):841--852.

\bibitem[\protect\astroncite{Tanaka}{2002}]{Tanaka2002}
Tanaka, K. (2002).
\newblock A unified approach to the measurement error problem in time series
  models.
\newblock {\em Econometric Theory}, 18(2):278--296.

\bibitem[\protect\astroncite{Tripodis and
  Buonaccorsi}{2009}]{TripodisBuonaccorsi2009}
Tripodis, Y. and Buonaccorsi, J.~P. (2009).
\newblock Prediction and forecasting in linear models with measurement error.
\newblock {\em Journal of statistical planning and inference},
  139(12):4039--4050.

\end{thebibliography}


\begin{sidewaystable}
\setlength\extrarowheight{1pt} 
\centering
\captionsetup{size=small}
\scriptsize

\begin{threeparttable}
\caption[]{Definition 1: The parameter estimation under different measurement error models: the AR(1) model with ``order-1 differencing" is used to fit the data of British Columbia, Ontario, Quebec and Alberta}\label{tab:par-Def-1}

\begin{tabular}{cccccccccccccccccc}
\hline
\multicolumn{1}{l}{}      & \multicolumn{1}{l}{} &           & \multicolumn{3}{c}{British Columbia} &  & \multicolumn{3}{c}{Ontario} &  & \multicolumn{3}{c}{Quebec} &  & \multicolumn{3}{c}{Alberta} \\ \cline{1-6} \cline{8-10} \cline{12-14} \cline{16-18} 
Method                    & Error Degree               & Parameter & EST        & SE      & p-value     &  & EST     & SE   & p-value  &  & EST     & SE   & p-value &  & EST     & SE   & p-value  \\ \hline
\multirow{2}{*}{Naive}    & \multirow{2}{*}{-}   & $\phi_0$  & -0.050     & 0.043     & 0.272       &  & -0.215  & 0.243  & 0.384    &  & -0.340  & 0.180  & 0.071   &  & -0.031  & 0.016  & 0.061    \\
                          &                      & $\phi_1$  & 0.138      & 0.214     & 0.533       &  & 0.215   & 0.157  & 0.183    &  & 0.012   & 0.124  & 0.923   &  & 0.052   & 0.144  & 0.721    \\ \hline
                          & Small                & $\phi_0$  & -0.027     & 0.025     & 0.313       &  & -0.113  & 0.134  & 0.406    &  & -0.183  & 0.111  & 0.112   &  & -0.017  & 0.009  & 0.088    \\
The Proposed Method       & ($\sigma_{e1}^2$)    & $\phi_1$  & 0.146      & 0.532     & 0.788       &  & 0.237   & 0.280  & 0.406    &  & 0.014   & 1.566  & 0.993   &  & 0.056   & 0.185  & 0.764    \\ \cline{2-18} 
with Additive Error       & Large                & $\phi_0$  & -0.027     & 0.025     & 0.298       &  & -0.097  & 0.263  & 0.715    &  & -0.181  & 0.100  & 0.083   &  & -0.014  & 0.073  & 0.845    \\
                          & ($\sigma_{e2}^2$)    & $\phi_1$  & 0.146      & 0.468     & 0.760       &  & 0.345   & 0.939  & 0.717    &  & 0.027   & 0.323  & 0.934   &  & 0.183   & 1.596  & 0.909    \\ \hline
                          & Small                & $\phi_0$  & -0.027     & 0.024     & 0.286       &  & -0.107  & 0.152  & 0.488    &  & -0.183  & 0.099  & 0.078   &  & -0.017  & 0.009  & 0.080    \\
The Proposed Method       & ($\sigma_{u1}^2$)    & $\phi_1$  & 0.151      & 0.236     & 0.535       &  & 0.275   & 0.238  & 0.260    &  & 0.016   & 0.166  & 0.923   &  & 0.060   & 0.180  & 0.740    \\ \cline{2-18} 
with Multiplicative Error & Large                & $\phi_0$  & -0.025     & 0.024     & 0.308       &  & -0.078  & 1.690  & 0.964    &  & -0.180  & 0.127  & 0.170   &  & -0.016  & 0.015  & 0.299    \\
                          & ($\sigma_{u2}^2$)    & $\phi_1$  & 0.192      & 0.300     & 0.535       &  & 0.476   & 3.955  & 0.905    &  & 0.031   & 1.327  & 0.981   &  & 0.087   & 0.360  & 0.812    \\ \hline
\end{tabular}
\begin{tablenotes}
\end{tablenotes}

\end{threeparttable}
\end{sidewaystable}

\begin{sidewaystable}
\setlength\extrarowheight{1pt} 
\centering
\captionsetup{size=small}
\scriptsize

\begin{threeparttable}
\caption[]{Definition 2: The parameter estimation under different measurement error models: the AR(2) model with ``no differencing" is used to fit the data of Ontario, the AR(1) model with ``order-1 differencing" is used to fit the data of Alberta, and the AR(2) model with ``order-1 differencing" is used to fit the data of British Columbia and Quebec.  }
\label{tab:par-Def-2}

\begin{tabular}{cccccccccccccccccc}
\hline
\multicolumn{1}{l}{}      & \multicolumn{1}{l}{}    &           & \multicolumn{3}{c}{British Columbia} &  & \multicolumn{3}{c}{Ontario} &  & \multicolumn{3}{c}{Quebec} &  & \multicolumn{3}{c}{Alberta} \\ \cline{1-6} \cline{8-10} \cline{12-14} \cline{16-18} 
Method                    & Error Degree            & Parameter & EST        & SE        & p-value     &  & EST     & SE     & p-value  &  & EST     & SE     & p-value &  & EST     & SE     & p-value  \\ \hline
\multirow{3}{*}{Naive}    & \multirow{3}{*}{-}      & $\phi_0$  & 0.062      & 0.034     & 0.097       &  & 2.126   & 1.388  & 0.138    &  & 0.225   & 0.058  & 0.001   &  & -0.013  & 0.022  & 0.561    \\
                          &                         & $\phi_1$  & -0.415     & 0.186     & 0.046       &  & 1.167   & 0.209  & $<$0.001 &  & -0.122  & 0.136  & 0.380   &  & -0.124  & 0.172  & 0.477    \\
                          &                         & $\phi_2$  & -0.254     & 0.185     & 0.195       &  & -0.370  & 0.140  & 0.014    &  & -0.309  & 0.092  & 0.003   &  & -       & -      & -        \\ \hline
                          &                         & $\phi_0$  & 0.034      & 0.020     & 0.114       &  & 1.146   & 0.759  & 0.144    &  & 0.174   & 0.042  & 0.000   &  & -0.007  & 0.012  & 0.567    \\
                          & Small ($\sigma_{e1}^2$) & $\phi_1$  & -0.432     & 0.201     & 0.053       &  & 1.173   & 0.216  & $<$0.001 &  & 0.124   & 0.032  & 0.001   &  & -0.131  & 0.185  & 0.486    \\
The Proposed Method       &                         & $\phi_2$  & -0.268     & 0.205     & 0.215       &  & -0.375  & 0.141  & 0.014    &  & -0.130  & 0.165  & 0.435   &  & -       & -      & -        \\ \cline{2-18} 
with Additive Error       &                         & $\phi_0$  & 0.036      & 0.024     & 0.164       &  & 1.138   & 0.747  & 0.140    &  & -0.327  & 0.096  & 0.002   &  & -0.007  & 0.012  & 0.554    \\
                          & Large ($\sigma_{e2}^2$) & $\phi_1$  & -0.497     & 0.265     & 0.085       &  & 1.189   & 0.239  & $<$0.001 &  & 0.162   & 0.044  & 0.001   &  & -0.158  & 0.247  & 0.529    \\
                          &                         & $\phi_2$  & -0.320     & 0.354     & 0.384       &  & -0.390  & 0.172  & 0.032    &  & 0.132   & 0.041  & 0.004   &  & -       & -      & -        \\ \hline
                          &                         & $\phi_0$  & 0.034      & 0.020     & 0.115       &  & 1.139   & 0.748  & 0.141    &  & -0.164  & 0.229  & 0.480   &  & -0.007  & 0.012  & 0.564    \\
                          & Small ($\sigma_{u1}^2$) & $\phi_1$  & -0.439     & 0.205     & 0.053       &  & 1.188   & 0.231  & $<$0.001 &  & -0.394  & 0.199  & 0.059   &  & -0.144  & 0.205  & 0.487    \\
The Proposed Method       &                         & $\phi_2$  & -0.273     & 0.204     & 0.205       &  & -0.389  & 0.162  & 0.024    &  & 0.128   & 0.042  & 0.006   &  & -       & -      & -        \\ \cline{2-18} 
with Multiplicative Error &                         & $\phi_0$  & 0.039      & 0.032     & 0.236       &  & 1.112   & 0.747  & 0.149    &  & 0.127   & 0.036  & 0.002   &  & -0.008  & 0.012  & 0.546    \\
                          & Large ($\sigma_{u2}^2$) & $\phi_1$  & -0.584     & 0.339     & 0.111       &  & 1.255   & 0.503  & 0.020    &  & -0.143  & 0.194  & 0.467   &  & -0.205  & 0.317  & 0.524    \\
                          &                         & $\phi_2$  & -0.393     & 0.322     & 0.245       &  & -0.451  & 0.510  & 0.384    &  & -0.353  & 0.111  & 0.004   &  & -       & -      & -        \\ \hline
\end{tabular}
\begin{tablenotes}
\end{tablenotes}

\end{threeparttable}
\end{sidewaystable}

\begin{sidewaystable}
\setlength\extrarowheight{1pt} 
\centering
\captionsetup{size=small}
\tiny

\begin{threeparttable}
\caption[]{The observed prediction error and expected prediction error for different lag order of autoregressive models}
\label{Tab:Table-Peh-AR}

\begin{tabular}{lllccccccccccccc}
\hline
                                &                                &                                            & \multicolumn{6}{c}{Observed Prediction Error}                      &  & \multicolumn{6}{c}{Expected Prediction Error}                      \\ \hline
Method                          & $\sigma_e^2$ (or $\sigma_u^2$) & Model                                      & Day 1 & Day 2 & Day 3 & Day 4 & Day 5 & $\sum_{h=1}^H{\rm OPE}(h)$ &  & Day 1 & Day 2 & Day 3 & Day 4 & Day 5 & $\sum_{h=1}^H{\rm EPE}(h)$ \\ \cline{1-9} \cline{11-16} 
                                &                                & \multicolumn{7}{l}{British Columbia}                                                                            &  &       &       &       &       &       &                            \\
Naive                           & -                              & \multicolumn{1}{c}{\multirow{5}{*}{\textbf{AR(1)}\tnote{a}}} & 0.015 & 0.015 & 0.032 & 0.043 & 0.020 & 0.126                      &  & 0.164 & 0.167 & 0.167 & 0.167 & 0.167 & 0.834                      \\
\multirow{2}{*}{Additive}       & Mild                           & \multicolumn{1}{c}{}                       & 0.010 & 0.005 & 0.011 & 0.011 & 0.000 & 0.037                      &  & 0.154 & 0.157 & 0.157 & 0.157 & 0.157 & 0.783                      \\
                                & Moderate                       & \multicolumn{1}{c}{}                       & 0.010 & 0.005 & 0.011 & 0.011 & 0.000 & 0.037                      &  & 0.154 & 0.157 & 0.157 & 0.157 & 0.157 & 0.784                      \\
\multirow{2}{*}{Multiplicative} & Mild                           & \multicolumn{1}{c}{}                       & 0.010 & 0.005 & 0.011 & 0.011 & 0.000 & 0.037                      &  & 0.044 & 0.044 & 0.044 & 0.044 & 0.044 & 0.222                      \\
                                & Moderate                       & \multicolumn{1}{c}{}                       & 0.010 & 0.005 & 0.011 & 0.011 & 0.000 & 0.037                      &  & 0.034 & 0.035 & 0.035 & 0.035 & 0.035 & 0.174                      \\ \cline{3-16} 
Naive                           & -                              & \multirow{5}{*}{AR(2)}                    & 0.016 & 0.014 & 0.031 & 0.042 & 0.019 & 0.122                      &  & 0.161 & 0.165 & 0.167 & 0.167 & 0.167 & 0.828                      \\
\multirow{2}{*}{Additive}       & Mild                           &                                            & 0.010 & 0.005 & 0.010 & 0.010 & 0.000 & 0.035                      &  & 0.151 & 0.155 & 0.157 & 0.157 & 0.157 & 0.777                      \\
                                & Moderate                       &                                            & 0.010 & 0.005 & 0.010 & 0.010 & 0.000 & 0.035                      &  & 0.151 & 0.155 & 0.157 & 0.157 & 0.157 & 0.778                      \\
\multirow{2}{*}{Multiplicative} & Mild                           &                                            & 0.010 & 0.005 & 0.010 & 0.010 & 0.000 & 0.035                      &  & 0.043 & 0.044 & 0.044 & 0.044 & 0.044 & 0.220                      \\
                                & Moderate                       &                                            & 0.010 & 0.005 & 0.010 & 0.010 & 0.000 & 0.034                      &  & 0.034 & 0.035 & 0.035 & 0.035 & 0.035 & 0.173                      \\ \hline
                                &                                & \multicolumn{7}{l}{Ontario}                                                                                     &  &       &       &       &       &       &                            \\
Naive                           & -                              & \multirow{5}{*}{\textbf{AR(1)}\tnote{a}}                    & 0.020 & 0.087 & 0.196 & 0.521 & 1.059 & 1.884                      &  & 2.527 & 2.643 & 2.649 & 2.649 & 2.649 & 13.117                     \\
\multirow{2}{*}{Additive}       & Mild                           &                                            & 0.001 & 0.004 & 0.007 & 0.056 & 0.175 & 0.243                      &  & 2.264 & 2.391 & 2.399 & 2.399 & 2.399 & 11.853                     \\
                                & Moderate                       &                                            & 0.000 & 0.000 & 0.000 & 0.023 & 0.110 & 0.134                      &  & 1.453 & 1.626 & 1.646 & 1.649 & 1.649 & 8.023                      \\
\multirow{2}{*}{Multiplicative} & Mild                           &                                            & 0.000 & 0.002 & 0.003 & 0.044 & 0.152 & 0.201                      &  & 0.558 & 0.599 & 0.603 & 0.603 & 0.603 & 2.965                      \\
                                & Moderate                       &                                            & 0.004 & 0.010 & 0.014 & 0.000 & 0.035 & 0.063                      &  & 0.270 & 0.331 & 0.345 & 0.348 & 0.348 & 1.642                      \\ \cline{3-16} 
Naive                           & -                              & \multicolumn{1}{c}{\multirow{5}{*}{AR(2)}} & 0.073 & 0.107 & 0.240 & 0.550 & 1.111 & 2.081                      &  & 2.517 & 2.648 & 2.648 & 2.649 & 2.649 & 13.111                     \\
\multirow{2}{*}{Additive}       & Mild                           & \multicolumn{1}{c}{}                       & 0.029 & 0.014 & 0.026 & 0.083 & 0.227 & 0.379                      &  & 2.256 & 2.398 & 2.398 & 2.399 & 2.399 & 11.851                     \\
                                & Moderate                       & \multicolumn{1}{c}{}                       & 0.045 & 0.008 & 0.031 & 0.063 & 0.221 & 0.368                      &  & 1.470 & 1.658 & 1.646 & 1.649 & 1.649 & 8.072                      \\
\multirow{2}{*}{Multiplicative} & Mild                           & \multicolumn{1}{c}{}                       & 0.034 & 0.012 & 0.027 & 0.076 & 0.222 & 0.370                      &  & 0.571 & 0.606 & 0.603 & 0.603 & 0.603 & 2.986                      \\
                                & Moderate                       & \multicolumn{1}{c}{}                       & 0.085 & 0.001 & 0.071 & 0.024 & 0.310 & 0.491                      &  & 0.454 & 0.469 & 0.415 & 0.390 & 0.375 & 2.103                      \\ \hline
                                &                                & \multicolumn{7}{l}{Quebec}                                                                                       &  &       &       &       &       &       &                            \\
Naive                           & -                              & \multicolumn{1}{c}{\multirow{5}{*}{\textbf{AR(1)}\tnote{a}}} & 0.163 & 0.607 & 1.357 & 2.289 & 3.294 & 7.709                      &  & 1.811 & 1.811 & 1.811 & 1.811 & 1.811 & 9.057                      \\
\multirow{2}{*}{Additive}       & Mild                           & \multicolumn{1}{c}{}                        & 0.061 & 0.216 & 0.479 & 0.778 & 1.053 & 2.587                      &  & 1.561 & 1.561 & 1.561 & 1.561 & 1.561 & 7.807                      \\
                                & Moderate                       & \multicolumn{1}{c}{}                        & 0.060 & 0.215 & 0.478 & 0.776 & 1.051 & 2.580                      &  & 0.811 & 0.811 & 0.811 & 0.811 & 0.811 & 4.057                      \\
\multirow{2}{*}{Multiplicative} & Mild                           & \multicolumn{1}{c}{}                        & 0.061 & 0.216 & 0.479 & 0.778 & 1.053 & 2.586                      &  & 0.399 & 0.399 & 0.399 & 0.399 & 0.399 & 1.995                      \\
                                & Moderate                       & \multicolumn{1}{c}{}                        & 0.060 & 0.215 & 0.477 & 0.776 & 1.050 & 2.578                      &  & 0.205 & 0.205 & 0.205 & 0.205 & 0.205 & 1.025                      \\ \cline{3-16} 
Naive                           & -                              & \multirow{5}{*}{AR(2)}                      & 0.129 & 0.524 & 1.226 & 2.115 & 3.085 & 7.079                      &  & 1.746 & 1.746 & 1.809 & 1.809 & 1.811 & 8.921                      \\
\multirow{2}{*}{Additive}       & Mild                           &                                             & 0.052 & 0.195 & 0.446 & 0.734 & 1.002 & 2.429                      &  & 1.375 & 1.375 & 1.447 & 1.447 & 1.451 & 7.096                      \\
                                & Moderate                       &                                             & 0.032 & 0.109 & 0.247 & 0.396 & 0.519 & 1.303                      &  & 0.413 & 0.413 & 0.407 & 0.407 & 0.402 & 2.043                      \\
\multirow{2}{*}{Multiplicative} & Mild                           &                                             & 0.051 & 0.190 & 0.438 & 0.723 & 0.988 & 2.390                      &  & 0.345 & 0.345 & 0.356 & 0.356 & 0.357 & 1.760                      \\
                                & Moderate                       &                                             & 0.038 & 0.141 & 0.333 & 0.560 & 0.774 & 1.847                      &  & 0.332 & 0.332 & 0.234 & 0.234 & 0.187 & 1.319                      \\ \hline
                                &                                & \multicolumn{7}{l}{Alberta}                                                                                      &  &       &       &       &       &       &                            \\
Naive                           & -                              & \multirow{5}{*}{ \textbf{AR(1)}\tnote{a}}                     & 0.002 & 0.007 & 0.027 & 0.055 & 0.070 & 0.160                      &  & 0.125 & 0.125 & 0.125 & 0.125 & 0.125 & 0.627                      \\
\multirow{2}{*}{Additive}       & Mild                           &                                             & 0.004 & 0.012 & 0.044 & 0.087 & 0.115 & 0.262                      &  & 0.115 & 0.115 & 0.115 & 0.115 & 0.115 & 0.577                      \\
                                & Moderate                       &                                             & 0.006 & 0.017 & 0.052 & 0.098 & 0.129 & 0.302                      &  & 0.035 & 0.035 & 0.035 & 0.035 & 0.035 & 0.177                      \\
\multirow{2}{*}{Multiplicative} & Mild                           &                                             & 0.004 & 0.012 & 0.044 & 0.087 & 0.115 & 0.263                      &  & 0.031 & 0.031 & 0.031 & 0.031 & 0.031 & 0.157                      \\
                                & Moderate                       &                                             & 0.005 & 0.013 & 0.045 & 0.089 & 0.118 & 0.270                      &  & 0.022 & 0.022 & 0.022 & 0.022 & 0.022 & 0.109                      \\ \cline{3-16} 
Naive                           & -                              & \multicolumn{1}{c}{\multirow{5}{*}{AR(2)}}  & 0.003 & 0.010 & 0.033 & 0.064 & 0.081 & 0.191                      &  & 0.122 & 0.122 & 0.125 & 0.125 & 0.125 & 0.621                      \\
\multirow{2}{*}{Additive}       & Mild                           & \multicolumn{1}{c}{}                        & 0.005 & 0.016 & 0.051 & 0.097 & 0.127 & 0.296                      &  & 0.112 & 0.112 & 0.115 & 0.115 & 0.115 & 0.570                      \\
                                & Moderate                       & \multicolumn{1}{c}{}                        & 0.006 & 0.018 & 0.056 & 0.104 & 0.136 & 0.320                      &  & 0.081 & 0.081 & 0.085 & 0.085 & 0.085 & 0.419                      \\
\multirow{2}{*}{Multiplicative} & Mild                           & \multicolumn{1}{c}{}                        & 0.005 & 0.016 & 0.052 & 0.099 & 0.129 & 0.301                      &  & 0.030 & 0.031 & 0.031 & 0.031 & 0.031 & 0.155                      \\
                                & Moderate                       & \multicolumn{1}{c}{}                        & 0.006 & 0.019 & 0.059 & 0.109 & 0.141 & 0.334                      &  & 0.022 & 0.022 & 0.022 & 0.022 & 0.022 & 0.109                      \\ \hline
\end{tabular}
\begin{tablenotes}
\item[a]{\scriptsize The selected model}
\end{tablenotes}

\end{threeparttable}
\end{sidewaystable}

\newpage
\thispagestyle{empty}

\setcounter{page}{1}
\setcounter{equation}{0}
\setcounter{section}{0}
\renewcommand\theequation{S.\arabic{equation}}
\renewcommand\thesection{S.\arabic{section}}


\renewcommand{\baselinestretch}{1.5}

\begin{center}
{\LARGE{\bf 
Supplementary Materials for ``Sensitivity Analysis of Error-Contaminated Time Series Data under Autoregressive Models with Application of COVID-19 Data"
}}
\end{center}
\baselineskip=22pt
\vskip 2mm


\appendix
\section{ Appendix}
\subsection{Regularity Conditions} \label{apx:Regularity-Cond}
\begin{itemize}
    \item[(R1)] The time series $\{X_t:t=1\ldots,T\}$ is stationary.
    \item[(R2)] The observed error-prone time series $\{X_t^*:t=1\ldots,T\}$ is stationary.
    \item[(R3)] For any $t \in \{1,\ldots, T\}$, $\frac{1}{T}\sum_{s=1}^T \gamma_{|s-t|} \to 0$ as $T \to \infty$.
    \item[(R4)] For any $p$, $\frac{1}{T}\sum_{t=1}^T\sum_{s=1}^TE\{(X_t-\mu)(X_{t+p}-\mu)(X_s-\mu)\}<\infty$. 
\end{itemize}

While the two process $\{X_t:t=1,\ldots, T\}$ and $\{X_t^*:t=1,\ldots, T\}$ are constrained by the measurement error model (\ref{eqn:measmodelClassic}) or (\ref{eqn:measmodelmultiplicative}), they can both be assumed to be stationary without inducing conflicting requirements on the associated processes. Obviously, the weak stationarity of $\{X_t:t=1,\ldots, T\}$ implies the weak stationarity of $\{X_t^*:t=1,\ldots, T\}$ if they are linked by (\ref{eqn:measmodelClassic}) or (\ref{eqn:measmodelmultiplicative}). 
Condition (R3) says that as the time series goes long enough, the average of the covariances between any paired variables is is negligible.  Condition (R4) requires the summation of the third moment of $X_t$ is $O(T)$, which is needed in Theorem 4 when $\phi_0\ne 0$; this condition can be satisfied if $E(\epsilon_t^3)=0$, for example. 

\subsection{ The proof of Theorem 1}  \label{apx:Pf-Thm1}
Applying the weak law of large numbers to $\widehat{\phi}_1^*$ given by (\ref{eqn:estimator-OLS}), we obtain that the estimator $\widehat{\phi}_1^*$  converges in probability to $\frac{\text{Cov}(X_t^*,X_{t-1}^*)}{\text{Var}(X_{t-1}^*)}$, which is denoted as $\phi_1^*$. Now we further examine $\phi_1^*$ by using the AR(1) model (\ref{eqn:timeseriresmodel}) and the measurement error model (\ref{eqn:measmodelClassic}): 
\begin{align*}
    \phi_1^* &= \frac{\text{Cov}(X_t^*,X_{t-1}^*)}{\text{Var}(X_{t-1}^*)} \\
    &=  \frac{\text{Cov}(\alpha_0+\alpha_1X_t+e_t,\alpha_0+\alpha_1X_{t-1}+e_{t-1})}{\text{Var}(\alpha_0+\alpha_1X_t+e_t)}  \\
    &=  \frac{\alpha_1^2 \text{Cov}(X_t,X_{t-1})}{\alpha_1^2 \text{Var}(X_t)+\text{Var}(e_t)}\\
    &=  \frac{\alpha_1^2 \text{Cov}(\phi_0+\phi_1X_{t-1}+\epsilon_t,X_{t-1})}{\alpha_1^2 \text{Var}(X_t)+\text{Var}(e_t)} \\
    &= \phi_1 \cdot \frac{\alpha_1^2 \text{Var}(X_{t-1})}{\alpha_1^2 \text{Var}(X_t)+Var(e_t)},
\end{align*}
where the second step is due to (\ref{eqn:measmodelClassic}), the third step is because of the independence among the $X_t$ and the $e_t$, and the fourth step is because of (\ref{eqn:timeseriresmodel}). Since the time series $\{X_t\}$ is stationary, it follows that $\text{Var}(X_t)=\text{Var}(X_{t-1})=\frac{\sigma_\epsilon^2}{1-\phi_1^2}$, and hence
\begin{align}\label{eqn:phistarproof-add}
    \phi_1^* &= \phi_1 \cdot \frac{\alpha_1^2 \sigma_\epsilon^2}{\alpha_1^2 \sigma_\epsilon^2+\sigma_e^2(1-\phi_1^2)} = \phi_1 \omega_1.
\end{align}

Next, applying the Slutsky's theorem to (\ref{eqn:estimator-OLS}), we have that as $T \to \infty$,
\begin{align*}
  \widehat{\phi}_0^* &\xrightarrow{p}  E(X_t^*) - \phi_1^* E(X_t^*),
\end{align*}
where the limit equals $\left(\alpha_0+\frac{\alpha_1\phi_0}{1-\phi_1}\right) (1-\phi_1\omega_1)$ by (\ref{eqn:phistarproof-add}) and the fact that  $E(X_t^*)=\alpha_0+\frac{\alpha_1\phi_0}{1-\phi_1}$. 

Finally, plugging the AR(1) model (\ref{eqn:timeseriresmodel}) into the measurement error model (\ref{eqn:working-p1}), we obtain that
\begin{equation} \label{eqn:X_tstar-add1}
       X_t^* = \alpha_0 + \alpha_1(\phi_0+\phi_1X_{t-1}+\epsilon_t) + e_t.
\end{equation}
On the other hand, plugging the measurement error model (\ref{eqn:measmodelClassic}) into the working model (\ref{eqn:working-p1}), we obtain that
\begin{align} \label{eqn:X_tstar-add2}
   X_t^* &= \phi_0^* + \phi_1^* (\alpha_0 + \alpha_1X_{t-1}+e_t) + \epsilon_t^*.
\end{align}
Then equating (\ref{eqn:X_tstar-add1}) and (\ref{eqn:X_tstar-add2}) that
\begin{equation*}
\epsilon^* = \alpha_0(1-\phi_1^*) + \alpha_1\phi_0 - \phi_0^* + \alpha_1(\phi_1-\phi_1^*)X_{t-1} + (1-\phi_1^*)e_t + \alpha_1\epsilon_t.
\end{equation*}
Consequently, by the independence assumption for $X_{t-1}$, $e_t$ and $\epsilon_t$, we obtain that
\begin{align*}
    Var(\epsilon_t^*) &= \phi_1^2\alpha_1^2 (1-\omega_1)^2 \text{Var}(X_{t-1}) + (1- \omega_1\phi_1)^2 \text{Var}(e_t) + \alpha_1^2 \text{Var}(\epsilon_t) \\ 
    &= \phi_1^2\alpha_1^2 (1-\omega_1)^2 \left(\frac{\sigma_\epsilon^2}{1-\phi_1^2}\right) + (1- \omega_1\phi_1)^2 \sigma_e^2 + \alpha_1^2 \sigma_\epsilon^2.
\end{align*}

\subsection{ The proof of Theorem 2} \label{apx:Pf-Thm2}

As noted in the beginning of \ref{apx:Pf-Thm1}, as $T\to\infty$, $\widehat{\phi}_1^* \xrightarrow{\,\,p\,\,} \phi_1^*$ where
\begin{align*}
\widehat{\phi}_1^* = \frac{\text{Cov}(X_t^*,X_{t-1}^*)}{\text{Var}(X_{t-1}^*)}.
\end{align*}

Now we further examine $\phi_1^*$ by using the AR(1) model (\ref{eqn:timeseriresmodel}) and the measurement error model (\ref{eqn:measmodelmultiplicative}):
\begin{align}\label{eqn:phi_1-pf}
    \phi_1^* 
    &= \frac{\text{Cov}(X_t^*,X_{t-1}^*)}{\text{Var}(X_{t-1}^*)} \nonumber\\
    &=  \frac{\text{Cov}(\beta_0u_tX_t,\beta_0u_{t-1}X_{t-1})}{\text{Var}(\beta_0u_{t-1}X_{t-1})}  \nonumber\\
    &=  \frac{\beta_0^2 \text{Cov}(u_tX_t,u_{t-1}X_{t-1})}{\beta_0^2 \text{Var}(u_{t-1}X_{t-1})} \nonumber\\
    &=  \frac{ \text{Cov}\{u_t(\phi_0 + \phi_1X_{t-1} + \epsilon_t),u_{t-1}X_{t-1}\}}{ \text{Var}(X_{t-1}u_{t-1})} \nonumber\\
    &=  \phi_1 \frac{ \text{Cov}(u_t X_{t-1},u_{t-1}X_{t-1})}{ \text{Var}(u_{t-1}X_{t-1})} \nonumber\\
    &=  \phi_1 \frac{ E(u_tu_{t-1} X_{t-1}^2) - E(u_{t}X_{t-1})E(u_{t-1}X_{t-1})}{ E(u_{t-1}^2X_{t-1}^2)-E^2(u_{t-1}X_{t-1})} \nonumber\\
    &=  \phi_1 \frac{ E(u_t)E(u_{t-1}) E(X_{t-1}^2) -  E(u_t)E(u_{t-1})E^2(X_{t-1})}{ E(u_{t-1}^2)E(X_{t-1}^2)-E^2(u_{t-1}X_{t-1})} \nonumber\\
    &=  \phi_1 \frac{ E(u_t)E(u_{t-1}) \text{Var}(X_{t-1})}{ \{\text{Var}(u_{t-1}) + E^2(u_{t-1})\}\{\text{Var}(X_{t-1}) + E^2(X_{t-1})\}-E^2(u_{t-1})E^2(X_{t-1})} \nonumber\\
    &= \phi_1 \frac{ \text{Var}(X_{t-1})}{ \{\text{Var}(u_{t-1}) + 1\}\{\text{Var}(X_{t-1}) + E^2(X_{t-1})\}-E^2(X_{t-1})} \nonumber\\
    &= \phi_1 \frac{ \text{Var}(X_{t-1})}{\text{Var}(u_{t-1})\text{Var}(X_{t-1}) + \text{Var}(u_{t-1})E^2(X_{t-1}) + \text{Var}(X_{t-1})},
\end{align}
where the second step is due to measurement error model (\ref{eqn:measmodelmultiplicative}), the seventh step is because $u_t$, $u_{t-1} $ and $X_{t-1}$ are mutually independent, and the second last step is due to $E(u_t)=1$. Since the time series $\{X_t\}$ is stationary, it follows that $E(X_t)=E(X_{t-1})=\frac{\phi_0}{1-\phi_1}$ and $\text{Var}(X_t)=\text{Var}(X_{t-1})=\frac{\sigma_\epsilon^2}{1-\phi_1^2}$. Hence (\ref{eqn:phi_1-pf}) becomes
\begin{align}\label{eqn:phistarproof-multi}
    \phi_1^*&=  \phi_1 \frac{ \text{Var}(X_{t-1})}{\text{Var}(u_{t-1})\text{Var}(X_{t-1}) + \text{Var}(u_{t-1})E^2(X_{t-1}) + \text{Var}(X_{t-1})} \nonumber\\
    &= \phi_1 \frac{ \frac{\sigma_\epsilon^2}{1-\phi_1^2}}{\sigma_u^2\frac{\sigma_\epsilon^2}{1-\phi_1^2} + \sigma_u^2 \left(\frac{\phi_0}{1-\phi_1}\right)^2 + \frac{\sigma_\epsilon^2}{1-\phi_1^2}} \nonumber\\
    &= \phi_1 \frac{\sigma_\epsilon^2}{\sigma_\epsilon^2\sigma_u^2+\sigma_\epsilon^2 +\sigma_u^2\phi_0^2\frac{1+\phi_1}{1-\phi_1}} = \phi_1\omega_2.
\end{align}

Next, applying the Slustky's Theorem to (\ref{eqn:estimator-OLS}) gives that as $T\to\infty$,
\begin{align*}
  \widehat{\phi}_0^* &\xrightarrow{\,\,\,p\,\,\,} \left(\frac{\beta_0\phi_0}{1-\phi_1}\right) (1-\phi_1\omega_2)
\end{align*}
by (\ref{eqn:phistarproof-multi}) as well as $E(X_t^*)=\frac{\beta_0\phi_0}{1-\phi_1}$. 

Finally plugging the AR(1) model (\ref{eqn:timeseriresmodel}) into the measurement error model (\ref{eqn:measmodelmultiplicative}), we obtain that
\begin{align} \label{eqn:X_tstar-multi1}
    X_t^* &= \beta_0(\phi_0+\phi_1X_{t-1}+\epsilon_t)u_t.
\end{align}
On the other hand, plugging the measurement error model (\ref{eqn:measmodelmultiplicative}) into the working model (\ref{eqn:working-p1}), we obtain that
\begin{align} \label{eqn:X_tstar-multi2}
   X_t^* &= \phi_0^* + \phi_1^* (\beta_0X_{t-1}u_{t-1}) + \epsilon_t^*.
\end{align}
Then equating (\ref{eqn:X_tstar-multi1}) and (\ref{eqn:X_tstar-multi2}) gives that
\begin{equation*}
    \epsilon^* = \beta_0\phi_0 u_t -\phi_0^* + \beta_0X_{t-1}(\phi_1 u_t - \omega_2\phi_1u_{t-1}) + \beta_0 u_t \epsilon_t.
\end{equation*}
yielding that
\begin{align*}
    Var(\epsilon_t^*) &= \phi_0^2\beta_0^2\text{Var}(u_t) +\beta_0^2\phi_1^2\text{Var}(X_{t-1}u_t) + \beta_0^2\omega_2^2\phi_1^2 \text{Var}(X_{t-1}u_{t-1})+ \beta_0^2 Var(u_t\epsilon_t) \\ 
    & =  \phi_0^2\beta_0^2\sigma_u^2  + (\beta_0^2\phi_1^2+\beta_0^2\omega_2^2\phi_1^2) \{E(X_{t-1}^2u_{t-1}^2)-E^2(X_t)E^2(u_{t-1})\}+ \beta_0^2 \{E(u_t^2)E(\epsilon_t^2)-E^2(u_t)E^2(\epsilon_t)\}\\
    & =  \phi_0^2\beta_0^2\sigma_u^2  + (\beta_0^2\phi_1^2+\beta_0^2\omega_2^2\phi_1^2)\{E(X_{t-1}^2)E(u_{t-1}^2)-E^2(X_t)E^2(u_{t-1})\}+ \beta_0^2  (\sigma_u^2+1)\sigma_\epsilon^2  \\
    & =  \beta_0^2\{\sigma_u^2\phi_0^2 + (1+\sigma_u^2)\sigma_\epsilon^2\} \\
    &  \qquad + \beta_0^2\phi_1^2(1+\omega_2^2)\left[\{\text{Var}(u_{t-1}) + E^2(u_{t-1})\}\{\text{Var}(X_{t-1}) + E^2(X_{t-1})\}-E^2(X_{t-1})\right]  \\
    & =  \beta_0^2\{\sigma_u^2\phi_0^2 + (1+\sigma_u^2)\sigma_\epsilon^2\} +  \beta_0^2\phi_1^2(1+\omega_2^2)\left[\{\text{Var}(u_{t-1}) + 1\}\{\text{Var}(X_{t-1}) + E^2(X_{t-1})\}-E^2(X_{t-1})\right]  \\
    & =  \beta_0^2\{\sigma_u^2\phi_0^2 + (1+\sigma_u^2)\sigma_\epsilon^2\} +  \beta_0^2\phi_1^2(1+\omega_2^2)\left\{\text{Var}(u_{t-1})\text{Var}(X_{t-1}) + \text{Var}(u_{t-1})E^2(X_{t-1}) + \text{Var}(X_{t-1})\right\}  \\
    & =  \beta_0^2\{\sigma_u^2\phi_0^2 + (1+\sigma_u^2)\sigma_\epsilon^2\} +  \beta_0^2\phi_1^2(1+\omega_2^2)\frac{Var(X_{t-1})}{\omega_2}  \\
    &= \beta_0^2\{\sigma_u^2\phi_0^2 + (1+\sigma_u^2)\sigma_\epsilon^2\}+\beta_0^2\phi_1^2\frac{1+\omega_2^2}{\omega_2}\frac{\sigma_\epsilon^2}{1-\phi_1^2},
\end{align*}
where the second step is because of the independence assumption as well as $E(u_{t-1}^2)=E(u_{t}^2)$ and $E(u_{t-1})=E(u_{t})$ such that $\text{Var}(X_{t-1}u_t) = \text{Var}(X_{t-1}u_{t-1})$, and the second last step is due to  $\omega_2=\frac{ \text{Var}(X_{t-1})}{\text{Var}(u_{t-1})\text{Var}(X_{t-1}) + \text{Var}(u_{t-1})E^2(X_{t-1}) + \text{Var}(X_{t-1})}$ in (\ref{eqn:phistarproof-multi}).

\subsection{ The proof of Theorem 3}

\underline{\bf Proof of Theorem~3(1):}\\

 For $k=1,\ldots,p$, applying the weak law of large numbers to $\widehat{\gamma}_k^*$, we obtain that as $T\to \infty$, the estimator $\widehat{\gamma}_k^*$  converges in probability to $\text{Cov} (X_t^*,X_{t-k}^*)$, denoted $\gamma_k^*$. 
 
 Next, we examine $\gamma_k$. By the form of measurement error model (\ref{eqn:measmodelClassic}), we have that for $0<k<t$,
\begin{align*}
    \text{Cov} (X_t^*,X_{t-k}^*) 
    &= \text{Cov} (\alpha_0 + \alpha_1 X_t + e_t, \alpha_0 + \alpha_1 X_{t-k}+e_{t-k}) \\
    &= \alpha_1^2 \text{Cov} (X_t,X_{t-k}) = \alpha_1^2 \gamma_k,
\end{align*}
and by (\ref{apx:gamma0star}), $\text{Var}(X_t^*) = \alpha_1^2 \gamma_0 + \sigma_e^2$, which is denoted as $\gamma_0^*$.

Thus, Theorem~3(1) follows.\\

\medskip

\noindent\underline{\bf Proof of Theorem~3(2)}:

First, by Theorem~3(1), we write
\begin{equation}\label{eqn:gammastarhat}
   \widehat{\gamma}^* = \alpha_1^2\gamma + o_p(1)
\end{equation}
and
\begin{equation*}\label{eqn:GAMMAstarhat}
    \widehat{\Gamma}^* = \alpha_1^2\Gamma+\sigma_e^2I_p + o_p(1),
\end{equation*}
where $\widehat{\Gamma}^* = \begin{pmatrix}
    \widehat{\gamma}_0^* & \cdots & \widehat{\gamma}_{p-1}^* \\
    \vdots   & \ddots & \vdots \\
    \widehat{\gamma}_{p-1}^*  & \cdots & \widehat{\gamma}_0^*
    \end{pmatrix}$. Then the naive estimator $\widehat{\phi}^*$ is obtained by replacing $\widehat{\gamma}_k$ in (\ref{eqn:esteqntrue}) with $\widehat{\gamma}_k^*$, 
\begin{equation}\label{apx:phistarhat}
  \widehat{\phi}^* =  \left\{\alpha_1^2 \Gamma + \sigma_e^2 I_p +o_p(1)\right\}^{-1}\left\{\alpha_1^2\gamma+o_p(1)\right\} 
  =  \alpha_1^2\left(\alpha_1^2 \Gamma + \sigma_e^2 I_p \right)^{-1}\gamma + o_p(1),
\end{equation}
and hence $\phi^*= \alpha_1^2\left(\alpha_1^2 \Gamma + \sigma_e^2 I_p \right)^{-1}\gamma$ such that $\widehat{\phi}^* \xrightarrow{\,\,p \,\,} \phi^*$ as $T \to \infty$.

Again, replacing $\widehat{\gamma}_k$ in (\ref{eqn:esteqntrue}) with $\widehat{\gamma}_k^*$ gives the naive estimator $\widehat{\phi}_0^*$
\begin{align*}
    \widehat{\phi}_0^* &=  \frac{1}{T-p}\sum_{t=p}^TX_t^* -  \left(\sum_{k=1}^p\widehat{\phi}_k^*\right) \left(\frac{1}{T-p} \sum_{t=p}^T X_{t-k}^*\right)\\
    &= E(X_t^*) - E(X_t^*) \sum_{k=1}^p \widehat{\phi}_k^* + o_p(1)\\
    &= \alpha_0 + \alpha_1 E(X_t) -\left\{\alpha_0 + \alpha_1 E(X_t) \right\}\sum_{k=1}^p \{\phi_k^*+o_p(1)\} + o_p(1)\\
    &= (1-\phi^{*\scalebox{0.5}{\rm T}}\cdot \mathds{1}_p)\left(\alpha_0 + \alpha_1 \mu \right) + o_p(1),
\end{align*}
where $\widehat{\phi}_k$ and $\phi_k$ are respectively the $k$th element of $\widehat{\phi}$ and $\phi$, the third step is because $\widehat{\phi}_k = \phi_k +o_p(1)$ by (\ref{apx:phistarhat}) as well as the model form (\ref{eqn:measmodelClassic}), and the last step is due to the stationarity of the time series $\{X_t\}$ such that $E(X_t)=\mu$.

Finally, noting that the native estimator  $\widehat{\sigma}_\epsilon^{2*}$ is given by $\widehat{\sigma}_\epsilon^{2*}= \widehat{\gamma}_0^* - 2 \widehat{\phi}^{*\scalebox{0.5}{T}}\widehat{\gamma}^* + \widehat{\phi}^{*\scalebox{0.5}{T}}\widehat{\Gamma}^*\widehat{\phi}^{*}$ by applying a version similar to (\ref{eqn:esteqntrue}), we obtain that
\begin{align*}
   \widehat{\sigma}_\epsilon^{2*}  &= \widehat{\gamma}_0^* - 2 \widehat{\phi}^{*\scalebox{0.5}{T}}\widehat{\gamma}^* + \widehat{\phi}^{*\scalebox{0.5}{T}}\widehat{\Gamma}^*\widehat{\phi}^{*}\\
    &= (\alpha_1^2 \gamma_0^2 + \sigma_e^2) -2\alpha_1^4\gamma^{\scalebox{0.5}{\rm T}}(\alpha_1^2 \Gamma + \sigma_e^2 I_p)^{-1} \gamma + \alpha_1^4 \gamma^{\scalebox{0.5}{\rm T}}(\alpha_1^2 \Gamma + \sigma_e^2 I_p)^{-1}(\alpha_1^2 \Gamma + \sigma_e^2 I_p)(\alpha_1^2 \Gamma + \sigma_e^2 I_p)^{-1}\gamma + o_p(1)\\
    &= \alpha_1^2\gamma_0 + \sigma_e^2 - \alpha_1^4 \gamma^{\scalebox{0.5}{\rm T}}(\alpha_1^2 \Gamma + \sigma_e^2 I_p)^{-1}\gamma  + o_p(1),
\end{align*}
where the second step is due to (\ref{apx:gamma0star}), (\ref{eqn:gammastarhat}) and (\ref{apx:phistarhat}).

\medskip

\noindent\underline{\bf Proof of Theorem~3(3):}

\noindent\underline{Step 1}: We show certain identities before proving Theorem~3(3).

1. By model (\ref{eqn:measmodelClassic}), we have that
\begin{align} \label{apx:The3.3.1}
    X_t^* - \widehat{\mu}^* &= \alpha_0 + \alpha_1 X_t + e_t - \frac{1}{T} \sum_{t=1}^T (\alpha_0 + \alpha_1 X_t + e_t ) \nonumber \\
     &= \alpha_1 \left( X_t - \frac{1}{T} \sum_{t=1}^T X_t \right) + \left(e_t   - \frac{1}{T} \sum_{t=1}^T e_t\right)   \nonumber \\
     &= \alpha_1 (X_t -\widehat{\mu}) + (e_t-\Bar{e}),
\end{align}
where the first step is because $\widehat{\mu}^* = \frac{1}{T} \sum_{t=1}^T X_t^*$ and in the last step $\bar{e}=\frac{1}{T} \sum_{t=1}^T e_t$.

2. For any $t$ and $s$, we have that 
\begin{align}\label{apx:Thm3-X_t2-X_se_s-Cov}
    & \text{Cov}\left\{(X_t-\widehat{\mu})^2,(X_s-\widehat{\mu})(e_s-\bar{e})\right\} \nonumber \\
    =&  E\{(X_t-\widehat{\mu})^2(X_s-\widehat{\mu})(e_s-\bar{e})\} -\{E(X_t-\widehat{\mu})^2\} E\{(X_s-\widehat{\mu})(e_s-\bar{e})\} \nonumber  \\
    =&  E\{(X_t-\widehat{\mu})^2(X_s-\widehat{\mu})\}E(e_s-\bar{e}) -\{E(X_t-\widehat{\mu})^2\} E(X_s-\widehat{\mu})E(e_s-\bar{e})\nonumber  \\
    =& 0,
\end{align}
where the second step is due to the independence of $e_t$ and $X_t$, and the last step is by $E(e_s-\bar{e})=0$.

3. By the independence of $e_t$ and $e_s$ for $t\ne s$, we have that
\begin{align}\label{apx:Thm3-X_t-e_t-Cov-tns}
     & \text{Cov}\left\{(X_t-\widehat{\mu})(e_t-\bar{e}),(X_s-\widehat{\mu})(e_s-\bar{e})\right\} \nonumber  \\
    =& E\{(X_t-\widehat{\mu})(e_t-\bar{e})(X_s-\widehat{\mu})(e_s-\bar{e})\} - E\{(X_t-\widehat{\mu})(e_t-\bar{e})\}E\{(X_s-\widehat{\mu})(e_s-\bar{e})\} \nonumber  \\
    =& E\{(X_t-\widehat{\mu})(X_s-\widehat{\mu})\}E\{(e_t-\bar{e})\}E\{(e_s-\bar{e})\} - E\{(X_t-\widehat{\mu})\}E\{(e_t-\bar{e})\}E\{(X_s-\widehat{\mu})\}E\{(e_s-\bar{e})\} \nonumber  \\
    =& 0,
\end{align}
where the second step is due to the independence of the $e_t$ and the $X_t$, and the last step is by \mbox{$E(e_s-\bar{e})=0$.}

4. For any $t$, we have that
\begin{align}\label{apx:Thm3-X_t-e_t-Var-tes}
     & \text{Var}\left\{(X_t-\widehat{\mu})(e_t-\bar{e})\right\} \nonumber \\
    =& E\{(X_t-\widehat{\mu})^2(e_t-\bar{e})^2\} - E^2\{(X_t-\widehat{\mu})(e_t-\bar{e})\}  \nonumber \\
    =& E\{(X_t-\widehat{\mu})^2\}E\{(e_t-\bar{e})^2\} - E^2\{(X_t-\widehat{\mu})\}E^2\{(e_t-\bar{e})\} \nonumber \\
    =& E\{(X_t-\widehat{\mu})^2\}E\{(e_t-\bar{e})^2\}.
\end{align}

5. For any $t$, we have
\begin{align}\label{apx:Thm3-X_t-mu2}
     & \lim\limits_{T\to\infty} E\{(X_t-\widehat{\mu})^2\} \nonumber \\
    =& \lim\limits_{T\to\infty} E\{(X_t-{\mu})^2 + (\mu-\widehat{\mu})^2 + 2(X_t-{\mu})(\mu-\widehat{\mu})\}  \nonumber \\
    =& \gamma_0 + \lim\limits_{T\to\infty} E\{  (\widehat{\mu}-\mu)^2 \} + 2 \lim\limits_{T\to\infty} E\{(X_t-{\mu})(\mu-\widehat{\mu}) \}  \nonumber \\
    =& \gamma_0 + \lim\limits_{T\to\infty} E\{  (\widehat{\mu}-\mu)^2 \} - 2 \lim\limits_{T\to\infty} E\left[(X_t-{\mu})\{\frac{1}{T}\sum_{s=1}^T(X_s - \mu)\} \right]  \nonumber \\
    =& \gamma_0 + \lim\limits_{T\to\infty} Var(\widehat{\mu}) - 2 \lim\limits_{T\to\infty} \frac{1}{T}\sum_{s=1}^T E\{(X_t-{\mu})(X_s-{\mu}) \} \nonumber \\
    =& \gamma_0 + 0 -  2 \lim\limits_{T\to\infty} \frac{1}{T}\sum_{s=1}^T \gamma_{|s-t|}  \nonumber \\
    =& \gamma_0,
\end{align}
where the third step is due to $\widehat{\mu}-\mu=\frac{1}{T}\sum_{s=1}(X_s-\mu)$, and the fourth step is because $E(\widehat{\mu}-\mu)=0$ by stationarity of the time series, the second last step is due to $\lim\limits_{T\to\infty} Var(\widehat{\mu})=0$ \citep[][Theorem 7.1.1.]{brockwell1991time}, and the last step due to Condition (R3).

6. Similar to (\ref{apx:Thm3-X_t-mu2}), we have that 
\begin{align}\label{apx:X_t-X_t-p-muhat}
    & \lim\limits_{T\to\infty} E\{(X_t-\widehat{\mu})(X_{t-p}-\widehat{\mu})\} \nonumber \\
   =& \lim\limits_{T\to\infty} E\{(X_t-\mu + \mu -\widehat{\mu})(X_{t-p}-\mu + \mu-\widehat{\mu})\}  \nonumber \\
   =& \lim\limits_{T\to\infty} \left[ E\{(X_t-\mu)(X_{t-p}-\mu)\} + E\{(\mu -\widehat{\mu})(X_{t-p}-\mu)\}
   + E\{(\mu -\widehat{\mu})(X_{t}-\mu)\} + E\{(\mu-\widehat{\mu})(\mu-\widehat{\mu})\} \right]  \nonumber \\
   =& \gamma_p + \lim\limits_{T\to\infty} \frac{1}{T}\sum_{s=1}^T (\gamma_{|t-s|} + \gamma_{|t-s-p|}) + \lim\limits_{T\to\infty} \text{Var}(\widehat{\mu}) \nonumber \\
   =& \gamma_p,
\end{align}
 where the last step is due to Condition (R3) and $\lim\limits_{T\to\infty} Var(\widehat{\mu})=0$ \citep[][Theorem 7.1.1]{brockwell1991time}.

7. For any $t$, we have 
\begin{align}\label{apx:Thm3-e_t2}
     &  E\{(e_t-\bar{e})^2\} \nonumber \\
    =& E\{e_t^2 - 2e_t\bar{e} + \bar{e}^2 \} \nonumber \\
    =&  \left\{ E(e_t^2) - \frac{2}{T}\sum_{s=1}^TE(e_te_s) + \frac{1}{T^2}\sum_{t=1}^T\sum_{s=1}^T E(e_te_s) \right\}\nonumber \\
    =& E(e_t^2) +  \left\{ - \frac{2}{T}E(e_te_t) + \frac{1}{T^2}\sum_{t=1}^TE(e_t^2) \right\} \nonumber \\
    =& \frac{T-1}{T} E(e_t^2) = \frac{T-1}{T}\sigma_e^2,
\end{align}
so $\lim\limits_{T\to\infty}E\{(e_t-\bar{e})^2\} = \sigma_e^2$.

8. By the independence of $e_t$ and $X_t$, for any $s$ and $t$, we have that
\begin{align}\label{apx:Thm3-X_t-e_t-3}
   & \text{Cov}\{(X_t-\widehat{\mu})(e_t-\bar{e}),(e_s-\bar{e})^2\} \nonumber \\
  =& E\{(X_t-\widehat{\mu})(e_t-\bar{e})(e_s-\bar{e})^2\} - E\{(X_t-\widehat{\mu})(e_t-\bar{e})\}E\{(e_s-\bar{e})^2\}  \nonumber  \\
  =& E(X_t-\widehat{\mu})E\{(e_t-\bar{e})(e_s-\bar{e})^2\}  -  E(X_t-\widehat{\mu})E(e_t-\bar{e})E(e_s-\bar{e})^2  \nonumber  \\
  =& 0,
\end{align}
where the last step is due to $E(X_t-\widehat{\mu}) = 0$ and $E(e_t-\bar{e})=0$.

9. For any $t\ne s$, $\text{Cov}\left\{(e_t-\bar{e})^2,(e_s-\bar{e})^2\right\}=0$;  and for $t=s$,
\begin{align}\label{apx:Thm3-e_t2-e_t2}
    & \text{Var}\{(e_t-\bar{e})^2\} \nonumber\\
   =& E\{(e_t-\bar{e})^4\} - E^2\{(e_t-\bar{e})^2\} \nonumber\\
   =& E(e_t^4) - 4 E(e_t^3\bar{e}) + 6 E(e_t^2\bar{e}^2) - 4 E(e_t\bar{e}^3) + E(\bar{e}_t^4)
    - \{E(e_t^2) - 2E(e_t\bar{e}) + E(\bar{e}^2)\}^2 \nonumber\\
   =& E(e_t^4) - \frac{4}{T} E(e_t^4) + \left[\frac{6(T-1)}{T^2} \{E(e_t^2)\}^2  + \frac{6}{T^2}E(e_t^4) \right]- \frac{4}{T^3} E(e_t^4) + \left[\frac{1}{T^3} E(e_t^4) + \frac{3(T-1)}{T^3} \{E(e_t^2)\}^2\right] \nonumber \\
    & \qquad - \left\{E(e_t^2) - \frac{2}{T}E(e_t^2) +\frac{1}{T}E(e_t^2)\right\}^2, 
\end{align}
so $\lim\limits_{T\to\infty}\text{Var}\{(e_t-\bar{e})^2\}= E(e_t^4) - \{E(e_t^2)\}^2 = E(e_t^4) - \sigma_e^4$. 

10. Similar to the derivation in  (\ref{apx:Thm3-e_t2-e_t2}), we can show $\text{Cov}\{(e_t-\bar{e})^2, (e_s-\bar{e})(e_{s+p}-\bar{e})\}=0$ for $s\ne t$ and $s\ne t-p$. For a given $t$,
\begin{align}\label{apx:Thm3-e_t-e_tp}
    & \text{Cov}\{(e_t-\bar{e})^2, (e_t-\bar{e})(e_{t+p}-\bar{e})\} \nonumber \\
   =& E\{(e_t-\bar{e})^3(e_{t+p}-\bar{e})\} - E\{(e_t-\bar{e})^2\}E\{(e_t-\bar{e})(e_{t+p}-\bar{e})\},
\end{align}
which can be derived analogously to the (\ref{apx:Thm3-e_t2-e_t2}) that $\lim_{T\to\infty} E\{(e_t-\bar{e})^3(e_{t+p}-\bar{e})\} - E\{(e_t-\bar{e})^2\}E\{(e_t-\bar{e})(e_{t+p}-\bar{e})\} = E\{e_t^3e_{t+p}\} - E\{e_t^2\} E\{e_te_{t+p}\}= 0$
 and similarly $\lim\limits_{T\to\infty}\text{Cov}\{(e_t-\bar{e})^2, (e_{t-p}-\bar{e})(e_{t}-\bar{e})\}=0$.

11. For any $t$,
\begin{align}\label{pf:p_01pr_p-r}
    &  \text{Cov}\left\{(X_t-\widehat{\mu})(e_{t+p}-\bar{e}),(X_{t+p-r}-\widehat{\mu})(e_{t+p}-\bar{e})\right\} \nonumber \\
    &= \left[E\left\{(X_{t}-\widehat{\mu})(X_{t+p-r}-\widehat{\mu})(e_{t+p}-\bar{e})^2\right\}
    - E(X_{t}-\widehat{\mu})E(X_{t+p-r}-\widehat{\mu})E^2(e_{t+p}-\bar{e}) \right] \nonumber\\
    &= E\left\{(X_{t}-\widehat{\mu})(X_{t+p-r}-\widehat{\mu})\right\}E\left\{(e_{t+p}-\bar{e})^2\right\}  \nonumber\\
    &= \gamma_{|p-r|}\left(\frac{T-1}{T}\right)\sigma_e^2,
\end{align}
where the second step is because of $E(X_{t}-\widehat{\mu})=0$ and the independence of $X_t$ and $e_t$, the third step is due to (\ref{apx:Thm3-e_t2}) and  (\ref{apx:X_t-X_t-p-muhat}). Hence,
\begin{equation*}
    \lim_{T\to\infty}\text{Cov}\left\{(X_t-\widehat{\mu})(e_{t+p}-\bar{e}),(X_{ t+p-r}-\widehat{\mu})(e_{t+p}-\bar{e})\right\} = \gamma_{|p-r|}\sigma_e^2.
\end{equation*} 

Similarly, 
\begin{equation*}
    \lim_{T\to\infty}\text{Cov}\left\{(X_{t+p}-\widehat{\mu})(e_t-\bar{e}),(X_{t-r}-\widehat{\mu})(e_t-\bar{e})\right\}=\gamma_{|p-r|}\sigma_e^2.
\end{equation*}

Then, similarly, 
\begin{align}\label{pf:p_01pr_p+r}
    & \text{Cov}\left\{(X_t-\widehat{\mu})(e_{t+p}-\bar{e}),(X_{t+p+r}-\widehat{\mu})(e_{t+p}-\bar{e})\right\} \nonumber\\
    &=  \left[ E\left\{(X_{t}-\widehat{\mu})(X_{t+p+r}-\widehat{\mu})(e_{t+p}-\bar{e})^2\right\}
    - E(X_{t}-\widehat{\mu})E(X_{t+p-r}-\widehat{\mu})E^2(e_{t+p}-\bar{e}) \right] \nonumber\\
    &=  E\left\{(X_{t}-\widehat{\mu})(X_{t+p+r}-\widehat{\mu})\right\}E\left\{(e_t-\bar{e})^2\right\} \nonumber\\
    &= \gamma_{p+r}\left(\frac{T-1}{T}\right)\sigma_e^2,
\end{align}
and hence $\lim_{T\to\infty}\text{Cov}\left\{(X_t-\widehat{\mu})(e_{t+p}-\bar{e}),(X_{t+p+r}-\widehat{\mu})(e_{t+p}-\bar{e})\right\}=\gamma_{p+r}\sigma_e^2$. Similarly,\\ $\lim_{T\to\infty}\text{Cov}\left\{(X_{t+p}-\widehat{\mu})(e_t-\bar{e}),(X_{t+r}-\widehat{\mu})(e_t-\bar{e})\right\} = \gamma_{p+r}\sigma_e^2$.

12. By independence assumption between $\{e_t\}$, if $t\ne s$ or $p\ne r$, we have that
\begin{equation}\label{apx:Thm3-et-etp-es-esr}
    \text{Cov}\left\{(e_t-\bar{e})(e_{t+p}-\bar{e}),(e_s-\bar{e})(e_{s+r}-\bar{e})\right\}=0.
\end{equation}
In addition, by (\ref{apx:Thm3-e_t2}), we have that
\begin{align}\label{apx:Thm3-Var-e_t-e_t+p}
& \text{Var}\left\{(e_t-\bar{e})(e_{t+p}-\bar{e})\right\} \nonumber \\
&=  E\left\{(e_t-\bar{e})^2(e_{t+p}-\bar{e})^2\right\} \nonumber \\
&= E\left\{(e_t-\bar{e})^2\right\}E\left\{(e_{t+p}-\bar{e})^2\right\}, \nonumber\\
&= \left(\frac{T-1}{T}\right)^2\sigma_e^4,
\end{align}
so $\lim_{T\to\infty}\text{Var}\left\{(e_t-\bar{e})(e_{t+p}-\bar{e})\right\}=\sigma_e^4$.

\hspace{0.2cm}

\noindent\underline{Step 2}: Now we prove the results in (3). \\

\noindent$1^{\circ}$. We first show the derivation of $q_{100}^*$ as follows:
\begin{align*}
    q_{100}^* &=  \lim_{T\to\infty} T \text{Cov}\left\{\frac{1}{T}\sum_{t=1}^T(X_t^*-\widehat{\mu}^*)^2,\frac{1}{T}\sum_{s=1}^T(X_s^*-\widehat{\mu}^*)^2\right\} \\
    &=  \lim_{T\to\infty} T \text{Cov}\left[\frac{1}{T}\sum_{t=1}^T\left\{\alpha_1^2(X_t-\widehat{\mu})^2+2\alpha_1(X_t-\widehat{\mu})(e_t-\bar{e}) +(e_t-\bar{e})^2 \right\},\right.\\
    & \left. \qquad \qquad \frac{1}{T}\sum_{s=1}^T\alpha_1^2(X_s-\widehat{\mu})^2+2\alpha_1(X_s-\widehat{\mu})(e_s-\bar{e}) +(e_s-\bar{e})^2\right] \\
    &= \alpha_1^4 \lim_{T\to\infty} T \text{Cov}\left\{\frac{1}{T}\sum_{t=1}^T(X_t-\widehat{\mu})^2,\frac{1}{T}\sum_{s=1}^T(X_s-\widehat{\mu})^2\right\} \\
    & \qquad \qquad + \lim_{T\to\infty} T \text{Cov}\left\{\frac{1}{T}\sum_{t=1}^T2\alpha_1(X_t-\widehat{\mu})(e_t-\bar{e}),\frac{1}{T}\sum_{s=1}^T2\alpha_1(X_s-\widehat{\mu})(e_s-\bar{e})\right\} \\
    & \qquad \qquad + \lim_{T\to\infty} T \text{Cov}\left\{\frac{1}{T}\sum_{t=1}^T(e_t-\bar{e})^2,\frac{1}{T}\sum_{s=1}^T(e_s-\bar{e})^2\right\} \\
   &= \alpha_1^4 q_{00} + \lim_{T\to\infty} \frac{4\alpha_1^2}{T} \sum_{t=1}^T\sum_{s=1}^T\text{Cov}\left\{(X_t-\widehat{\mu})(e_t-\bar{e}),(X_s-\widehat{\mu})(e_s-\bar{e})\right\} \\
    & \qquad \qquad + \lim_{T\to\infty} \frac{1}{T}\sum_{t=1}^T\sum_{s=1}^T \text{Cov}\left\{(e_t-\bar{e})^2,(e_s-\bar{e})^2\right\} \\
   &= \alpha_1^4 q_{00}  + \lim_{T\to\infty} \frac{4\alpha_1^2}{T} \sum_{t=1}^T\text{Cov}\left\{(X_t-\widehat{\mu})(e_t-\bar{e}),(X_t-\widehat{\mu})(e_t-\bar{e})\right\} \\
   & \qquad \qquad + \lim_{T\to\infty} \frac{1}{T}\sum_{t=1}^T \text{Cov}\left\{(e_t-\bar{e})^2,(e_t-\bar{e})^2\right\} \\
   &= \alpha_1^4 q_{00}  + 4\alpha_1^2E\left\{(X_t-\widehat{\mu})^2(e_t-\bar{e})^2\right\} +  E(e_t^4) - \left\{E(e_t^2)\right\}^2 \\
   &= \alpha_1^4 q_{00} + 4\alpha_1^2\gamma_0\sigma_e^2 + E(e_t^4) -\sigma_e^4,
\end{align*}
where the second step is due to (\ref{apx:The3.3.1}), the third step is because of (\ref{apx:Thm3-X_t2-X_se_s-Cov}), (\ref{apx:Thm3-X_t-e_t-3}), and the definition $q_{00}=\lim_{T\to\infty} T \text{Cov}\left\{\frac{1}{T}\sum_{t=1}^T(X_t-\widehat{\mu})^2,\frac{1}{T}\sum_{s=1}^T(X_s-\widehat{\mu})^2\right\}$, the fifth step is due to (\ref{apx:Thm3-X_t-e_t-Cov-tns}) and (\ref{apx:Thm3-e_t2-e_t2}), and the sixth step is because (\ref{apx:Thm3-X_t-e_t-Var-tes}) and (\ref{apx:Thm3-e_t2-e_t2}), and the last step is because (\ref{apx:Thm3-e_t2}) and (\ref{apx:Thm3-X_t-e_t-3}).


\noindent$2^{\circ}$. We derive the value of $q_{10p}^*$:
\begin{align*}
     q_{10p}^* &=  \lim_{T\to\infty} T \text{Cov}\left\{\frac{1}{T}\sum_{t=1}^T(X_t^*-\widehat{\mu}^*)^2,\frac{1}{T-p}\sum_{s=1}^{T-p}(X_s^*-\widehat{\mu}^*)(X_{s+p}^*-\widehat{\mu}^*)\right\} \\
      &=  \lim_{T\to\infty} T \text{Cov}\left[\frac{1}{T}\sum_{t=1}^T\left\{\alpha_1^2(X_t-\widehat{\mu})^2+2\alpha_1(X_t-\widehat{\mu})(e_t-\bar{e}) +(e_t-\bar{e})^2 \right\},\right.\\
    &  \qquad \qquad \frac{1}{T-p}\sum_{s=1}^{T-p}\alpha_1^2(X_s-\widehat{\mu})(X_{s+p}-\widehat{\mu})+\alpha_1(X_s-\widehat{\mu})(e_{s+p}-\bar{e}) \\
    &  \qquad \qquad +\alpha_1(X_{s+p}-\widehat{\mu})(e_{s}-\bar{e}) +(e_s-\bar{e})(e_{s+p}-\bar{e})\Bigg] \\
    &= \alpha_1^4 \lim_{T\to\infty} T \text{Cov}\left\{\frac{1}{T}\sum_{t=1}^T(X_t-\widehat{\mu})^2,\frac{1}{T-p}\sum_{s=1}^{T-p}(X_s-\widehat{\mu})(X_{s+p}-\widehat{\mu})\right\} \\
    & \qquad \qquad + \lim_{T\to\infty} T \text{Cov}\left\{\frac{1}{T}\sum_{t=1}^T2\alpha_1(X_t-\widehat{\mu})(e_t-\bar{e}),\frac{1}{T-p}\sum_{s=1}^{T-p}\alpha_1(X_s-\widehat{\mu})(e_{s+p}-\bar{e})\right\} \\
    & \qquad \qquad + \lim_{T\to\infty} T \text{Cov}\left\{\frac{1}{T}\sum_{t=1}^T2\alpha_1(X_t-\widehat{\mu})(e_t-\bar{e}),\frac{1}{T-p}\sum_{s=1}^{T-p}\alpha_1(X_{s+p}-\widehat{\mu})(e_{s}-\bar{e})\right\} \\
    & \qquad \qquad + \lim_{T\to\infty} T \text{Cov}\left\{\frac{1}{T}\sum_{t=1}^T(e_t-\bar{e})^2,\frac{1}{T-p}\sum_{s=1}^{T-p}(e_s-\bar{e})(e_{s+p}-\bar{e})\right\} \\
   &= \alpha_1^4q_{0p} + \lim_{T\to\infty} \frac{2\alpha_1^2}{T-p} \sum_{t=1}^T\sum_{s=1}^{T-p}\text{Cov}\left\{(X_t-\widehat{\mu})(e_t-\bar{e}),(X_s-\widehat{\mu})(e_{s+p}-\bar{e_s})\right\} \\
   & \qquad \qquad + \lim_{T\to\infty} \frac{2\alpha_1^2}{T-p} \sum_{t=1}^{T}\sum_{s=1}^{T-p}\text{Cov}\left\{(X_t-\widehat{\mu})(e_t-\bar{e}),(X_{s+p}-\widehat{\mu})(e_{s}-\bar{e})\right\} \\
   &=\alpha_1^4 q_{0p}  + \lim_{T\to\infty} \frac{2\alpha_1^2}{T-p} \sum\limits_{\substack{t=p \\ (s=t-p)}}^T\text{Cov}\left\{(X_t-\widehat{\mu})(e_t-\bar{e}),(X_{t-p}-\widehat{\mu})(e_t-\bar{e})\right\} \\
   & \qquad \qquad + \lim_{T\to\infty} \frac{2\alpha_1^2}{T-p} \sum\limits_{\substack{t=1 \\ (s=t)}}^{T-p}\text{Cov}\left\{(X_t-\widehat{\mu})(e_t-\bar{e}),(X_{t+p}-\widehat{\mu})(e_t-\bar{e})\right\} \\
   &=\alpha_1^4 q_{0p}  + 2\alpha_1^2E\left\{(X_t-\widehat{\mu})(X_{t-p}-\widehat{\mu})(e_t-\bar{e})^2\right\}  + 2\alpha_1^2E\left\{(X_t-\widehat{\mu})(X_{t+p}-\widehat{\mu})(e_t-\bar{e})^2\right\}  \\
   &=\alpha_1^4 q_{0p} + 4\alpha_1^2\gamma_p\sigma_e^2,
\end{align*}
where the second step is due to (\ref{apx:The3.3.1}), the third step is because of (\ref{apx:Thm3-X_t2-X_se_s-Cov}) and  (\ref{apx:Thm3-X_t-e_t-3}), the fourth step is by definition that $q_{0p}=\lim_{T\to\infty} T \text{Cov}\left\{\frac{1}{T}\sum_{t=1}^T(X_t-\widehat{\mu})^2,\frac{1}{T-p}\sum_{s=1}^{T-p}(X_s-\widehat{\mu})(X_{s+p}-\widehat{\mu})\right\}$ and (\ref{apx:Thm3-e_t-e_tp}), the fifth step is due to (\ref{apx:Thm3-X_t-e_t-Cov-tns}), and the last step is result from (\ref{apx:Thm3-e_t2}) and (\ref{apx:X_t-X_t-p-muhat}).

\noindent$3^{\circ}$. We derive $q_{1pr}^*$ for $p>0$, $r>0$ and $p\ne r$:
\begin{align}\label{pf:q_1prstar}
   q_{1pr}^* &=  \lim_{T\to\infty} T \text{Cov}\left\{\frac{1}{T-p}\sum_{t=1}^{T-p}(X_t^*-\widehat{\mu}^*)(X_{t+p}^*-\widehat{\mu}^*),\frac{1}{T-r}\sum_{s=1}^{T-r}(X_s^*-\widehat{\mu}^*)(X_{s+r}^*-\widehat{\mu}^*)\right\} \nonumber \\
    &=  \lim_{T\to\infty} T \text{Cov}\Bigg[\frac{1}{T-p}\sum_{t=1}^{T-p}\left\{\alpha_1^2(X_t-\widehat{\mu})(X_{t+p}-\widehat{\mu}) +\alpha_1(X_{t}-\widehat{\mu})(e_{t+p}-\bar{e}) \right. \nonumber\\
    & \left. \qquad \qquad +\alpha_1(X_{t+p}-\widehat{\mu})(e_{t}-\bar{e}) +(e_t-\bar{e})(e_{t+p}-\bar{e})  \right\},\nonumber\\
    &  \qquad \qquad \frac{1}{T-r}\sum_{s=1}^{T-r}\alpha_1^2(X_s-\widehat{\mu})(X_{s+r}-\widehat{\mu})+\alpha_1(X_s-\widehat{\mu})(e_{s+r}-\bar{e}) \nonumber\\
    &  \qquad \qquad +\alpha_1(X_{s+r}-\widehat{\mu})(e_{s}-\bar{e}) +(e_s-\bar{e})(e_{s+r}-\bar{e})\Bigg] \nonumber\\
    &= \alpha_1^4 \lim_{T\to\infty} T \text{Cov}\left\{\frac{1}{T-p}\sum_{t=1}^{T-p}(X_t-\widehat{\mu})(X_{t+p}-\widehat{\mu}),\frac{1}{T-r}\sum_{s=1}^{T-r}(X_s-\widehat{\mu})(X_{s+r}-\widehat{\mu})\right\} \nonumber\\
    & \qquad \qquad + \lim_{T\to\infty} T \text{Cov}\left\{\frac{1}{T-p}\sum_{t=1}^{T-p}\alpha_1(X_t-\widehat{\mu})(e_{t+p}-\bar{e}),\frac{1}{T-r}\sum_{s=1}^{T-r}\alpha_1(X_s-\widehat{\mu})(e_{s+r}-\bar{e})\right\} \nonumber\\
    & \qquad \qquad + \lim_{T\to\infty} T \text{Cov}\left\{\frac{1}{T-p}\sum_{t=1}^{T-p}\alpha_1(X_t-\widehat{\mu})(e_{t+p}-\bar{e}),\frac{1}{T-r}\sum_{s=1}^{T-r}\alpha_1(X_{s+r}-\widehat{\mu})(e_{s}-\bar{e})\right\} \nonumber\\
    & \qquad \qquad + \lim_{T\to\infty} T \text{Cov}\left\{\frac{1}{T-p}\sum_{t=1}^{T-p}\alpha_1(X_{t+p}-\widehat{\mu})(e_t-\bar{e}),\frac{1}{T-r}\sum_{s=1}^{T-r}\alpha_1(X_s-\widehat{\mu})(e_{s+r}-\bar{e})\right\}  \nonumber\\
    & \qquad \qquad + \lim_{T\to\infty} T \text{Cov}\left\{\frac{1}{T-p}\sum_{t=1}^{T-p}\alpha_1(X_{t+p}-\widehat{\mu})(e_t-\bar{e}),\frac{1}{T-r}\sum_{s=1}^{T-r}\alpha_1(X_{s+r}-\widehat{\mu})(e_{s}-\bar{e})\right\} \nonumber\\
   &=\alpha_1^4 q_{pr}  + \alpha_1^2\lim_{T\to\infty} \frac{T}{(T-p)(T-r)} \sum\limits_{\substack{t=\text{max}(1,r-p+1) \\ (s=t+p-r)}}^{T-p}\text{Cov}\left\{(X_t-\widehat{\mu})(e_{t+p}-\bar{e}),(X_{t+p-r}-\widehat{\mu})(e_{t+p}-\bar{e})\right\}\nonumber\\
   & \qquad + \alpha_1^2\lim_{T\to\infty} \frac{T}{(T-p)(T-r)}  \sum\limits_{\substack{t=1 \\ (s=t+p)}}^{T-p-r}\text{Cov}\left\{(X_t-\widehat{\mu})(e_{t+p}-\bar{e}),(X_{t+p+r}-\widehat{\mu})(e_{t+p}-\bar{e})\right\} \nonumber\\
   & \qquad + \alpha_1^2\lim_{T\to\infty} \frac{T}{(T-p)(T-r)}  \sum\limits_{\substack{t=r+1 \\ (s=t-r)}}^{T-p}\text{Cov}\left\{(X_{t+p}-\widehat{\mu})(e_t-\bar{e}),(X_{t-r}-\widehat{\mu})(e_{t}-\bar{e})\right\} \nonumber\\
   & \qquad + \alpha_1^2\lim_{T\to\infty} \frac{T}{(T-p)(T-r)}  \sum\limits_{\substack{t=1 \\ (s=t)}}^{T-\max(p,r)}\text{Cov}\left\{(X_{t+p}-\widehat{\mu})(e_t-\bar{e}),(X_{t+r}-\widehat{\mu})(e_t-\bar{e})\right\} \nonumber \\
   &= \alpha_1^4 q_{pr} + 2\alpha_1^2 \sigma_e^2 (\gamma_{|p-r|}+\gamma_{p+r}),
\end{align}
where the second step is due to (\ref{apx:The3.3.1}),  the third step is because of (\ref{apx:Thm3-X_t2-X_se_s-Cov}) and  a similar version to (\ref{apx:Thm3-X_t-e_t-3}), the fourth step is because (\ref{apx:Thm3-et-etp-es-esr}) and by the definition that\\ $q_{pr}=  \lim_{T\to\infty} T \text{Cov}\left\{\frac{1}{T-p}\sum_{t=1}^{T-p}(X_t-\widehat{\mu})(X_{t+p}-\widehat{\mu}),\frac{1}{T-r}\sum_{s=1}^{T-r}(X_s-\widehat{\mu})(X_{s+r}-\widehat{\mu})\right\}$, and the last step is from (\ref{pf:p_01pr_p-r}) and (\ref{pf:p_01pr_p+r}).

\noindent$4^{\circ}$. Finally, we present the derivation of $q_{1pp}^*$ for $p\ne 0$,
\begin{align}\label{pf:q_1ppstar}
   q_{1pp}^* &=  \lim_{T\to\infty} T \text{Cov}\left\{\frac{1}{T-p}\sum_{t=1}^{T-p}(X_t^*-\widehat{\mu}^*)(X_{t+p}^*-\widehat{\mu}^*),\frac{1}{T-p}\sum_{s=1}^{T-p}(X_s^*-\widehat{\mu}^*)(X_{s+p}^*-\widehat{\mu}^*)\right\} \nonumber \\
    &=  \lim_{T\to\infty} T \text{Cov}\Bigg[\frac{1}{T-p}\sum_{t=1}^{T-p}\left\{\alpha_1^2(X_t-\widehat{\mu})(X_{t+p}-\widehat{\mu}) +\alpha_1(X_{t}-\widehat{\mu})(e_{t+p}-\bar{e}) \right. \nonumber\\
    & \left. \qquad \qquad +\alpha_1(X_{t+p}-\widehat{\mu})(e_{t}-\bar{e}) +(e_t-\bar{e})(e_{t+p}-\bar{e})  \right\},\nonumber\\
    &  \qquad \qquad \frac{1}{T-p}\sum_{s=1}^{T-p}\alpha_1^2(X_s-\widehat{\mu})(X_{s+p}-\widehat{\mu})+\alpha_1(X_s-\widehat{\mu})(e_{s+p}-\bar{e}) \nonumber\\
    &  \qquad \qquad +\alpha_1(X_{s+p}-\widehat{\mu})(e_{s}-\bar{e}) +(e_s-\bar{e})(e_{s+p}-\bar{e})\Bigg] \nonumber\\
    &= \alpha_1^4 \lim_{T\to\infty} T \text{Cov}\left\{\frac{1}{T-p}\sum_{t=1}^{T-p}(X_t-\widehat{\mu})(X_{t+p}-\widehat{\mu}),\frac{1}{T-p}\sum_{s=1}^{T-p}(X_s-\widehat{\mu})(X_{s+p}-\widehat{\mu})\right\} \nonumber\\
    & \qquad \qquad + \lim_{T\to\infty} T \text{Cov}\left\{\frac{1}{T-p}\sum_{t=1}^{T-p}\alpha_1(X_t-\widehat{\mu})(e_{t+p}-\bar{e}),\frac{1}{T-p}\sum_{s=1}^{T-p}\alpha_1(X_s-\widehat{\mu})(e_{s+p}-\bar{e})\right\} \nonumber\\
    & \qquad \qquad + \lim_{T\to\infty} T \text{Cov}\left\{\frac{1}{T-p}\sum_{t=1}^{T-p}\alpha_1(X_t-\widehat{\mu})(e_{t+p}-\bar{e}),\frac{1}{T-p}\sum_{s=1}^{T-p}\alpha_1(X_{s+p}-\widehat{\mu})(e_{s}-\bar{e})\right\} \nonumber\\
    & \qquad \qquad + \lim_{T\to\infty} T \text{Cov}\left\{\frac{1}{T-p}\sum_{t=1}^{T-p}\alpha_1(X_{t+p}-\widehat{\mu})(e_t-\bar{e}),\frac{1}{T-p}\sum_{s=1}^{T-p}\alpha_1(X_s-\widehat{\mu})(e_{s+p}-\bar{e})\right\}  \nonumber\\
    & \qquad \qquad + \lim_{T\to\infty} T \text{Cov}\left\{\frac{1}{T-p}\sum_{t=1}^{T-p}\alpha_1(X_{t+p}-\widehat{\mu})(e_t-\bar{e}),\frac{1}{T-p}\sum_{s=1}^{T-p}\alpha_1(X_{s+p}-\widehat{\mu})(e_{s}-\bar{e})\right\} \nonumber\\
    & \qquad \qquad + \lim_{T\to\infty} T \text{Cov}\left\{\frac{1}{T-p}\sum_{t=1}^{T-p}(e_t-\bar{e})(e_{t+p}-\bar{e}),\frac{1}{T-p}\sum_{s=1}^{T-p}(e_s-\bar{e})(e_{s+p}-\bar{e})\right\} \nonumber\\
   &=\alpha_1^4 q_{pp} + \alpha_1^2\lim_{T\to\infty} \frac{T}{(T-p)^2} \sum\limits_{\substack{t=1 \\ s=t}}^{T-p}\text{Cov}\left\{(X_t-\widehat{\mu})(e_{t+p}-\bar{e}),(X_{t}-\widehat{\mu})(e_{t+p}-\bar{e})\right\}\nonumber\\
   & \qquad + \alpha_1^2\lim_{T\to\infty} \frac{T}{(T-p)^2} \sum\limits_{\substack{t=1 \\ s=t+p}}^{T-2p}\text{Cov}\left\{(X_t-\widehat{\mu})(e_{t+p}-\bar{e}),(X_{t+2p}-\widehat{\mu})(e_{t+p}-\bar{e})\right\} \nonumber\\
   & \qquad + \alpha_1^2\lim_{T\to\infty} \frac{T}{(T-p)^2} \sum\limits_{\substack{t=1+p \\ s=t-p}}^{T-p}\text{Cov}\left\{(X_{t+p}-\widehat{\mu})(e_t-\bar{e}),(X_{t-p}-\widehat{\mu})(e_{t}-\bar{e})\right\} \nonumber\\
   & \qquad + \alpha_1^2\lim_{T\to\infty} \frac{T}{(T-p)^2} \sum\limits_{\substack{t=1 \\ s=t}}^T\text{Cov}\left\{(X_{t+p}-\widehat{\mu})(e_t-\bar{e}),(X_{t+p}-\widehat{\mu})(e_t-\bar{e})\right\}\nonumber\\
   & \qquad + \alpha_1^2\lim_{T\to\infty} \frac{T}{(T-p)^2} \text{Var}\left\{(e_t-\bar{e})(e_{t+p}-\bar{e})\right\} = \alpha_1^4 q_{pp} + 2\alpha_1^2 \sigma_e^2 (\gamma_{0}+\gamma_{2p})  + \sigma_e^4,
\end{align}
where  the second step is due to (\ref{apx:The3.3.1}),  the third step is because of (\ref{apx:Thm3-X_t2-X_se_s-Cov}) and  a similar version to (\ref{apx:Thm3-X_t-e_t-3}), the fourth step is because (\ref{apx:Thm3-et-etp-es-esr}) and by the definition that\\ $q_{pp}=  \lim_{T\to\infty} T \text{Cov}\left\{\frac{1}{T-p}\sum_{t=1}^{T-p}(X_t-\widehat{\mu})(X_{t+p}-\widehat{\mu}),\frac{1}{T-p}\sum_{s=1}^{T-p}(X_s-\widehat{\mu})(X_{s+p}-\widehat{\mu})\right\}$, the last step is because of (\ref{apx:Thm3-Var-e_t-e_t+p}), and (\ref{pf:p_01pr_p-r}) and (\ref{pf:p_01pr_p+r}) with $q=p$.

\subsection{ The proof of Theorem 4 }

\underline{\bf Proof of Theorem~4(1):}

For $k=1,\ldots,p$, applying the weak law of large numbers to $\widehat{\gamma}_k^*$, we obtain that as $T\to \infty$, the estimator $\widehat{\gamma}_k^*$  converges in probability to $\text{Cov} (X_t^*,X_{t-k}^*)$, which is denoted as $\gamma_k^*$. 

 Next, we examine $\gamma_k$. By the form of measurement error model (\ref{eqn:measmodelmultiplicative}), we have that for $0<k<t$,
\begin{align*}
  & \text{Cov} (X_t^*,X_{t-k}^*)\\
  &= \text{Cov} (\beta_0 X_t u_t, \beta_0 X_{t-k}u_{t-k}) \\
    &= \beta_0^2 \{E(X_tu_tX_{t-k}u_{t-k})-E(X_tu_t)E(X_{t-k}u_{t-k})\}\\
    &= \beta_0^2 \{E(u_t)E(u_{t-k})\text{Cov}(X_t,X_{t-k})\} \\
    &= \beta_0^2 \{\text{Cov}(X_t,X_{t-k})\}= \beta_0^2 \gamma_k,
\end{align*}
and by (\ref{apx:Thm4-gamma0star}), $\text{Var}(X_t^*)= \beta_0^2 \left\{(\sigma_u^2+1) \gamma_0 + \sigma_u^2  \mu^2\right\}$, which is denoted as $\gamma_0^*$. Thus, Theorem~4(1) follows.\\

\medskip

\noindent\underline{\bf Proof of Theorem~4(2)}:

First, by Theorem~4(1), we write
\begin{equation*}
   \widehat{\gamma}^* = \beta_0^2 \gamma + o_p(1)
\end{equation*}
and
\begin{align*}
    \widehat{\Gamma}^* &= \begin{pmatrix}
    \beta_0^2(\sigma_u^2+1)\gamma_0 +\beta_0\sigma_u^2\mu^2 & \beta_0^2 \gamma_1 & \cdots & \beta_0^2\gamma_{p-1} \\
    \vdots &  & \ddots & \vdots \\
    \beta_0^2\gamma_{p-1} & \beta_0^2\gamma_{p-2}   & \cdots & \beta_0^2(\sigma_u^2+1)\gamma_0 +\beta_0\sigma_u^2\mu^2
    \end{pmatrix} + o_p(1)\\
    &= \beta_0^2 \left\{ \Gamma + \sigma_u^2 (\gamma_0 + \mu^2) I_p\right\} + o_p(1).
\end{align*}
where $\widehat{\Gamma}^* = \begin{pmatrix}
    \widehat{\gamma}_0^* & \cdots & \widehat{\gamma}_{p-1}^* \\
    \vdots   & \ddots & \vdots \\
    \widehat{\gamma}_{p-1}^*  & \cdots & \widehat{\gamma}_0^*
    \end{pmatrix}$. Then the naive estimator $\widehat{\phi}^*$ is obtained by replacing $\widehat{\gamma}_k$ in (\ref{eqn:esteqntrue}) with $\widehat{\gamma}_k^*$, 
\begin{equation}\label{apx:Th4-phihat-star}
    \widehat{\phi}^* = [ \beta_0^2 \left\{ \Gamma + \sigma_u^2 (\gamma_0 + \mu^2) I_p\right\} + o_p(1)]^{-1} \{\beta_0^2 \gamma + o_p(1)\}  = \left\{ \Gamma + \sigma_u^2 (\gamma_0+\mu^2) I_p\right\}^{-1}\gamma + o_p(1),
\end{equation}
and hence $\phi^*= \left\{ \Gamma + \sigma_u^2 (\gamma_0+\mu^2) I_p\right\}^{-1}\gamma$ such that $\widehat{\phi}^* \xrightarrow{\,\,p \,\,} \phi^*$ as $T \to \infty$.

Again, by replacing $\widehat{\gamma}_k$ in (\ref{eqn:esteqntrue}) with $\widehat{\gamma}_k^*$ gives the naive estimator $\widehat{\phi}_0^*$
\begin{align*}
    \widehat{\phi}_0^* &=  \frac{1}{T-p}\sum_{t=p}^TX_t^* - \left( \sum_{k=1}^p\widehat{\phi}_k^* \right) \left(\frac{1}{T-p} \sum_{t=p}^T X_{t-k}^*\right)\\
    &= E(X_t^*) - E(X_t^*) \sum_{k=1}^p \widehat{\phi}_k^* + o_p(1) \\
    &= \beta_0 E(X_t) - \beta_0 E(X_t)\sum_{k=1}^p \{\phi_k^*+o_p(1)\} + o_p(1) \\
    &= \beta_0 (1-\phi^{*\scalebox{0.5}{\rm T}} \mathds{1}_p)\mu + o_p(1),
\end{align*}
where $\widehat{\phi}_k$ and $\phi_k$ are respectively the $k$th element of $\widehat{\phi}$ and $\phi$, the third step is because $\widehat{\phi}_k = \phi_k +o_p(1)$ by (\ref{apx:Th4-phihat-star}) as well as the model form (\ref{eqn:measmodelmultiplicative}), and the last step is due to the stationarity of the time series $\{X_t\}$ such that $E(X_t)=\mu$.

Finally, noting that the native estimator  $\widehat{\sigma_\epsilon}^{*2}$ is given by $\widehat{\sigma}_\epsilon^{*2}= \widehat{\gamma}_0^* - 2 \widehat{\phi}^{*\scalebox{0.5}{T}}\widehat{\gamma}^* + \widehat{\phi}^{*\scalebox{0.5}{T}}\widehat{\Gamma}^*\widehat{\phi}^{*}$ by applying a version similar to (\ref{eqn:esteqntrue}), we obtain that
\begin{align*}
    \widehat{\sigma}_\epsilon^{*2}  =& \widehat{\gamma}_0^* - 2 \widehat{\phi}^{*\scalebox{0.5}{T}}\widehat{\gamma}^* + \widehat{\phi}^{*\scalebox{0.5}{T}}\widehat{\Gamma}^*\widehat{\phi}^{*}\\
    =& \beta_0^2 \left\{(\sigma_u^2+1) \gamma_0 + \sigma_u^2  \mu^2\right\} 
    -2\beta_0^2\gamma^{\scalebox{0.5}{\rm T}}\{\Gamma + \sigma_u^2 (\gamma_0+\mu^2) I\}^{-1} \gamma \\ & + \beta_0^2 \gamma^{\scalebox{0.5}{\rm T}}\{\Gamma + \sigma_u^2 (\gamma_0+\mu^2) I\}^{-1}\{\Gamma + \sigma_u^2 (\gamma_0+\mu^2) I\}\{\Gamma + \sigma_u^2 (\gamma_0+\mu^2) I\}^{-1}\gamma + o_p(1) \\
    =& \beta_0^2 \left\{(\sigma_u^2+1) \gamma_0 + \sigma_u^2  \mu^2\right\} - \beta_0^2\gamma^{\scalebox{0.5}{\rm T}}\{\Gamma + \sigma_u^2 (\gamma_0+\mu^2) I\}^{-1} \gamma + o_p(1) .
\end{align*}

\medskip

\noindent\underline{\bf Proof of Theorem~4(3)}:

\noindent\underline{Step 1}: We show that as $T \to \infty$,
\begin{equation}\label{apx:Thm4-Step-1}
    \sqrt{T}\left(\frac{1}{T}\sum_{t=1}^{T} (X_t^*-\mu^*)(X_{t+p}^*-\mu^*)- \frac{1}{T-p}\sum_{t=1}^{T-p}(X_t^*-\widehat{\mu}^*)(X_{t+p}^*-\widehat{\mu}^*)\right)=o_p(1).
\end{equation}

With some simple algebra,
\begin{align}\label{apx:equiv}
    & \sqrt{T}\left\{\frac{1}{T}\sum_{t=1}^{T} (X_t^*-\mu^*)(X_{t+p}^*-\mu^*) - \frac{1}{T-p}\sum_{t=1}^{T-p}(X_t^*-\widehat{\mu}^*)(X_{t+p}^*-\widehat{\mu}^*)\right\} \nonumber\\
    &= \sqrt{T}\left\{\frac{1}{T}\sum_{t=1}^{T} (X_t^*-\mu^*)(X_{t+p}^*-\mu^*) - \frac{1}{T-p}\sum_{t=1}^{T-p}(X_t^*-\mu^*+\mu^*-\widehat{\mu}^*)(X_{t+p}^*-\mu^*+\mu^*-\widehat{\mu}^*)\right\} \nonumber\\
    &= \sqrt{T}\left\{ \frac{1}{T}\sum_{t=1}^{T} (X_t^*-\mu^*)(X_{t+p}^*-\mu^*) - \frac{1}{T-p}\sum_{t=1}^{T-p} (X_t^*-\mu^*)(X_{t+p}^*-\mu^*)  \right. \nonumber\\
 & \qquad  \left. -\frac{1}{T-p}\sum_{t=1}^{T-p} (X_t^*-\mu^*)(\mu^*-\widehat{\mu}^*) - \frac{1}{T-p}\sum_{t=1}^{T-p} (X_{t+p}^*-\mu^*)(\mu^*-\widehat{\mu}^*) - \frac{1}{T-p}\sum_{t=1}^{T-p}(\mu^*-\widehat{\mu}^*)^2\right\}  \nonumber\\
    &= \sqrt{T} \left(\frac{T-p}{T} -1\right) \frac{1}{T-p} \sum_{t=1}^{T-p} (X_t^*-\mu^*)(X_{t+p}^* -\mu^*) + \frac{1}{\sqrt{T}} \sum_{t=T-p+1}^T (X_t^*-\mu^*)(X_{t+p}^* -\mu^*)  \nonumber\\
 & \qquad  + \sqrt{T} (\widehat{\mu}^*-\mu^*) \left( \frac{1}{T-p} \sum_{t=1}^{T-p} X_t^* + \frac{1}{T-p} \sum_{t=1}^{T-p} X_{t+p}^* - \widehat{\mu}^* - \mu^* \right) \\
 &\triangleq I_1 + I_2 + I_3. \nonumber
\end{align}

Now we examine each term in (\ref{apx:equiv}) as $T \to \infty$ separately. First,
\begin{align}\label{eqn:thm4-pf1-I1}
 I_1 &= -\frac{p}{\sqrt{T}}  \frac{1}{T-p} \sum_{t=1}^{T-p} (X_t^*-\mu^*)(X_{t+p}^* -\mu^*) \nonumber \\
 &= -\frac{p}{\sqrt{T}}  \{ \gamma_p^* +o_p(1)\} = o_p(1) \qquad \text{ as } \quad  T \to \infty.
\end{align}

Next, we examine the second term $I_2$ in (\ref{apx:equiv}). Since $T^{-\frac{1}{2}}E[\sum_{t=T-p+1}^T (X_t^*-\mu^*)(X_{t+p}^* -\mu^*)]\le T^{-\frac{1}{2}}p\text{Var}(X_t)$ \citep[][p.230]{brockwell1991time} and $T^{-\frac{1}{2}}p\text{Var}(X_t) \to 0$ as $T \to \infty$, we have that
\begin{equation} \label{eqn:thm4-pf1-I2}
   I_2 = \frac{1}{\sqrt{T}} \sum_{t=T-p+1}^T (X_t^*-\mu^*)(X_{t+p}^* -\mu^*) =  o_p(1).
\end{equation}

Finally, we examine $I_3$ in (\ref{apx:equiv}).
\begin{align}\label{eqn:thm4-pf1-I3-1}
 & \frac{1}{T-p} \sum_{t=1}^{T-p} X_{t+p}^* - \widehat{\mu}^* \nonumber \\
 =& \frac{1}{T-p} \sum_{t=1}^{T-p} X_{t+p}^* - \frac{1}{T}\sum_{t=1}^{p}X_t^* - \frac{1}{T}\sum_{t=p+1}^{T}X_t^* \nonumber  \\
 =& \frac{1}{T-p} \sum_{t=1}^{T-p} X_{t+p}^* - \frac{1}{T}\sum_{t=1}^{p}X_t^* - \frac{1}{T}\sum_{t=1}^{T-p}X_{t+p}^*  \nonumber \\
 =& (\frac{1}{T-p}-\frac{1}{T})\sum_{t=1}^{T-p} X_{t+p}^*
 - \frac{1}{T}\sum_{t=1}^pX_t^* \nonumber \\
 =& o_p(1) \qquad \text{ as }  \quad T \to \infty,
\end{align}
 where $\widehat{\mu}^*=\frac{1}{T}\sum_{t=1}^TX_t^*$, and $\frac{1}{T}\sum_{t=1}^pX_t^*=o_p(1)$ because $E(\frac{1}{T}\sum_{t=1}^pX_t^*)=\frac{1}{T}pE(X_t) \to 0$ as $T\to\infty$. In addition, by the weak law of large numbers, 
\begin{equation}\label{eqn:thm4-pf1-I3-2}
     \frac{1}{T-p} \sum_{t=1}^{T-p} X_t^* - \mu^* \xrightarrow{\,\,p\,\,} 0 \qquad \text{ as } \quad  {T \to \infty}.
\end{equation}
By condition (R2) and the central limit theorem for strictly stationary $p$-dependent sequences \citep[][Theorem 6.4.2]{brockwell1991time}, we have 
\begin{equation}\label{eqn:thm4-pf1-I3-3}
    \sqrt{T}(\widehat{\mu}^*-\mu^*) = O_p(1).
\end{equation}
Therefore, applying (\ref{eqn:thm4-pf1-I1}), (\ref{eqn:thm4-pf1-I2}), (\ref{eqn:thm4-pf1-I3-1}), (\ref{eqn:thm4-pf1-I3-2}) and (\ref{eqn:thm4-pf1-I3-3}) yields (\ref{apx:Thm4-Step-1}).\\

\noindent\underline{Step 2}: We show that as $T\to\infty$, the asymptotic covariance matrix of
$\sqrt{T} \left\{(\widehat{\gamma}_0^* ,\widehat{\gamma}^{*\rm\tiny T})^{\rm\tiny T} -({\gamma}_0^*, {\gamma}^{*\rm\tiny T})^{\rm\tiny T} \right\}$ equals
    \begin{equation*}
        \lim_{T\to\infty}  \text{Cov}\left\{\frac{1}{\sqrt{T}}\sum_{t=1}^T(X_t^*-\mu^*)(X_{t+r}^*-\mu^*),\frac{1}{\sqrt{T}}\sum_{s=1}^T(X_s^*-\mu^*)(X_{s+q}^*-\mu^*)\right\}.
    \end{equation*}

For $k\le p$
\begin{align*}
 & \sqrt{T}(\widehat{\gamma}_k - \gamma_k) \\ 
 =&  \sqrt{T}\left\{\frac{1}{T-k}\sum_{t=1}^{T-k}(X_t^*-\widehat{\mu}^*)(X_{t+k}^*-\widehat{\mu}^*) - \gamma_k \right\} \\
  =&  \sqrt{T}\left\{ \frac{1}{T}\sum_{t=1}^{T} (X_t^*-\mu^*)(X_{t+k}^*-\mu^*) - \gamma_k \right\} \\
  & + \sqrt{T}\left\{\frac{1}{T-k}\sum_{t=1}^{T-k}(X_t^*-\widehat{\mu}^*)(X_{t+k}^*-\widehat{\mu}^*)  - \frac{1}{T}\sum_{t=1}^{T} (X_t^*-\mu^*)(X_{t+k}^*-\mu^*)\right\}\\
  =& \left\{ \frac{1}{\sqrt{T}}\sum_{t=1}^{T} (X_t^*-\mu^*)(X_{t+k}^*-\mu^*) - \gamma_k \right\} + o_p(1),
\end{align*}
where the last step is due to (\ref{apx:Thm4-Step-1}).

Hence, the $(r,q)$ element of matrix $\lim\limits_{T\to\infty} \text{Var}\left(\sqrt{T} \left\{(\widehat{\gamma}_0^* ,\widehat{\gamma}^{*\rm\tiny T})^{\rm\tiny T} -({\gamma}_0^*, {\gamma}^{*\rm\tiny T})^{\rm\tiny T} \right\}\right)$ is given by 
\begin{align*}
 \lim_{T\to\infty}  \text{Cov}\left\{\frac{1}{\sqrt{T}}\sum_{t=1}^T(X_t^*-\mu^*)(X_{t+r}^*-\mu^*),\frac{1}{\sqrt{T}}\sum_{s=1}^T(X_s^*-\mu^*)(X_{s+q}^*-\mu^*)\right\}.\\
\end{align*}

\noindent\underline{Step 3:} We show certain identities to be used for proving Theorem~4(3):

1. By model (\ref{eqn:measmodelmultiplicative}), we have that
\begin{align} \label{apx:The4-base}
    X_t^* - {\mu}^* &= \beta_0 X_t u_t - \beta_0 \mu \nonumber \\
     &= \beta_0 X_t u_t - \beta_0 u_t \mu + \beta_0 u_t \mu - \beta_0 \mu   \nonumber \\
     &= \beta_0 \{u_t (X_t - \mu) + \mu(u_t-1)\}
\end{align}
where the first step is because ${\mu}^* = E(\beta_0X_tu_t)= \beta_0E(X_t)E(u_t)=\beta_0\mu$.

2. 
 We have that
\begin{align}\label{apx:Thm4-stm-11-main}
     &  \lim_{T\to\infty} \frac{1}{T}\sum_{t=1}^T\sum_{s=1}^T \text{Cov}\left\{u_t^2(X_t-\mu)^2,u_s^2(X_s-\mu)^2\right\}  \nonumber \\
     =&  \lim_{T\to\infty} \frac{1}{T}\sum_{t=1}^T\sum_{s=1}^T \left[E\{u_t^2u_s^2(X_t-\mu)^2(X_s-\mu)^2\}-E(u_t^2)E(u_s^2)E\{(X_t-\mu)^2\}E\{(X_s-\mu)^2\}\right],  \nonumber \\
    =&  \lim_{T\to\infty} \frac{1}{T}\sum_{t=1}^T\sum\limits_{\substack{s=1 \\ s\ne t}}^T \left[E(u_t^2u_s^2)E\{(X_t-\mu)^2(X_s-\mu)^2\}-E(u_t^2)E(u_s^2)E\{(X_t-\mu)^2\}E\{(X_s-\mu)^2\}\right] \nonumber \\
     &\qquad + \lim_{T\to\infty} \frac{1}{T}\sum\limits_{\substack{t=1 \\ s=t}}^T \left[E(u_t^4)E\{(X_t-\mu)^4\}-E^2(u_t^2)E^2\{(X_t-\mu)^2\}\right], \nonumber \\
    =& \lim_{T\to\infty} \frac{1}{T}\sum_{t=1}^T\sum\limits_{\substack{s=1 \\ s\ne t}}^T \left[E(u_t^2)E(u_s^2)\text{Cov}\{(X_t-\mu)^2,(X_s-\mu)^2\}\right] + \lim_{T\to\infty} \frac{1}{T}\sum\limits_{\substack{t=1 \\ s=t}}^T E^2(u_t^2)\text{Var}\{(X_t-\mu)^2\} \nonumber \\
    &\qquad + \lim_{T\to\infty} \frac{1}{T}\sum\limits_{\substack{t=1 \\ s=t}}^T \left\{E(u_t^4)-E^2(u_t^2)\right\}E\{(X_t-\mu)^4\} \nonumber \\
    =& \lim_{T\to\infty} \frac{1}{T}\sum_{t=1}^T\sum\limits_{s=1}^T \left[E(u_t^2)E(u_s^2)\text{Cov}\{(X_t-\mu)^2,(X_s-\mu)^2\}\right] + \lim_{T\to\infty} \frac{1}{T}\sum\limits_{t=1}^T \left\{E(u_t^4)-E^2(u_t^2)\right\}E\{(X_t-\mu)^4\} \nonumber \\
    =& (\sigma_u^2+1)^2q_{00} +  \{E(u_t^4)-(\sigma_u^2+1)^2\}E\{(X_t-\mu)^4\},
\end{align}
where the second and third step is due to the independence between $u_t$ and $X_t$. In the last step, we use the definition $q_{00}= \lim_{T\to\infty} \frac{1}{T}\sum\limits_{t=1}^T\sum\limits_{s=1}^T \text{Cov}\left\{(X_t-\mu)^2,(X_s-\mu)^2\right\}$, $E(u_t^2)=\sigma_u^2+1$, and the fact that $E(u_t^4)$ and $E\{(X_t-\mu)^4\}$ are time-independent which are derived from Conditions (R1) and (R2) together with independence between $u_t$ and $X_t$.

3.  Similar to the derivation in (\ref{apx:Thm4-stm-11-main}), now we derive the summation of $\text{Cov}\{\beta_0^2u_t^2(X_t-\mu)^2, \beta_0^2u_su_{s+p}(X_{s}-\mu)(X_{s+p}-\mu)\}$ for $p>0$,
\begin{align}\label{apx:Thm4-q0p-basic}
   &  \lim_{T\to\infty} \frac{1}{T}\sum_{t=1}^T \sum_{s=1}^T \text{Cov}\{\beta_0^2u_t^2(X_t-\mu)^2, \beta_0^2u_su_{s+p}(X_{s}-\mu)(X_{s+p}-\mu)\} \nonumber\\
   &= \lim_{T\to\infty} \frac{\beta_0^4}{T}\sum_{t=1}^T \sum_{s=1}^T \left[ E(u_t^2u_su_{s+p})E\{(X_t-\mu)^2(X_{s}-\mu)(X_{s+p}-\mu)\} \right. \nonumber\\
     & \left. -E(u_t^2)E(u_s)E(u_{s+p})E(X_t-\mu)^2E\{(X_{s}-\mu)(X_{s+p}-\mu)\}\right] \nonumber\\
   &= \lim_{T\to\infty} \frac{\beta_0^4}{T}\sum_{t=1}^T \sum_{s=1}^T E(u_t^2)E(u_s)E(u_{s+p}) \text{Cov}\{(X_t-\mu)^2,(X_s-\mu)(X_{s+p}-\mu)\}\nonumber\\
   & \qquad + \lim_{T\to\infty} \frac{\beta_0^4}{T} \sum\limits_{\substack{t=1 \\ s=t}}^T \left\{E(u_t^3)E(u_{t+p})-E(u_t^2)E(u_t)E(u_{t+p})\right\}E\{(X_t-\mu)^3(X_{t+p}-\mu)\} \nonumber\\
   & \qquad + \lim_{T\to\infty} \frac{\beta_0^4}{T} \sum\limits_{\substack{t=1 \\ s=t-p}}^T \left\{E(u_t^3)E(u_{t-p})-E(u_t^2)E(u_t)E(u_{t-p})\right\}E\{(X_t-\mu)^3(X_{t-p}-\mu)\} \nonumber\\
   &= \lim_{T\to\infty} \frac{\beta_0^4}{T}\sum_{t=1}^T \sum_{s=1}^T (\sigma_u^2+1) \text{Cov}\{(X_t-\mu)^2,(X_s-\mu)(X_{s+p}-\mu)\} \nonumber \\
   &\qquad + \beta_0^4 \left\{E(u_t^3)-E(u_t^2)\right\} E\{(X_t-\mu)^3(X_{t+p}-\mu)\} \nonumber\\
   &\qquad + \beta_0^4 \left\{E(u_t^3)-E(u_t^2)\right\} E\{(X_t-\mu)^3(X_{t-p}-\mu)\}, \nonumber\\
   &=\beta_0^4 q_{0p} (\sigma_u^2+1) + \beta_0^4 \left\{E(u_t^3)-(\sigma_u^2+1)\right\} \left[ E\{(X_t-\mu)^3(X_{t+p}-\mu)\} + E\{(X_t-\mu)^3(X_{t-p}-\mu)\}\right],
\end{align}
where the first step is because $X_t$ and $u_t$ are independent, and the second last step is due to $E(u_t^2)=Var(u_t)+E(u_t^2)=\sigma_u^2+1$ and is derived similar to the second and third step in (\ref{apx:Thm4-stm-11-main}), and the last step is because of the definition that $q_{0p}= \lim_{T\to\infty}\frac{1}{T}\sum\limits_{t=1}^T \sum\limits_{s=1}^T \text{Cov}\{(X_t-\mu)^2,(X_s-\mu)(X_{s+p}-\mu)\}$ and the fact that $E\{(X_t-\mu)^3(X_{t+p}-\mu)\}$, $E\{(X_t-\mu)^3(X_{t-p}-\mu)\}$ and $E(u_t^3)$ are time-independent, derived from Conditions (R1) and (R2) together with the independence between $u_t$ and $X_t$.

4. Analogous to the derivation in (\ref{apx:Thm4-stm-11-main}) and (\ref{apx:Thm4-q0p-basic}), we derive the summation of  $\text{Cov}\{u_tu_{t+p}(X_t-\mu)(X_{t+p}-\mu),u_su_{s+r}(X_s-\mu)(X_{s+r}-\mu)\}$ for $p>0$, $r>0$ and $p \ne r$,
\begin{align} \label{apx:Thm4-qpq-basic}
     & \beta_0^4 \lim_{T\to\infty} \frac{1}{T} \sum_{t=1}^T\sum_{s=1}^T \text{Cov}\{u_tu_{t+p}(X_t-\mu)(X_{t+p}-\mu),u_su_{s+r}(X_s-\mu)(X_{s+r}-\mu)\}\nonumber\\
   =&  \beta_0^4 \lim_{T\to\infty} \frac{1}{T}\sum_{t=1}^T\sum_{s=1}^T E(u_tu_{t+p}u_su_{s+r}) \text{Cov}\{(X_t-\mu)(X_{t+p}-\mu),(X_{s}-\mu)(X_{s+r}-\mu)\}\nonumber\\
    &\qquad + \beta_0^4 \lim_{T\to\infty} \frac{1}{T}\sum\limits_{\substack{t=1 \\ s=t}}^T \left\{E(u_t^2)E(u_{t+p})E(u_{t+r})-1\right\} E\{(X_t-\mu)^2(X_{t+p}-\mu)(X_{t+r}-\mu)\} \nonumber\\
    &  \qquad + \beta_0^4 \lim_{T\to\infty} \frac{1}{T}\sum\limits_{\substack{t=1 \\ s=t+p}}^T \left\{E(u_{t+p}^2)E(u_{t})E(u_{t+p+r})-1\right\} E\{(X_t-\mu)(X_{t+p}-\mu)^2(X_{t+p+r}-\mu)\} \nonumber\\
    &  \qquad + \beta_0^4 \lim_{T\to\infty} \frac{1}{T}\sum\limits_{\substack{t=1 \\ s=t-r}}^T \left\{E(u_{t}^2)E(u_{t+p})E(u_{t-r})-1\right\} E\{(X_{t-r}-\mu)(X_{t}-\mu)^2(X_{t+p}-\mu)\} \nonumber\\
    &  \qquad + \beta_0^4 \lim_{T\to\infty} \frac{1}{T}\sum\limits_{\substack{t=1 \\ s=t+p-r}}^T \left\{E(u_{t+p}^2)E(u_{t})E(u_{t+p-r})-1\right\} E\{(X_t-\mu)(X_{t+p-r}-\mu)X_{t+p}-\mu)^2\} \nonumber\\
   =& \beta_0^4 q_{pr}  + \beta_0^4 \sigma_u^2 E\{(X_t-\mu)^2(X_{t+p}-\mu)(X_{t+r}-\mu)\} + \beta_0^4 \sigma_u^2  E\{(X_t-\mu)(X_{t+p}-\mu)^2(X_{t+p+r}-\mu)\} \nonumber \\
   &\qquad +\beta_0^4 \sigma_u^2 E\{(X_{t-r}-\mu)(X_{t}-\mu)^2(X_{t+p}-\mu)\} + \beta_0^4 \sigma_u^2  E\{(X_t-\mu)(X_{t+p-r}-\mu)(X_{t+p}-\mu)^2\},
\end{align}
where the third step is derived analogously to the second step of (\ref{apx:Thm4-q0p-basic}), and $E(u_tu_{t+p}u_{s}u_{s+r})=1$, and the last step is due to the definition $q_{pr}= \lim_{T\to\infty}\frac{1}{T}\sum\limits_{t=1}^T \sum\limits_{s=1}^T \text{Cov}\{(X_t-\mu)(X_{t+p}-\mu),(X_{s}-\mu)(X_{s+r}-\mu)\}$ and the fact that $E\{(X_t-\mu)^2(X_{t+p}-\mu)(X_{t+r}-\mu)\}$, $E\{(X_t-\mu)(X_{t+p}-\mu)^2(X_{t+p+r}-\mu)\}$, $E\{(X_{t-r}-\mu)(X_{t}-\mu)^2(X_{t+p}-\mu)\}$, and $E\{(X_t-\mu)(X_{t+p}-\mu)^2(X_{t+2p}-\mu)\}$ are time-independent derived from Conditions (R1) and (R2).

5. Similar to the derivation in (\ref{apx:Thm4-stm-11-main}), (\ref{apx:Thm4-q0p-basic}), and (\ref{apx:Thm4-qpq-basic}), we derive the summation of  $\text{Cov}\{u_tu_{t+p}(X_t-\mu)(X_{t+p}-\mu),u_su_{s+p}(X_s-\mu)(X_{s+p}-\mu)\}$ for $p>0$,
\begin{align} \label{apx:Thm4-qpp-basic}
    & \beta_0^4 \lim_{T\to\infty} \frac{1}{T} \sum_{t=1}^T\sum_{s=1}^T \text{Cov}\{u_tu_{t+p}(X_t-\mu)(X_{t+p}-\mu),u_su_{s+p}(X_s-\mu)(X_{s+p}-\mu)\}  \nonumber \\
    &=  \beta_0^4 \lim_{T\to\infty} \frac{1}{T}\sum_{t=1}^T\sum_{s=1}^T E(u_t)E(u_{t+p})E(u_s)E(u_{s+p}) \text{Cov}\{(X_t-\mu)(X_{t+p}-\mu),(X_{s}-\mu)(X_{s+p}-\mu)\nonumber\\
    &  \qquad + \beta_0^4 \lim_{T\to\infty} \frac{1}{T}\sum\limits_{\substack{t=1 \\ s=t}}^T \left\{E(u_t^2)E(u_{t+p}^2)-1\right\} \text{Var}\{(X_t-\mu)(X_{t+p}-\mu)\} \nonumber\\
    &  \qquad + \beta_0^4 \lim_{T\to\infty} \frac{1}{T}\sum\limits_{\substack{t=1 \\ s=t+p}}^T \left\{E(u_{t+p}^2)E(u_{t})E(u_{t+2p})-1\right\} E\{(X_t-\mu)(X_{t+p}-\mu)^2(X_{t+2p}-\mu)\} \nonumber\\
    &  \qquad + \beta_0^4 \lim_{T\to\infty} \frac{1}{T}\sum\limits_{\substack{t=1 \\ s=t-p}}^T \left\{E(u_{t}^2)E(u_{t-p})E(u_{t+p})-1\right\} E\{(X_{t-p}-\mu)(X_{t}-\mu)^2(X_{t+p}-\mu)\} \nonumber\\
   &= \beta_0^4 q_{pp}  + \beta_0^4 (\sigma_u^4 + 2\sigma_u^2)  \text{Var}\{(X_t-\mu)(X_{t+p}-\mu)\} + 2\beta_0^4  E\{(X_t-\mu)(X_{t+p}-\mu)^2(X_{t+2p}-\mu)\}, 
\end{align}
where the last step is by the definition $q_{pp} = \lim_{T\to\infty}\frac{1}{T}\sum\limits_{t=1}^T \sum\limits_{s=1}^T \text{Cov}\{(X_t-\mu)(X_{t+p}-\mu),(X_s-\mu)(X_{s+p}-\mu)\}$ and $E\{(X_t-\mu)(X_{t+p}-\mu)^2(X_{t+2p}-\mu)\} = E\{(X_{t-p}-\mu)(X_{t}-\mu)^2(X_{t+p}-\mu)\}$ due to the stationarity of the time series and the fact that $\text{Var}\{(X_t-\mu)(X_{t+p}-\mu)\}$ and $E\{(X_t-\mu)(X_{t+p}-\mu)^2(X_{t+2p}-\mu)\}$ are time-independent, resulting from the Conditions (R1) and (R2).

6. For any $t$, $s$ and $p$,  we have that
\begin{align}\label{apx:Thm4-stm-12}
   & \text{Cov}\{(X_{t}-\mu)(X_{t-p}-\mu),(X_{s}-\mu)\} \nonumber \\
  =& E\{(X_{t}-\mu)(X_{t-p}-\mu)(X_{s}-\mu)\} - E\{(X_{t}-\mu)(X_{t-p}-\mu)\}E(X_{s}-\mu) \nonumber\\
  =& E\{(X_{t}-\mu)(X_{t-p}-\mu)(X_{s}-\mu)\}, 
\end{align}
where the last step is because $E(X_{s}-\mu)=0$.

7. For any $t$ and $s$, we have that
\begin{align}\label{apx:Thm4-stm-22}
    & \text{Cov}\{u_t(u_t-1)(X_t-\mu),u_s(u_s-1)(X_s-\mu)\} \nonumber \\
   =& E\{u_t(u_t-1)(X_t-\mu)u_s(u_s-1)(X_s-\mu)\} - E\{u_t(u_t-1)(X_t-\mu)\}E\{u_s(u_s-1)(X_s-\mu)\} \nonumber \\
   =& E\{u_t(u_t-1)(X_t-\mu)u_s(u_s-1)(X_s-\mu)\} \nonumber \\
   =& E\{u_t(u_t-1)u_s(u_s-1)\}E\{(X_t-\mu)(X_s-\mu)\},
\end{align}
where the second step is because of the independence between  $u_t$ and $X_t$ and that $E(X_t-\mu)=0$. Then, $E\{u_t(u_t-1)u_s(u_s-1)\} = \sigma_u^4$ for $t\ne s$ and $E\{u_t^2(u_t-1)^2\}=E(u_t^4)-2E(u_t^3) + \sigma_u^2 +1$ for any $t$.

By (\ref{apx:Thm4-stm-22}), we have that
\begin{align}\label{apx:Thm4-stm-22-2}
  &  \lim_{T\to\infty} \frac{1}{T}\sum_{t=1}^T\sum_{s=1}^T \text{Cov}\left\{u_t(u_t-1)(X_t-\mu), u_s(u_s-1)(X_s-\mu)\right\}  \nonumber \\
    =&  \lim_{T\to\infty} \frac{1}{T}\sum_{t=1}^T\sum_{s=1}^T E\{u_t(u_t-1)u_s(u_s-1)\}E\{(X_t-\mu)(X_s-\mu)\}  \nonumber \\  
=& \lim_{T\to\infty} \frac{1}{T}\sum_{t=1}^T\sum_{s=1}^T \sigma_u^4 E\{(X_t-\mu)(X_s-\mu)\} + \lim_{T\to\infty} \frac{1}{T}\sum\limits_{\substack{t=1 \\ s= t}}^T \left\{ E(u_t^4)-2E(u_t^3) + \sigma_u^2 +1 - \sigma_u^4 \right\} E\{(X_t-\mu)^2\} \nonumber \\
=&\sigma_u^4 \sum_{h=-\infty}^\infty \gamma_h + \left\{ E(u_t^4)-2E(u_t^3) + \sigma_u^2 +1 - \sigma_u^4 \right\} \gamma_0,
\end{align}
where the last is because $\lim_{T\to\infty} \frac{1}{T}\sum_{t=1}^T\sum_{s=1}^T E\{(X_t-\mu)(X_s-\mu)\} = \sum_{h=-\infty}^\infty \gamma_h$ \citep[][Theorem 7.1.1]{brockwell1991time}.

8. For any $t$, $s$ and $p>0$, we have that
\begin{align}\label{apx:Thm4-stm-22-3}
    & \text{Cov}\{u_t(u_t-1)(X_t-\mu),u_{s+p}(u_s-1)(X_{s+p}-\mu)\} \nonumber \\
   =& E\{u_t(u_t-1)(X_t-\mu)u_{s+p}(u_s-1)(X_{s+p}-\mu)\} - E\{u_t(u_t-1)(X_t-\mu)\}E\{u_{s+p}(u_s-1)(X_{s+p}-\mu)\} \nonumber \\
   =& E\{u_t(u_t-1)u_{s+p}(u_s-1)\}E\{(X_t-\mu)(X_{s+p}-\mu)\} \nonumber \\
   =& E\{u_t(u_t-1)u_{s+p}(u_s-1)\} \gamma_{|s+p-t|},
\end{align}
where the second step is because of the independence between  $u_t$ and $X_t$ and that $E(X_t-\mu)=0$. Then, $E\{u_t(u_t-1)u_{s+p}(u_s-1)\} = 0$ for $t\ne s$ and $E\{u_t(u_t-1)^2u_{t+p}\}=E\{u_t(u_t-1)^2\}=E\{(u_t-1)^3\}+\sigma_u^2$ for any $s=t$.

9. By independence of $u_t$ and $u_s$, for $t\ne s$, we have that
\begin{equation}\label{apx:Thm4-stm-13-1}
    \text{Cov}\{u_t^2 (X_t - \mu)^2,  (u_s-1)^2\} = 0,
\end{equation}
and for any $t$,
\begin{align}\label{apx:Thm4-stm-13-2}
  &  \text{Cov}\{u_t^2 (X_t - \mu)^2,  (u_t-1)^2\} \nonumber \\
 =& E\{u_t^2(u_t-1)^2(X_t - \mu)^2\} - E\{u_t^2(X_t - \mu)^2\} E\{(u_t-1)^2\} \nonumber \\
 =& \left[E\{u_t^2(u_t-1)^2\} -E(u_t^2) E(u_t-1)^2 \right]E\{(X_t - \mu)^2\} \nonumber \\
 =& \left\{ E(u_t^4)-2E(u_t^3) + \sigma_u^2 +1 - \sigma_u^4 - \sigma_u^2 \right\} \gamma_0 \nonumber \\
 =& \left\{ E(u_t^4)-2E(u_t^3) +1 - \sigma_u^4  \right\} \gamma_0.
\end{align}

10. By independence of $u_t$ and $u_s$, for $s\ne t$, $s \ne t+p$ and any $p$, we have that
\begin{equation}\label{apx:Thm4-stm-13-3}
    \text{Cov}\{u_t u_{t+p} (X_t - \mu)(X_{t+p}-\mu),  (u_s-1)^2\} = 0.
\end{equation}
For any $t$  and $p>0$,
\begin{align}\label{apx:Thm4-stm-13-4}
  &  \text{Cov}\{u_t u_{t+p} (X_t - \mu)(X_{t+p}-\mu),  (u_t-1)^2\} \nonumber \\
 =& E\{u_t u_{t+p}(u_t-1)^2(X_t - \mu)(X_{t+p}-\mu)\} - E\{u_t u_{t+p}(X_t - \mu)(X_{t+p}-\mu)\} E\{(u_t-1)^2\} \nonumber \\
 =& \left[E\{u_t u_{t+p}(u_t-1)^2\} -E(u_t u_{t+p}) E\{(u_t-1)^2\} \right]E\{(X_t - \mu)(X_{t+p}-\mu)\} \nonumber \\
 =& E\left\{ (u_t-1)^3\right\} \gamma_p,
\end{align}
and
\begin{equation*}
  \text{Cov}\{u_t u_{t-p} (X_t - \mu)(X_{t-p}-\mu),  (u_t-1)^2\} = E\left\{ (u_t-1)^3\right\} \gamma_p.
\end{equation*}

11. For any $t$ and $s$, and $r\ne p$ and $r>0$, we have that 
\begin{equation}\label{apx:Thm4-stm-13-5}
    \text{Cov}\{u_t u_{t+p} (X_t - \mu)(X_{t+p}-\mu),  (u_s-1)(u_{s+r}-1)\} = 0.
\end{equation}

By independence of $u_t$ and $u_s$, for $t\ne s$ and any $p$, we have that
\begin{equation}\label{apx:Thm4-stm-13-6}
    \text{Cov}\{u_t u_{t+p} (X_t - \mu)(X_{t+p}-\mu),  (u_s-1)(u_{s+p}-1)\} = 0,
\end{equation}
and for any $t$ and $p>0$,
\begin{align}\label{apx:Thm4-stm-13-7}
  &  \text{Cov}\{u_t u_{t+p} (X_t - \mu)(X_{t+p}-\mu),  (u_t-1)(u_{t+p}-1)\} \nonumber \\
 =& E\{u_t u_{t+p}(u_t-1)(t_{t+p}-1)(X_t - \mu)(X_{t+p}-\mu)\} \nonumber \\
 & - E\{u_t u_{t+p}(X_t - \mu)(X_{t+p}-\mu)\} E\{(u_t-1)(u_{t+p}-1)\} \nonumber \\
 =& E\{u_t (u_t-1)\}E\{u_{t+p}(u_{t+p}-1)\}E\{(X_t - \mu)(X_{t+p}-\mu)\} \nonumber \\
 =& \sigma_u^4 \gamma_p.
\end{align}

12. For any $t$, we have that
\begin{align}\label{apx:Thm4-stm-23-1}
  & \text{Cov}\{u_t(u_t-1)(X_t - \mu), (u_s-1)^2\} \nonumber \\
  &= E\{u_t(u_t-1)(X_t - \mu)(u_s-1)^2\} - E\{u_t(u_t-1)(X_t - \mu)\}E\{(u_s-1)^2\} \nonumber\\
  &= \left[E\{u_t(u_t-1)(u_s-1)^2\} - E\{u_t(u_t-1)\}E\{(u_s-1)^2\}\right]E(X_t - \mu) =0,
\end{align}
where the last step is because $E(X_t-\mu)=0$.

13. By independence assumption between $\{u_t\}$, if $t\ne s$ or $p\ne r$, we have that
\begin{equation}\label{apx:Thm4-ut-utp-usr}
    \text{Cov}\left\{(u_t-1)(u_{t+p}-1),(u_s-1)(u_{s+r}-1)\right\}=0.
\end{equation}
In addition, for any $t$ and $p$ we have that
\begin{align}\label{apx:Thm4-Var-e_u-e_u+p}
& \text{Var}\left\{(u_t-1)(u_{t+p}-1)\right\} \nonumber \\
&=  E\left\{(u_t-1)^2(u_{t+p}-1)^2\right\} \nonumber \\
&= E\left\{(u_t-1)^2\right\}E\left\{(u_{t+p}-1)^2\right\} \nonumber\\
&= \sigma_u^4,
\end{align}
and for any $t$, we have that
\begin{align}\label{apx:Thm4-Var-u_t-1-4}
    & \text{Var}(u_t-1)^2\nonumber \\
  = & E\{(u_t-1)^4\} - E^2\{(u_t-1)^2\} \nonumber \\
  = & E\{(u_t-1)^4\} - \sigma_u^4. 
\end{align}

\ \\~\\

\noindent\underline{Step 4}: Now we prove the results in (3).

\noindent$1^{\circ}$. We first show the derivation of $q_{200}^*$ as follows:
\begin{align*} 
    q_{200}^* &=  \lim_{T\to\infty} T \text{Cov}\left\{\frac{1}{T}\sum_{t=1}^T(X_t^*-\mu^*)^{2},\frac{1}{T}\sum_{s=1}^T(X_s^{*}-\mu^*)^2\right\} \\
    &=  \lim_{T\to\infty} \frac{\beta_0^4 }{T} \sum_{t=1}^T\sum_{s=1}^T\text{Cov}\left\{u_t^2 (X_t - \mu)^2 + 2\mu u_t(u_t-1)(X_t - \mu) + \mu^2(u_t-1)^2,\right. \\ 
    & \qquad \left. u_s^2 (X_s - \mu)^2 + 2\mu u_s(u_s-1)(X_s - \mu) + \mu^2(u_s-1)^2\right\} \\
    &=  \lim_{T\to\infty} \frac{\beta_0^4 }{T} \sum_{t=1}^T\sum_{s=1}^T\text{Cov}\left\{u_t^2 (X_t - \mu)^2,  u_s^2 (X_s - \mu)^2\right\} \\
    &\qquad + \lim_{T\to\infty} \frac{4\mu\beta_0^4 }{T} \sum_{t=1}^T\sum_{s=1}^T\text{Cov}\left\{u_t^2 (X_t - \mu)^2,  u_s(u_s-1)(X_s - \mu)\right\} \\
    &\qquad + \lim_{T\to\infty} \frac{2\mu^2\beta_0^4 }{T} \sum_{t=1}^T\sum_{s=1}^T\text{Cov}\left\{u_t^2 (X_t - \mu)^2,  (u_s-1)^2\right\} \\
    &\qquad +  \lim_{T\to\infty} \frac{4\mu^2\beta_0^4 }{T} \sum_{t=1}^T\sum_{s=1}^T\text{Cov}\left\{u_t(u_t-1)(X_t - \mu),  u_s(u_s-1)(X_s - \mu)\right\} \\
    &\qquad +  \lim_{T\to\infty} \frac{\mu^4\beta_0^4 }{T} \sum_{t=1}^T\sum_{s=1}^T\text{Cov}\left\{ (u_t-1)^2,   (u_s-1)^2\right\} \\
    &= \beta_0^4(\sigma_u^2+1)^2q_{0} +  \beta_0^4\{E(u_t^4)-(\sigma_u^2+1)^2\}E\{(X_t-\mu)^4\} \nonumber \\
    & \qquad + 4\mu\beta_0^4\sigma_u^2(\sigma_u^2+1) v_{00} + 4\mu\beta_0^4\{E(u_t^4) - E(u_t^3)-\sigma_u^2(\sigma_u^2+1)\} 
    E\{(X_t-\mu)^3\} \nonumber \\
    & \qquad+ 2\mu^2\beta_0^4 \left\{ E(u_t^4)-2E(u_t^3) +1 - \sigma_u^4  \right\} \gamma_0 \nonumber \\
    & \qquad + 4\mu^2\beta_0^4 \left[\sigma_u^4 \sum_{h=-\infty}^\infty \gamma_h + \left\{ E(u_t^4)-2E(u_t^3) + \sigma_u^2 +1 - \sigma_u^4 \right\} \gamma_0\right] \nonumber \\
    & \qquad + \mu^4\beta_0^4 \left[E\{(u_t-1)^4\} - \sigma_u^4\right], 
\end{align*}
where the second step is due to (\ref{apx:The4-base}), the third step is because of (\ref{apx:Thm4-stm-23-1}),  the last step is by (\ref{apx:Thm4-stm-11-main}),   (\ref{apx:Thm4-stm-12}), (\ref{apx:Thm4-stm-22-2}), (\ref{apx:Thm4-stm-13-1}),   (\ref{apx:Thm4-stm-13-2}), and (\ref{apx:Thm4-Var-u_t-1-4}).

\noindent$2^{\circ}$. Then we derive the value of $q_{20p}^*$:
{\small
\begin{align*}
     q_{20p}^* &=  \lim_{T\to\infty} T \text{Cov}\left\{\frac{1}{T}\sum_{t=1}^T(X_t^*-\mu^*)^{2},\frac{1}{T}\sum_{s=1}^T(X_s^*-\mu^*)(X_{s+p}^*-\mu^*)\right\} \\
     &=  \lim_{T\to\infty} \frac{\beta_0^4 }{T} \sum_{t=1}^T\sum_{s=1}^T\text{Cov}\left\{u_t^2 (X_t - \mu)^2 + 2\mu u_t(u_t-1)(X_t - \mu) + \mu^2(u_t-1)^2,\right. \\ 
    & \qquad \left. u_su_{s+p} (X_s - \mu)(X_{s+p} - \mu) + \mu u_s(u_{s+p}-1)(X_s - \mu) + \mu u_{s+p}(u_s-1)(X_{s+p} - \mu) + \mu^2(u_s-1)(u_{s+p}-1)\right\} \\
     &=  \lim_{T\to\infty} \frac{\beta_0^4 }{T} \sum_{t=1}^T\sum_{s=1}^T\text{Cov}\left\{u_t^2 (X_t - \mu)^2,  u_su_{s+p} (X_s - \mu)(X_{s+p} - \mu)\right\} \\
    &\qquad +  \lim_{T\to\infty} \frac{\mu\beta_0^4 }{T} \sum_{t=1}^T\sum_{s=1}^T\text{Cov}\left\{u_t^2 (X_t - \mu)^2,   u_s(u_{s+p}-1)(X_s - \mu) +  u_{s+p}(u_s-1)(X_{s+p} - \mu)\right\} \\
     &\qquad +  \lim_{T\to\infty} \frac{2\mu\beta_0^4 }{T} \sum_{t=1}^T\sum_{s=1}^T\text{Cov}\left\{u_su_{s+p} (X_s - \mu)(X_{s+p} - \mu),  u_t(u_t-1)(X_t - \mu)\right\} \\
    &\qquad + \lim_{T\to\infty} \frac{\mu^2\beta_0^4 }{T} \sum_{t=1}^T\sum_{s=1}^T\left[\text{Cov}\left\{u_t^2 (X_t - \mu)^2,  (u_s-1)(u_{s+p}-1)\right\} \right. \\
    &\qquad + \left. \text{Cov}\left\{(u_t-1)^2,u_su_{s+p} (X_s - \mu)(X_{s+p} - \mu)\right\}\right] \\
    &\qquad +  \lim_{T\to\infty} \frac{2\mu^2\beta_0^4 }{T} \sum_{t=1}^T\sum_{s=1}^T\text{Cov}\left\{u_t(u_t-1)(X_t - \mu),   u_s(u_{s+p}-1)(X_s - \mu)\right\} \\
    &\qquad +  \lim_{T\to\infty} \frac{2\mu^2\beta_0^4 }{T} \sum_{t=1}^T\sum_{s=1}^T\text{Cov}\left\{u_t(u_t-1)(X_t - \mu),   u_{s+p}(u_s-1)(X_{s+p} - \mu)\right\} \\
    &\qquad +  \lim_{T\to\infty} \frac{\mu^4\beta_0^4 }{T} \sum_{t=1}^T\sum_{s=1}^T\text{Cov}\left\{ (u_t-1)^2,   (u_s-1)(u_{s+p}-1)\right\} \\
   &=\beta_0^4 q_{0p} (\sigma_u^2+1) + \beta_0^4 \left\{E(u_t^3)-(\sigma_u^2+1)\right\} \left[ E\{(X_t-\mu)^3(X_{t+p}-\mu)\} + E\{(X_t-\mu)^3(X_{t-p}-\mu)\}\right] \\
   & \qquad + \mu\beta_0^4 E\{u_t^3 - u_t^2\}\left[E\{(X_t-\mu)^2(X_{t-p}-\mu)\}+E\{(X_t-\mu)^2(X_{t+p}-\mu)\}\right] \\
   & \qquad + 2\mu\beta_0^4\sigma_u^2 v_{0p} + 2\mu\beta_0^4 E\{u_t^3 - u_t^2-\sigma_u^2\}\left[E\{(X_t-\mu)^2(X_{t-p}-\mu)\}+E\{(X_t-\mu)^2(X_{t+p}-\mu)\}\right] \\
   & \qquad + 2\mu^2\beta_0^4 E(u_t-1)^3\gamma_p + 4\mu^2\beta_0^4 \left\{E(u_t-1)^3+\sigma_u^2\right\}\gamma_p + \mu^4\beta_0^4 \sigma_u^4 \\
   &=\beta_0^4 q_{p} (\sigma_u^2+1) + \beta_0^4 \left\{E(u_t^3)-(\sigma_u^2+1)\right\} \left[ E\{(X_t-\mu)^3(X_{t+p}-\mu)\} + E\{(X_t-\mu)^3(X_{t-p}-\mu)\}\right] \\
   & \qquad + 2\mu\beta_0^4\sigma_u^2 v_{p} + \mu\beta_0^4 E\{3u_t^3 - 3u_t^2-2\sigma_u^2\}\left[E\{(X_t-\mu)^2(X_{t-p}-\mu)\}+E\{(X_t-\mu)^2(X_{t+p}-\mu)\}\right] \\
   & \qquad + 6\mu^2\beta_0^4 E(u_t-1)^3 \gamma_p + 4\mu^2\beta_0^4\sigma_u^2 \gamma_p,
\end{align*}
}where the second step is by (\ref{apx:The4-base}), the third step is because (\ref{apx:Thm4-stm-12}) and (\ref{apx:Thm4-stm-23-1}),  and the second last step is because (\ref{apx:Thm4-q0p-basic}), (\ref{apx:Thm4-stm-13-5}), (\ref{apx:Thm4-stm-13-4}), (\ref{apx:Thm4-stm-22-3}), and (\ref{apx:Thm4-ut-utp-usr}).

\noindent$3^{\circ}$. Then we derive the value of $q_{2pr}^*$ for $r\ne p$
{\footnotesize
\begin{align}\label{pf:q_2prstar}
   q_{2pr}^* &=  \lim_{T\to\infty} T \text{Cov}\left\{\frac{1}{T}\sum_{t=1}^T(X_t^*-\mu^*)(X_{t+p}^*-\mu^*),\frac{1}{T}\sum_{s=1}^T(X_s^*-\mu^*)(X_{s+r}^*-\mu^*)\right\} \nonumber \\
    &=  \lim_{T\to\infty} \frac{\beta_0^4 }{T} \sum_{t=1}^T\sum_{s=1}^T\text{Cov}\left\{ u_tu_{t+p} (X_t - \mu)(X_{t+p} - \mu) \right. \nonumber \\ 
    & \qquad + \mu u_t(u_{t+p}-1)(X_t - \mu)  + \mu u_{t+p}(u_t-1)(X_{t+p} - \mu) + \mu^2(u_t-1)(u_{t+p}-1), \nonumber \\ 
    & \qquad \left. u_su_{s+r} (X_s - \mu)(X_{s+r} - \mu) + \mu u_s(u_{s+r}-1)(X_s - \mu) + \mu u_{s+r}(u_s-1)(X_{s+r} - \mu) + \mu^2(u_s-1)(u_{s+r}-1)\right\} \nonumber\\
    &= \lim_{T\to\infty} \frac{\beta_0^4 }{T} \sum_{t=1}^T\sum_{s=1}^T\text{Cov}\left\{ u_tu_{t+p} (X_t - \mu)(X_{t+p} - \mu), u_su_{s+r} (X_s - \mu)(X_{s+r} - \mu)\right\} \nonumber \\
    & \qquad +  \lim_{T\to\infty} \frac{\mu\beta_0^4 }{T} \sum_{t=1}^T\sum_{s=1}^T\text{Cov}\left\{ u_su_{s+r} (X_s - \mu)(X_{s+r} - \mu),  u_t(u_{t+p}-1)(X_t - \mu)\right\} \nonumber \\
    & \qquad +  \lim_{T\to\infty} \frac{\mu\beta_0^4 }{T} \sum_{t=1}^T\sum_{s=1}^T\text{Cov}\left\{ u_su_{s+r} (X_s - \mu)(X_{s+r} - \mu),  u_{t+p}(u_t-1)(X_{t+p} - \mu)\right\} \nonumber \\
    & \qquad +  \lim_{T\to\infty} \frac{\mu\beta_0^4 }{T} \sum_{t=1}^T\sum_{s=1}^T\text{Cov}\left\{ u_tu_{t+p} (X_t - \mu)(X_{t+p} - \mu), u_s(u_{s+r}-1)(X_s - \mu)\right\} \nonumber \\
    & \qquad +  \lim_{T\to\infty} \frac{\mu\beta_0^4 }{T} \sum_{t=1}^T\sum_{s=1}^T\text{Cov}\left\{ u_tu_{t+p} (X_t - \mu)(X_{t+p} - \mu), u_{s+r}(u_s-1)(X_{s+r} - \mu)\right\} \nonumber \\
    & \qquad +   \lim_{T\to\infty} \frac{2\mu^2\beta_0^4}{T} \sum_{t=1}^T\sum_{s=1}^T\text{Cov}\left\{ u_tu_{t+p} (X_t - \mu)(X_{t+p} - \mu), (u_s-1)(u_{s+r}-1)\right\} \nonumber \\
    & \qquad +   \lim_{T\to\infty} \frac{\mu^2\beta_0^4 }{T} \sum_{t=1}^T\sum_{s=1}^T\text{Cov}\left\{u_t(u_{t+p}-1)(X_t - \mu) , u_s(u_{s+r}-1)(X_s - \mu)\right\} \nonumber \\
    & \qquad +   \lim_{T\to\infty} \frac{\mu^2\beta_0^4 }{T} \sum_{t=1}^T\sum_{s=1}^T\text{Cov}\left\{u_t(u_{t+p}-1)(X_t - \mu) , u_{s+r}(u_s-1)(X_{s+r} - \mu)\right\} \nonumber \\
    & \qquad +   \lim_{T\to\infty} \frac{\mu^2\beta_0^4 }{T} \sum_{t=1}^T\sum_{s=1}^T\text{Cov}\left\{u_{t+p}(u_t-1)(X_{t+p} - \mu) , u_s(u_{s+r}-1)(X_s - \mu)\right\} \nonumber \\
    & \qquad +   \lim_{T\to\infty} \frac{\mu^2\beta_0^4 }{T} \sum_{t=1}^T\sum_{s=1}^T\text{Cov}\left\{u_{t+p}(u_t-1)(X_{t+p} - \mu) , u_{s+r}(u_s-1)(X_{s+r} - \mu)\right\} \nonumber \\
    &\qquad +  \lim_{T\to\infty} \frac{\mu^4\beta_0^4 }{T} \sum_{t=1}^T\sum_{s=1}^T\text{Cov}\left\{ (u_t-1)(u_{t+p}-1),   (u_s-1)(u_{s+q}-1)\right\}, \nonumber \\
    &= \beta_0^4 q_{pr}  + \beta_0^4 \sigma_u^2 \left[ E\{(X_t-\mu)^2(X_{t+p}-\mu)(X_{t+r}-\mu)\} +   E\{(X_t-\mu)(X_{t+p}-\mu)^2(X_{t+p+r}-\mu)\} \right. \nonumber \\
   &\qquad \left. + E\{(X_{t-r}-\mu)(X_{t}-\mu)^2(X_{t+p}-\mu)\} +   E\{(X_t-\mu)(X_{t+p-r}-\mu)(X_{t+p}-\mu)^2\} \right] \nonumber \\ 
    &\qquad + \mu \beta_0^4 \sigma_u^2 \left[ E\{(X_t-\mu)(X_{t+p}-\mu)(X_{t+r}-\mu)\} +   E\{(X_t-\mu)(X_{t+p}-\mu)(X_{t+p+r}-\mu)\} \right. \nonumber \\
   &\qquad \left. + E\{(X_{t-r}-\mu)(X_{t}-\mu)(X_{t+p}-\mu)\} +   E\{(X_t-\mu)(X_{t+p-r}-\mu)(X_{t+p}-\mu)\} \right] \nonumber \\  
   &\qquad + 2\mu^2\beta_0^4\sigma_u^2(\gamma_{|p-r|}+\gamma_{p+r}), 
 \end{align}
}where the second step is by (\ref{apx:The4-base}), the third step is because (\ref{apx:Thm4-stm-12}) and (\ref{apx:Thm4-stm-23-1}), and the second last step is because (\ref{apx:Thm4-q0p-basic}), (\ref{apx:Thm4-stm-13-5}), (\ref{apx:Thm4-stm-22-3}), and (\ref{apx:Thm4-ut-utp-usr}).

\noindent$4^{\circ}$. Finally, similar to the derivation of $q_{2pq}^*$, now we derive the value of $q_{2pp}^*$
{\footnotesize
\begin{align}\label{pf:q_2prstar}
   q_{2pp}^* &=  \lim_{T\to\infty} T \text{Cov}\left\{\frac{1}{T}\sum_{t=1}^T(X_t^*-\mu^*)(X_{t+p}^*-\mu^*),\frac{1}{T}\sum_{s=1}^T(X_s^*-\mu^*)(X_{s+p}^*-\mu^*)\right\} \nonumber \\
   &=  \lim_{T\to\infty} \frac{\beta_0^4 }{T} \sum_{t=1}^T\sum_{s=1}^T\text{Cov}\left\{ u_tu_{t+p} (X_t - \mu)(X_{t+p} - \mu) + \mu u_t(u_{t+p}-1)(X_t - \mu)  \right. \nonumber \\ 
   & \qquad + \mu u_{t+p}(u_t-1)(X_{t+p} - \mu) + \mu^2(u_t-1)(u_{t+p}-1), \nonumber \\ 
   & \qquad \left. u_su_{s+r} (X_s - \mu)(X_{s+p} - \mu) + \mu u_s(u_{s+p}-1)(X_s - \mu) + \mu u_{s+p}(u_s-1)(X_{s+p} - \mu) + \mu^2(u_s-1)(u_{s+p}-1)\right\} \nonumber\\
   &= \lim_{T\to\infty} \frac{\beta_0^4 }{T} \sum_{t=1}^T\sum_{s=1}^T\text{Cov}\left\{ u_tu_{t+p} (X_t - \mu)(X_{t+p} - \mu), u_su_{s+p} (X_s - \mu)(X_{s+p} - \mu)\right\} \nonumber \\
    & \qquad +   \lim_{T\to\infty} \frac{2\mu^2\beta_0^4 }{T} \sum_{t=1}^T\sum_{s=1}^T\text{Cov}\left\{ u_tu_{t+p} (X_t - \mu)(X_{t+p} - \mu), (u_s-1)(u_{s+p}-1)\right\} \nonumber \\
    & \qquad +  \lim_{T\to\infty} \frac{\mu\beta_0^4 }{T} \sum_{t=1}^T\sum_{s=1}^T\text{Cov}\left\{ u_su_{s+p} (X_s - \mu)(X_{s+p} - \mu),  u_t(u_{t+p}-1)(X_t - \mu)\right\} \nonumber \\
    & \qquad +  \lim_{T\to\infty} \frac{\mu\beta_0^4 }{T} \sum_{t=1}^T\sum_{s=1}^T\text{Cov}\left\{ u_su_{s+p} (X_s - \mu)(X_{s+p} - \mu),  u_{t+p}(u_t-1)(X_{t+p} - \mu)\right\} \nonumber \\
    & \qquad +  \lim_{T\to\infty} \frac{\mu\beta_0^4 }{T} \sum_{t=1}^T\sum_{s=1}^T\text{Cov}\left\{ u_tu_{t+p} (X_t - \mu)(X_{t+p} - \mu), u_s(u_{s+p}-1)(X_s - \mu)\right\} \nonumber \\
    & \qquad +  \lim_{T\to\infty} \frac{\mu\beta_0^4 }{T} \sum_{t=1}^T\sum_{s=1}^T\text{Cov}\left\{ u_tu_{t+p} (X_t - \mu)(X_{t+p} - \mu), u_{s+p}(u_s-1)(X_{s+p} - \mu)\right\} \nonumber \\
    & \qquad +   \lim_{T\to\infty} \frac{\mu^2\beta_0^4 }{T} \sum_{t=1}^T\sum_{s=1}^T\text{Cov}\left\{u_t(u_{t+p}-1)(X_t - \mu) , u_s(u_{s+p}-1)(X_s - \mu)\right\} \nonumber \\
    & \qquad +   \lim_{T\to\infty} \frac{\mu^2\beta_0^4 }{T} \sum_{t=1}^T\sum_{s=1}^T\text{Cov}\left\{u_t(u_{t+p}-1)(X_t - \mu) , u_{s+p}(u_s-1)(X_{s+p} - \mu)\right\} \nonumber \\
    & \qquad +   \lim_{T\to\infty} \frac{\mu^2\beta_0^4 }{T} \sum_{t=1}^T\sum_{s=1}^T\text{Cov}\left\{u_{t+p}(u_t-1)(X_{t+p} - \mu) , u_s(u_{s+p}-1)(X_s - \mu)\right\} \nonumber \\
    & \qquad +   \lim_{T\to\infty} \frac{\mu^2\beta_0^4 }{T} \sum_{t=1}^T\sum_{s=1}^T\text{Cov}\left\{u_{t+p}(u_t-1)(X_{t+p} - \mu) , u_{s+p}(u_s-1)(X_{s+p} - \mu)\right\} \nonumber \\
    &\qquad +  \lim_{T\to\infty} \frac{\mu^4\beta_0^4 }{T} \sum_{t=1}^T\sum_{s=1}^T\text{Cov}\left\{ (u_t-1)(u_{t+p}-1),   (u_s-1)(u_{s+p}-1)\right\}, \nonumber \\
   &= \beta_0^4 q_{pp}  + \beta_0^4 (\sigma_u^4 + 2\sigma_u^2)  \text{Var}\{(X_t-\mu)(X_{t+p}-\mu)\} + 2\beta_0^4  E\{(X_t-\mu)(X_{t+p}-\mu)^2(X_{t+2p}-\mu)\} \\
   &\qquad + \mu \beta_0^4 \sigma_u^2 \left[ E\{(X_t-\mu)(X_{t+p}-\mu)^2\} +   2E\{(X_t-\mu)(X_{t+p}-\mu)(X_{t+2p}-\mu)\} +  E\{(X_t-\mu)^2(X_{t+p}-\mu)\} \right] \nonumber \\  
   &\qquad + 2\mu^2\beta_0^4 \sigma_u^4 \gamma_p + 2\mu^2\beta_0^4 \sigma_u^2 (\gamma_0 + \gamma_{2p}) + \mu^4\beta_0^4\sigma_u^4, \nonumber
\end{align}
}where the second step is by (\ref{apx:The4-base}), the third step is because (\ref{apx:Thm4-stm-12}) and (\ref{apx:Thm4-stm-23-1}), and the last step is because (\ref{apx:Thm4-qpp-basic}), (\ref{apx:Thm4-stm-13-6}), (\ref{apx:Thm4-stm-13-7}) and (\ref{apx:Thm4-Var-e_u-e_u+p}).


\newpage

\section{Tables}

\makeatletter
\renewcommand{\tablename}{Supplementary Table \@gobble}
\makeatother

\makeatletter
\renewcommand{\figurename}{Supplementary Figure \@gobble}
\makeatother


\begin{table}[htbp]
\centering
\scriptsize
\caption{The results of the augmented Dickey-Fuller test \label{tab:ADFtest3}}
\begin{tabular}{llccccccccccc}
\hline
                              &                  & \multicolumn{2}{c}{British Columbia} &  & \multicolumn{2}{c}{Ontario}     &  & \multicolumn{2}{c}{Quebec}      &  & \multicolumn{2}{c}{Alberta}     \\ \cline{3-4} \cline{6-7} \cline{9-10} \cline{12-13} 
Definition                    & Transformation   & TSV    & p-value   &  & TSV & p-value &  & TSV & p-value &  & TSV & p-value \\ \hline
\multirow{2}{*}{Definition 1} & $X_t$            & -8.346                   & $<$0.01   &  & -1.527                & 0.755   &  & -1.813                & 0.645   &  & -2.850                & 0.245   \\
                              & $X_{t+1}- X_{t}$ & -6.974                   & $<$0.01   &  & -5.522                & $<$0.01 &  & -3.880                & 0.027   &  & -3.516                & 0.059   \\
\multirow{2}{*}{Definition 2} & $X_t$            & -1.208                   & 0.878     &  & -4.294                & $<$0.01 &  & -2.018                & 0.566   &  & -1.768                & 0.662   \\
                              & $X_{t+1}- X_{t}$ & -3.336                   & 0.084     &  & -2.599                & 0.342   &  & -3.340                & 0.084   &  & -3.296                & 0.090   \\
\multirow{2}{*}{Definition 3} & $X_t$            & -1.325                   & 0.833     &  & -2.264                & 0.471   &  & 0.098                 & 0.999   &  & -2.688                & 0.307   \\
                              & $X_{t+1}- X_{t}$ & -3.590                   & 0.048     &  & -4.584                & $<$0.01 &  & -2.209                & 0.492   &  & -2.008                & 0.569   \\ \hline
\end{tabular}
\end{table}

\begin{table}[htbp]
\centering
\footnotesize
\caption{The results of the augmented Dickey-Fuller test \label{tab:deathdef}}
\begin{tabular}{lccccclccccc}
\hline
             & \multicolumn{2}{c}{British Columbia}                                    &  & \multicolumn{2}{c}{Ontario} &  & \multicolumn{2}{c}{Quebec}                                     &                      & \multicolumn{2}{c}{Alberta}                                    \\ \cline{2-3} \cline{5-6} \cline{8-9} \cline{11-12} 
Definition   & Differencing                           & lag $p$                        &  & Differencing     & lag $p$  &  & \multicolumn{1}{l}{Differencing} & \multicolumn{1}{l}{lag $p$} & \multicolumn{1}{l}{} & \multicolumn{1}{l}{Differencing} & \multicolumn{1}{l}{lag $p$} \\ \hline
Definition 1 & 1 degree & 1 &  & 1 degree         & 1        &  & 1 degree                         & 1                           &                      & 1 degree                         & 1                           \\
             & no differencing                        & 2                              &  & -                & -        &  & - & -                           &                      & -                                & -                           \\ 
Definition 2 & 1 degree                               & 2                              &  & no differencing  & 2        &  & 1 degree                         & 2                           &                      & 1 degree                         & 1                           \\
Definition 3 & 1 degree                               & 1                              &  & 1 degree         & 4        &  & -                                & -                           &                      & -                                & -                           \\
\hline
\end{tabular}
\end{table}

\begin{sidewaystable}
\setlength\extrarowheight{1pt} 
\centering
\captionsetup{size=small}
\small

\begin{threeparttable}
\caption[]{The parameter values of $\sigma_e^2$ or $\sigma_u^2$ for the measurement error model (\ref{eqn:measmodelClassic}) or (\ref{eqn:measmodelmultiplicative}) that are used for sensitivity analyses. }
\label{tab:par-sigmas}

\begin{tabular}{cccccccccccccc}
\hline
Definition                    & Error Model                   &  & \multicolumn{2}{c}{British Columbia} &  & \multicolumn{2}{c}{Ontario} &  & \multicolumn{2}{c}{Quebec} &  & \multicolumn{2}{c}{Alberta} \\ \hline
\multirow{6}{*}{Definition 1} &                               &  & \multicolumn{2}{c}{AR(1)}            &  & \multicolumn{2}{c}{AR(1)}   &  & \multicolumn{2}{c}{AR(1)}  &  & \multicolumn{2}{c}{AR(1)}   \\
                              & Additive ($\sigma_e^2$)       &  & 0.1             & 0.2                &  & 0.5         & 1             &  & 0.5        & 1             &  & 0.1        & 0.3            \\
                              & Multiplicative ($\sigma_u^2$) &  & 0.3             & 0.6                &  & 0.5         & 1             &  & 0.5        & 1             &  & 0.4        & 0.8            \\
                              &                               &  & \multicolumn{2}{c}{AR(2)\tnote{*}}           &  & \multicolumn{2}{c}{-}       &  & \multicolumn{2}{c}{-}      &  & \multicolumn{2}{c}{-}       \\
                              & Additive ($\sigma_e^2$)       &  & 0.1             & 0.2                &  & -           & -             &  & -          & -             &  & -          & -              \\
                              & Multiplicative ($\sigma_u^2$) &  & 0.01            & 0.02               &  & -           & -             &  & -          & -             &  & -          & -              \\
 \hline
\multirow{3}{*}{Definition 2} &                               &  & \multicolumn{2}{c}{AR(2)}            &  & \multicolumn{2}{c}{AR(2)\tnote{*}}  &  & \multicolumn{2}{c}{AR(2)}  &  & \multicolumn{2}{c}{AR(1)}   \\
                              & Additive ($\sigma_e^2$)       &  & 0.05            & 0.1                &  & 0.05        & 0.1           &  & 0.1        & 0.2           &  & 0.05       & 0.1            \\
                              & Multiplicative ($\sigma_u^2$) &  & 0.2             & 0.5                &  & 0.005       & 0.01          &  & 0.3        & 0.6           &  & 0.4        & 0.8            \\ \hline
\multirow{3}{*}{Definition 3} &                               &  & \multicolumn{2}{c}{AR(2)}            &  & \multicolumn{2}{c}{AR(4)}   &  & \multicolumn{2}{c}{-}      &  & \multicolumn{2}{c}{-}       \\
                              & Additive ($\sigma_e^2$)       &  & 0.03            & 0.06               &  & 0.02        & 0.05           &  & -          & -             &  & -          & -              \\
                              & Multiplicative ($\sigma_u^2$) &  & 0.3             & 0.6                &  & 0.1         & 0.2           &  & -          & -             &  & -          & -              \\ \hline
\end{tabular}
\begin{tablenotes}
\item[*]{\scriptsize The time series with no differencing}
\end{tablenotes}

\end{threeparttable}
\end{sidewaystable}

\begin{table}[htbp]
\centering
\tiny
\caption{Definition 3:  The parameter estimation under different measurement error models: the AR(1) model with ``order-1 differencing" is used to fit the data of British Columbia and the AR(4) model with ``order-1 differencing" is used to fit the data of Ontario.}\label{tab:par-Def-3}
\begin{tabular}{cccccccccc}
\hline
                          &                                          &           & \multicolumn{3}{c}{British Columbia} &  & \multicolumn{3}{c}{Ontario} \\ \cline{1-6} \cline{8-10} 
Method                    & Error Degree                             & Parameter & EST        & SE        & p-value     &  & EST     & SE     & p-value  \\ \hline
\multirow{5}{*}{Naive}    & \multirow{5}{*}{-}                       & $\phi_0$  & 0.105      & 0.038     & 0.018       &  & 0.379   & 0.057  & $<$0.001 \\
                          &                                          & $\phi_1$  & -0.207     & 0.077     & 0.020       &  & -0.086  & 0.099  & 0.391    \\
                          &                                          & $\phi_2$  & -          & -         & -           &  & -0.287  & 0.106  & 0.012    \\
                          &                                          & $\phi_3$  & -          & -         & -           &  & -0.301  & 0.094  & 0.004    \\
                          &                                          & $\phi_4$  & -          & -         & -           &  & -0.284  & 0.078  & 0.001    \\ \hline
                          & \multirow{5}{*}{Small ($\sigma_{e1}^2$)} & $\phi_0$  & 0.057      & 0.021     & 0.021       &  & 0.206   & 0.031  & $<$0.001 \\
                          &                                          & $\phi_1$  & -0.213     & 0.086     & 0.029       &  & -0.088  & 0.100  & 0.383    \\
                          &                                          & $\phi_2$  & -          & -         & -           &  & -0.290  & 0.109  & 0.014    \\
                          &                                          & $\phi_3$  & -          & -         & -           &  & -0.303  & 0.094  & 0.003    \\
The Proposed Method       &                                          & $\phi_4$  & -          & -         & -           &  & -0.287  & 0.081  & 0.002    \\ \cline{2-10} 
with Additive Error       & \multirow{5}{*}{Large ($\sigma_{e2}^2$)} & $\phi_0$  & 0.058      & 0.021     & 0.017       &  & 0.212   & 0.036  & $<$0.001 \\
                          &                                          & $\phi_1$  & -0.234     & 0.147     & 0.137       &  & -0.102  & 0.123  & 0.417    \\
                          &                                          & $\phi_2$  & -          & -         & -           &  & -0.306  & 0.139  & 0.037    \\
                          &                                          & $\phi_3$  & -          & -         & -           &  & -0.318  & 0.107  & 0.006    \\
                          &                                          & $\phi_4$  & -          & -         & -           &  & -0.308  & 0.093  & 0.003    \\ \hline
                          & \multirow{5}{*}{Small ($\sigma_{u1}^2$)} & $\phi_0$  & 0.058      & 0.023     & 0.027       &  & 0.210   & 0.033  & $<$0.001 \\
                          &                                          & $\phi_1$  & -0.244     & 0.090     & 0.019       &  & -0.097  & 0.107  & 0.375    \\
                          &                                          & $\phi_2$  & -          & -         & -           &  & -0.300  & 0.117  & 0.016    \\
                          &                                          & $\phi_3$  & -          & -         & -           &  & -0.312  & 0.098  & 0.004    \\
The Proposed Method       &                                          & $\phi_4$  & -          & -         & -           &  & -0.300  & 0.087  & 0.002    \\ \cline{2-10} 
with Multiplicative Error & \multirow{5}{*}{Large ($\sigma_{u2}^2$)} & $\phi_0$  & 0.066      & 0.035     & 0.087       &  & 0.230   & 0.058  & 0.001    \\
                          &                                          & $\phi_1$  & -0.401     & 0.219     & 0.092       &  & -0.139  & 0.183  & 0.454    \\
                          &                                          & $\phi_2$  & -          & -         & -           &  & -0.347  & 0.213  & 0.116    \\
                          &                                          & $\phi_3$  & -          & -         & -           &  & -0.354  & 0.159  & 0.035    \\
                          &                                          & $\phi_4$  & -          & -         & -           &  & -0.361  & 0.149  & 0.023    \\ \hline
\end{tabular}
\end{table}

\begin{landscape}
\begin{figure}[!p] 
\centering
\includegraphics[width=1.2\textwidth]{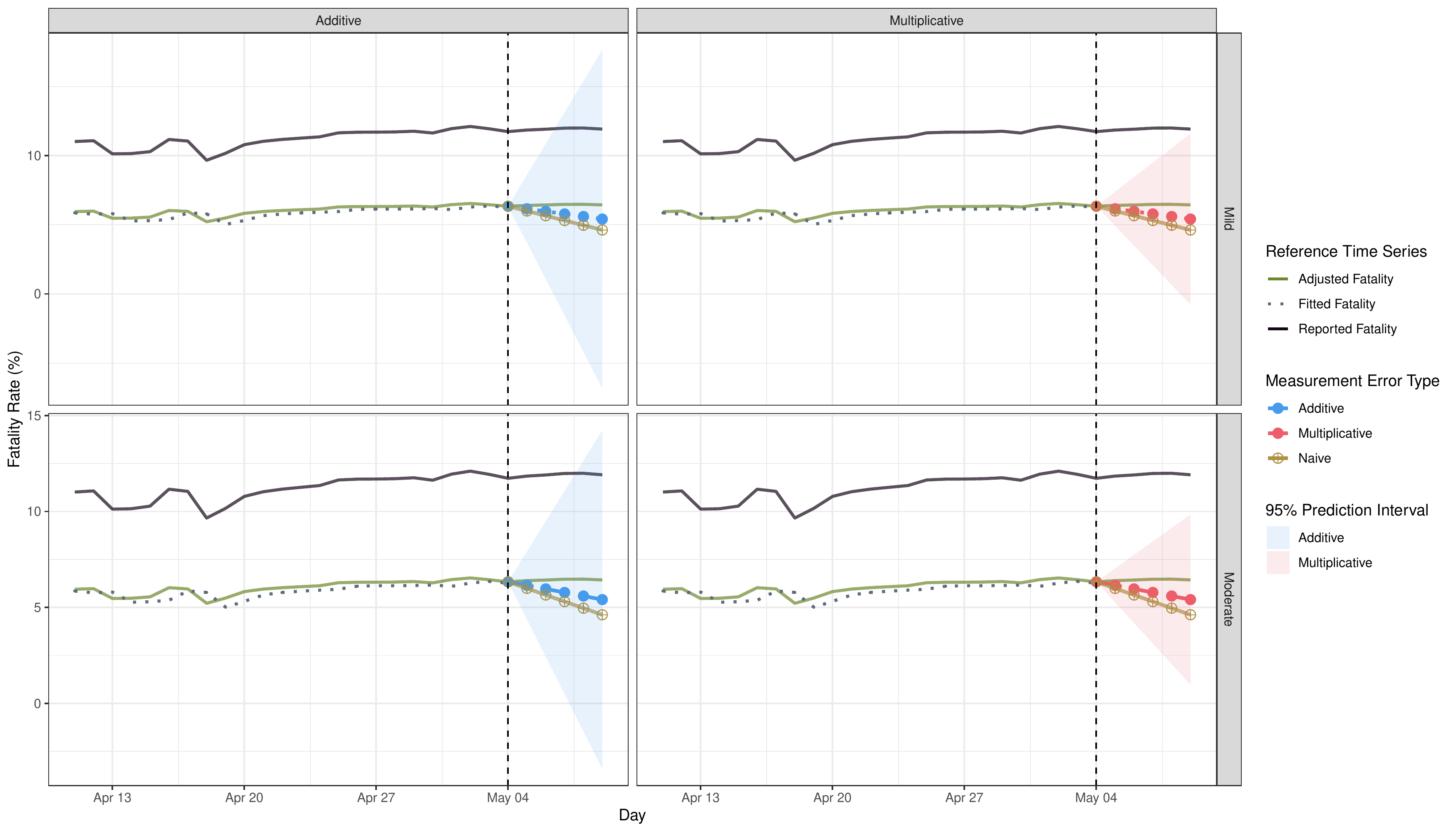}
\caption{ Quebec by Definition 1 (AR(1), order-1 differencing): A 5-day forecasting of the true mortality rate (May 5 - May 9) based on the additive (in blue) or multiplicative (in red) versus the naive model (in dark yellow); the reported mortality rates  (in black) and the adjusted true mortality rate accounting for the asymptomatic cases (in green).}\label{fig:QCfata3}
\end{figure}

\end{landscape}

\begin{landscape}
\begin{figure}[!p]
\centering
\includegraphics[width=1.2\textwidth]{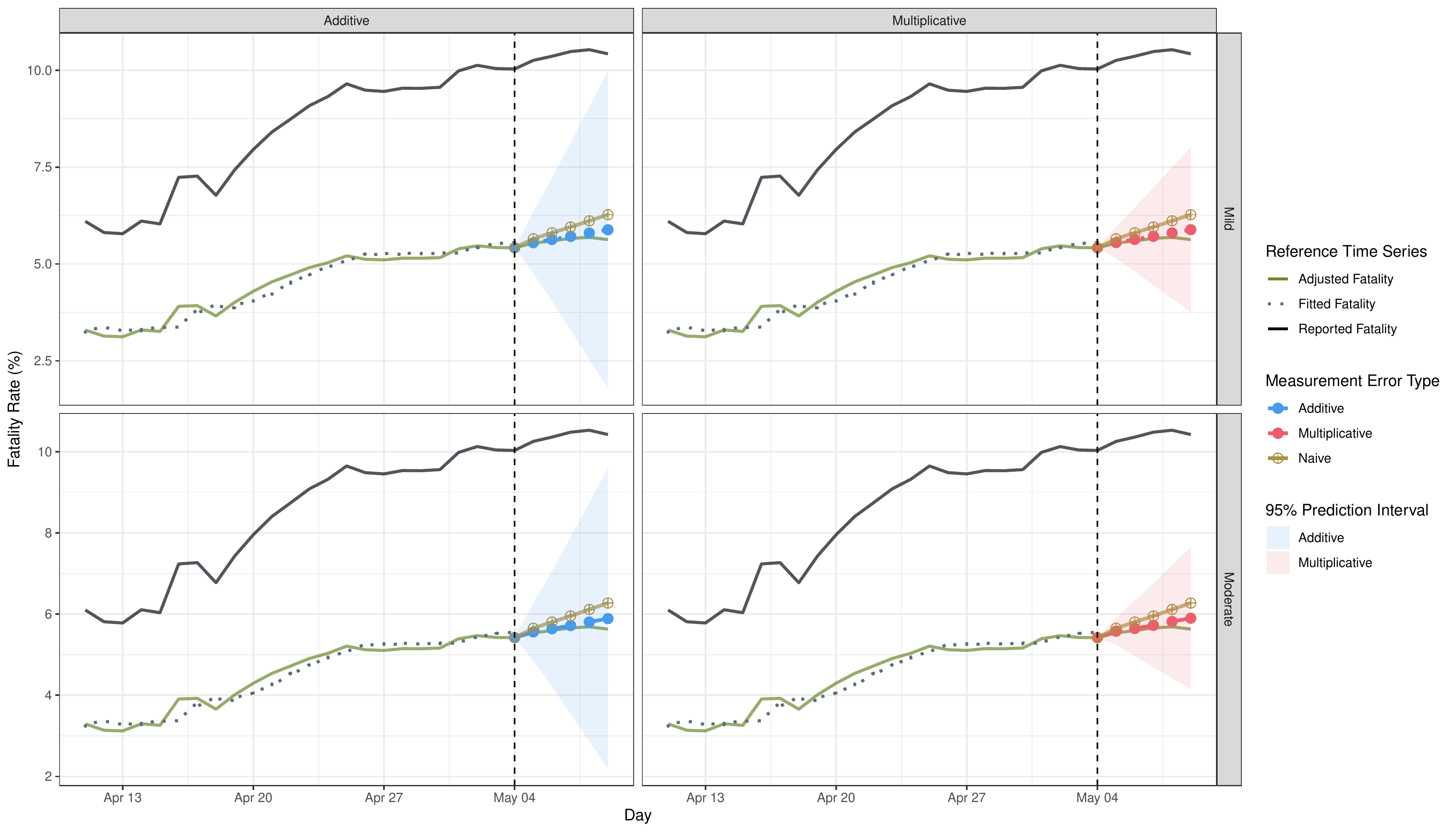}
\caption{ Quebec by Definition 2 (AR(2), order-1 differencing): A 5-day forecasting of the true mortality rate (May 5 - May 9) based on the additive (in blue) or multiplicative (in red) versus the naive model (in dark yellow); the reported mortality rates  (in black) and the adjusted true mortality rate accounting for the asymptomatic cases (in green).}\label{fig:QCfata4}
\end{figure}
\end{landscape}

\begin{landscape}
\begin{figure}[!p] 
\centering
\includegraphics[width=1.2\textwidth]{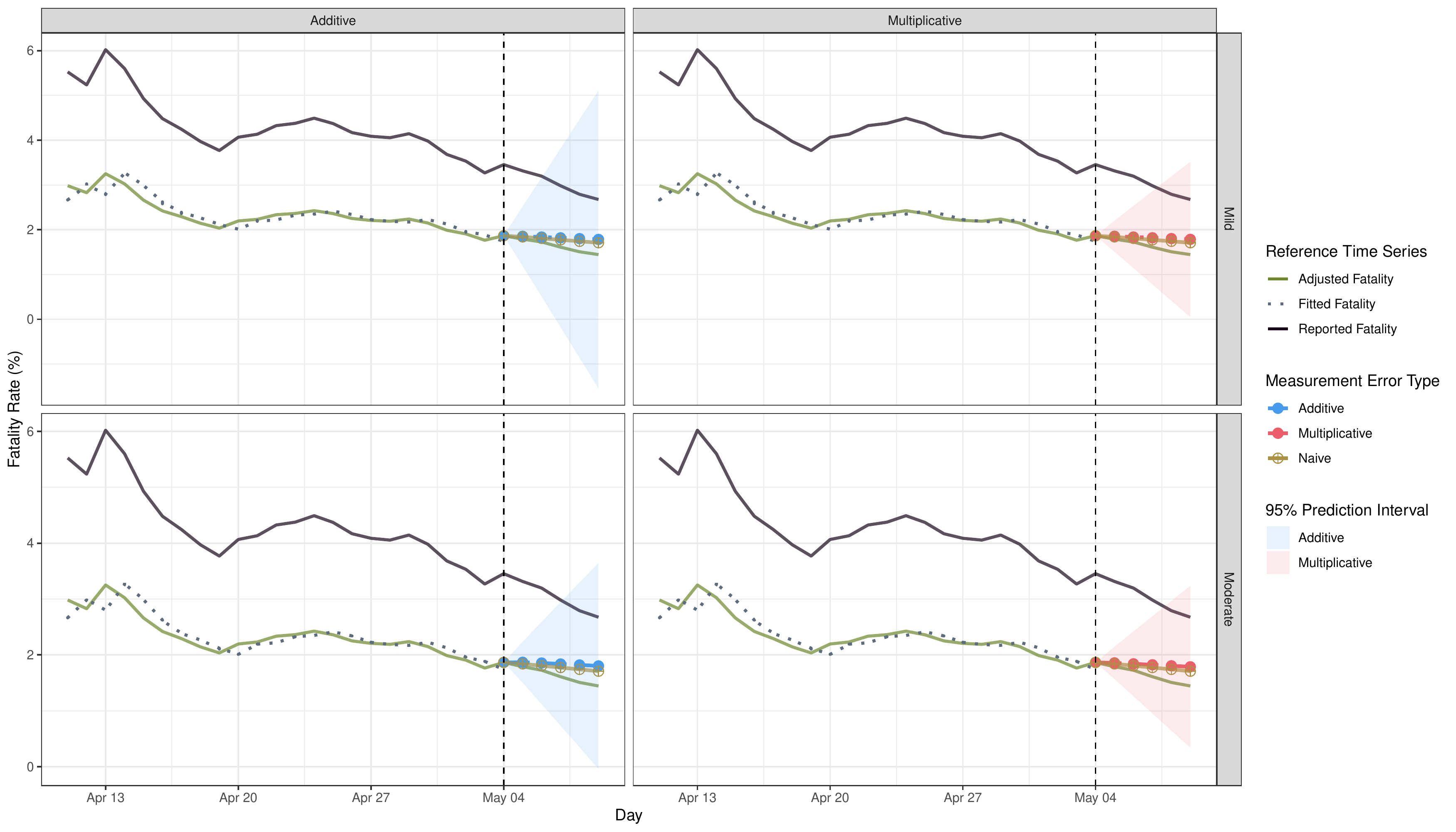}
\caption{ Alberta by Definition 1 (AR(1), order-1 differencing): A 5-day forecasting of the true mortality rate (May 5 - May 9) based on the additive (in blue) or multiplicative (in red) versus the naive model (in dark yellow); the reported mortality rates  (in black) and the adjusted true mortality rate accounting for the asymptomatic cases (in green).}\label{fig:ABfata3}
\end{figure}

\end{landscape}

\begin{landscape}
\begin{figure}[!p]
\centering
\includegraphics[width=1.2\textwidth]{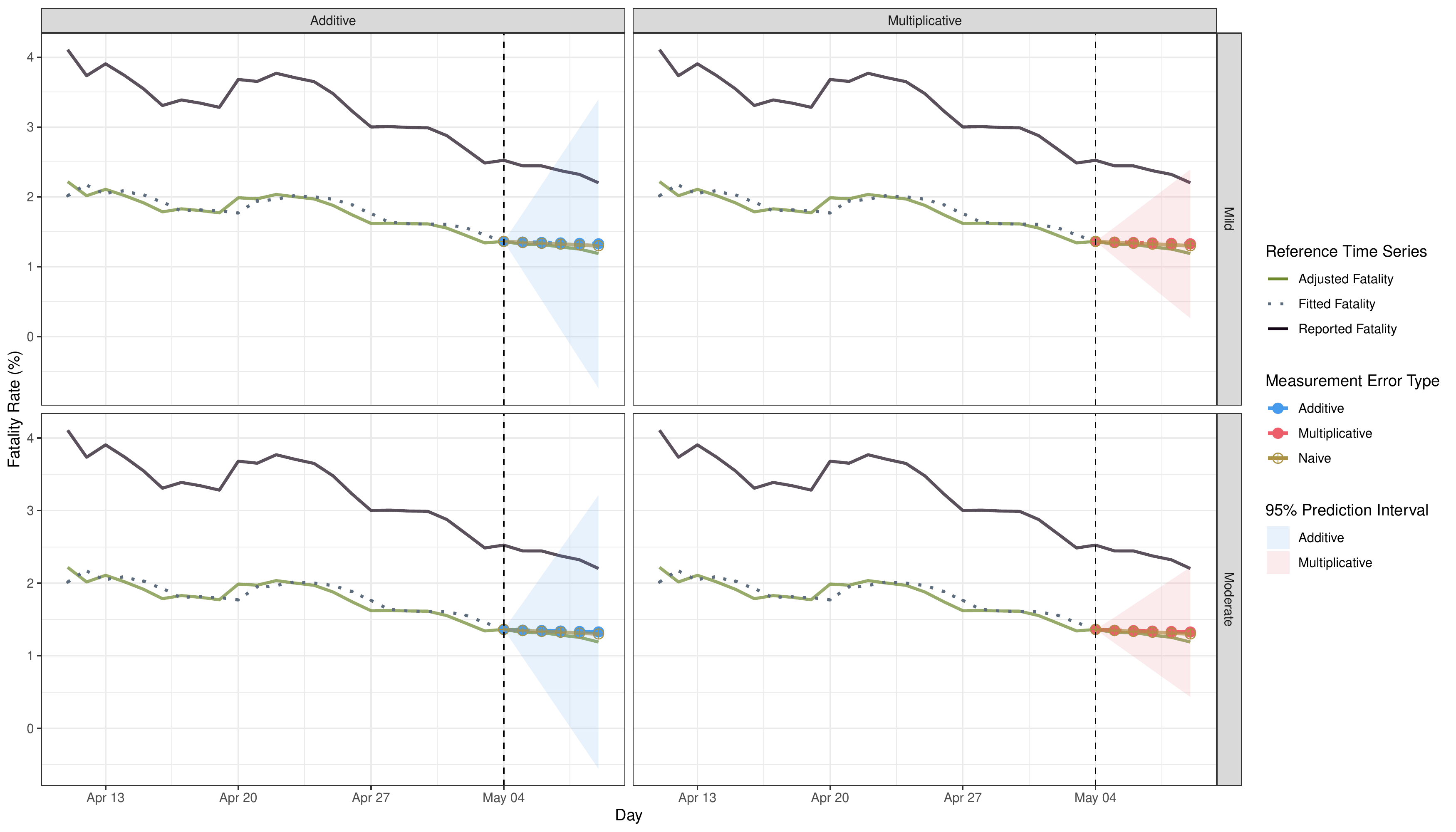}
\caption{ Alberta by Definition 2 (AR(1), order-1 differencing): A 5-day forecasting of the true mortality rate (May 5 - May 9) based on the additive (in blue) or multiplicative (in red) versus the naive model (in dark yellow); the reported mortality rates  (in black) and the adjusted true mortality rate accounting for the asymptomatic cases (in green).}\label{fig:ABfata4}
\end{figure}
\end{landscape}

\begin{sidewaystable}
\setlength\extrarowheight{1pt} 
\centering
\captionsetup{size=small}
\tiny

\begin{threeparttable}
\caption[]{Definition 1: The observed prediction error and expected prediction error for different definition of death rates.}
\label{Tab:Table-Peh-Def-1}

\begin{tabular}{ccccccccccccccc}
\hline
                                & \multicolumn{1}{l}{}                               & \multicolumn{6}{c}{Observed Prediction Error}                      &  & \multicolumn{6}{c}{Expected Prediction Error}                      \\ \hline
Method                          & \multicolumn{1}{l}{$\sigma_e^2$ (or $\sigma_u^2$)} & Day 1 & Day 2 & Day 3 & Day 4 & Day 5 & $\sum_{h=1}^H{\rm OPE}(h)$ &  & Day 1 & Day 2 & Day 3 & Day 4 & Day 5 & $\sum_{h=1}^H{\rm EPE}(h)$ \\ \hline
                                & \multicolumn{14}{c}{Definition 1}                                                                                                                                                               \\ \cline{3-15} 
                                &                                                    & \multicolumn{13}{l}{British Columbia}                                                                                                      \\
Naive                           & -                                                  & 0.017 & 0.006 & 0.017 & 0.058 & 0.081 & 0.178                      &  & 0.069 & 0.081 & 0.081 & 0.082 & 0.083 & 0.396                      \\
\multirow{2}{*}{Additive}       & Mild                                               & 0.011 & 0.001 & 0.005 & 0.027 & 0.035 & 0.078                      &  & 0.066 & 0.078 & 0.078 & 0.080 & 0.080 & 0.382                      \\
                                & Moderate                                           & 0.012 & 0.001 & 0.005 & 0.027 & 0.036 & 0.080                      &  & 0.057 & 0.070 & 0.070 & 0.072 & 0.073 & 0.342                      \\
\multirow{2}{*}{Multiplicative} & Mild                                               & 0.011 & 0.001 & 0.005 & 0.027 & 0.035 & 0.078                      &  & 0.020 & 0.023 & 0.023 & 0.023 & 0.023 & 0.111                      \\
                                & Moderate                                           & 0.013 & 0.001 & 0.004 & 0.029 & 0.037 & 0.083                      &  & 0.015 & 0.019 & 0.018 & 0.019 & 0.019 & 0.090                      \\ \cline{3-15} 
                                &                                                    & \multicolumn{13}{l}{Ontario}                                                                                                               \\
Naive                           & -                                                  & 0.830 & 0.077 & 3.409 & 0.360 & 8.264 & 12.940                     &  & 0.612 & 1.446 & 2.048 & 2.372 & 2.514 & 8.991                      \\
\multirow{2}{*}{Additive}       & Mild                                               & 0.004 & 0.116 & 0.002 & 0.161 & 0.004 & 0.288                      &  & 0.607 & 1.440 & 2.046 & 2.373 & 2.517 & 8.983                      \\
                                & Moderate                                           & 0.004 & 0.119 & 0.001 & 0.172 & 0.007 & 0.304                      &  & 0.591 & 1.422 & 2.040 & 2.378 & 2.531 & 8.963                      \\
\multirow{2}{*}{Multiplicative} & Mild                                               & 0.004 & 0.119 & 0.001 & 0.171 & 0.007 & 0.302                      &  & 0.176 & 0.420 & 0.602 & 0.704 & 0.754 & 2.655                      \\
                                & Moderate                                           & 0.003 & 0.132 & 0.000 & 0.225 & 0.029 & 0.389                      &  & 0.169 & 0.418 & 0.630 & 0.775 & 0.888 & 2.879                      \\ \cline{3-15} 
                                &                                                    & \multicolumn{13}{l}{Quebec}                                                                                                                \\
Naive                           & -                                                  & 0.163 & 0.607 & 1.357 & 2.289 & 3.294 & 7.709                      &  & 1.811 & 1.811 & 1.811 & 1.811 & 1.811 & 9.057                      \\
\multirow{2}{*}{Additive}       & Mild                                               & 0.061 & 0.216 & 0.479 & 0.778 & 1.053 & 2.587                      &  & 1.561 & 1.561 & 1.561 & 1.561 & 1.561 & 7.807                      \\
                                & Moderate                                           & 0.060 & 0.215 & 0.478 & 0.776 & 1.051 & 2.580                      &  & 0.811 & 0.811 & 0.811 & 0.811 & 0.811 & 4.057                      \\
\multirow{2}{*}{Multiplicative} & Mild                                               & 0.061 & 0.216 & 0.479 & 0.778 & 1.053 & 2.586                      &  & 0.399 & 0.399 & 0.399 & 0.399 & 0.399 & 1.995                      \\
                                & Moderate                                           & 0.060 & 0.215 & 0.477 & 0.776 & 1.050 & 2.578                      &  & 0.205 & 0.205 & 0.205 & 0.205 & 0.205 & 1.025                      \\ \cline{3-15} 
                                &                                                    & \multicolumn{13}{l}{Alberta}                                                                                                               \\
Naive                           & -                                                  & 0.002 & 0.007 & 0.027 & 0.055 & 0.070 & 0.160                      &  & 0.125 & 0.125 & 0.125 & 0.125 & 0.125 & 0.627                      \\
\multirow{2}{*}{Additive}       & Mild                                               & 0.004 & 0.012 & 0.044 & 0.087 & 0.115 & 0.262                      &  & 0.115 & 0.115 & 0.115 & 0.115 & 0.115 & 0.577                      \\
                                & Moderate                                           & 0.006 & 0.017 & 0.052 & 0.098 & 0.129 & 0.302                      &  & 0.035 & 0.035 & 0.035 & 0.035 & 0.035 & 0.177                      \\
\multirow{2}{*}{Multiplicative} & Mild                                               & 0.004 & 0.012 & 0.044 & 0.087 & 0.115 & 0.263                      &  & 0.031 & 0.031 & 0.031 & 0.031 & 0.031 & 0.157                      \\
                                & Moderate                                           & 0.005 & 0.013 & 0.045 & 0.089 & 0.118 & 0.270                      &  & 0.022 & 0.022 & 0.022 & 0.022 & 0.022 & 0.109                      \\ \hline
\end{tabular}

\end{threeparttable}
\end{sidewaystable}

\begin{sidewaystable}
\setlength\extrarowheight{1pt} 
\centering
\captionsetup{size=small}
\tiny

\begin{threeparttable}
\caption[]{Definition 2: The observed prediction error and expected prediction error for different definition of death rates.}
\label{Tab:Table-Peh-Def-2}

\begin{tabular}{ccccccccccccccc}
\hline
                                & \multicolumn{1}{l}{}                               & \multicolumn{6}{c}{Observed Prediction Error}                      &  & \multicolumn{6}{c}{Expected Prediction Error}                      \\ \cline{1-8} \cline{10-15} 
Method                          & \multicolumn{1}{l}{$\sigma_e^2$ (or $\sigma_u^2$)} & Day 1 & Day 2 & Day 3 & Day 4 & Day 5 & $\sum_{h=1}^H{\rm OPE}(h)$ &  & Day 1 & Day 2 & Day 3 & Day 4 & Day 5 & $\sum_{h=1}^H{\rm EPE}(h)$ \\ \hline
                                & \multicolumn{14}{c}{Definition 2}                                                                                                                                                               \\ \cline{3-15} 
                                &                                                    & \multicolumn{13}{l}{British Columbia}                                                                                                      \\
Naive                           & -                                                  & 0.015 & 0.015 & 0.032 & 0.043 & 0.020 & 0.126                      &  & 0.164 & 0.167 & 0.167 & 0.167 & 0.167 & 0.834                      \\
\multirow{2}{*}{Additive}       & Mild                                               & 0.010 & 0.005 & 0.011 & 0.011 & 0.000 & 0.037                      &  & 0.154 & 0.157 & 0.157 & 0.157 & 0.157 & 0.783                      \\
                                & Moderate                                           & 0.010 & 0.005 & 0.011 & 0.011 & 0.000 & 0.037                      &  & 0.154 & 0.157 & 0.157 & 0.157 & 0.157 & 0.784                      \\
\multirow{2}{*}{Multiplicative} & Mild                                               & 0.010 & 0.005 & 0.011 & 0.011 & 0.000 & 0.037                      &  & 0.044 & 0.044 & 0.044 & 0.044 & 0.044 & 0.222                      \\
                                & Moderate                                           & 0.010 & 0.005 & 0.011 & 0.011 & 0.000 & 0.037                      &  & 0.034 & 0.035 & 0.035 & 0.035 & 0.035 & 0.174                      \\ \cline{3-15} 
                                &                                                    & \multicolumn{13}{l}{Ontario}                                                                                                               \\
Naive                           & -                                                  & 0.020 & 0.087 & 0.196 & 0.521 & 1.059 & 1.884                      &  & 2.527 & 2.643 & 2.649 & 2.649 & 2.649 & 13.117                     \\
\multirow{2}{*}{Additive}       & Mild                                               & 0.001 & 0.004 & 0.007 & 0.056 & 0.175 & 0.243                      &  & 2.264 & 2.391 & 2.399 & 2.399 & 2.399 & 11.853                     \\
                                & Moderate                                           & 0.000 & 0.000 & 0.000 & 0.023 & 0.110 & 0.134                      &  & 1.453 & 1.626 & 1.646 & 1.649 & 1.649 & 8.023                      \\
\multirow{2}{*}{Multiplicative} & Mild                                               & 0.000 & 0.002 & 0.003 & 0.044 & 0.152 & 0.201                      &  & 0.558 & 0.599 & 0.603 & 0.603 & 0.603 & 2.965                      \\
                                & Moderate                                           & 0.004 & 0.010 & 0.014 & 0.000 & 0.035 & 0.063                      &  & 0.270 & 0.331 & 0.345 & 0.348 & 0.348 & 1.642                      \\ \cline{3-15} 
                                &                                                    & \multicolumn{13}{l}{Quebec}                                                                                                                \\
Naive                           & -                                                  & 0.013 & 0.044 & 0.086 & 0.183 & 0.413 & 0.739                      &  & 0.174 & 0.176 & 0.191 & 0.192 & 0.193 & 0.926                      \\
\multirow{2}{*}{Additive}       & Mild                                               & 0.000 & 0.001 & 0.003 & 0.013 & 0.065 & 0.081                      &  & 0.163 & 0.165 & 0.181 & 0.182 & 0.183 & 0.874                      \\
                                & Moderate                                           & 0.000 & 0.002 & 0.003 & 0.014 & 0.068 & 0.087                      &  & 0.130 & 0.133 & 0.149 & 0.151 & 0.153 & 0.716                      \\
\multirow{2}{*}{Multiplicative} & Mild                                               & 0.000 & 0.002 & 0.003 & 0.013 & 0.066 & 0.084                      &  & 0.044 & 0.045 & 0.049 & 0.049 & 0.050 & 0.236                      \\
                                & Moderate                                           & 0.002 & 0.003 & 0.004 & 0.017 & 0.073 & 0.098                      &  & 0.030 & 0.030 & 0.033 & 0.034 & 0.034 & 0.162                      \\ \cline{3-15} 
                                &                                                    & \multicolumn{13}{l}{Alberta}                                                                                                               \\
Naive                           & -                                                  & 0.001 & 0.000 & 0.002 & 0.003 & 0.012 & 0.019                      &  & 0.047 & 0.047 & 0.047 & 0.047 & 0.047 & 0.236                      \\
\multirow{2}{*}{Additive}       & Mild                                               & 0.001 & 0.001 & 0.003 & 0.007 & 0.019 & 0.031                      &  & 0.044 & 0.045 & 0.045 & 0.045 & 0.045 & 0.223                      \\
                                & Moderate                                           & 0.001 & 0.001 & 0.003 & 0.006 & 0.019 & 0.031                      &  & 0.036 & 0.037 & 0.037 & 0.037 & 0.037 & 0.185                      \\
\multirow{2}{*}{Multiplicative} & Mild                                               & 0.001 & 0.001 & 0.003 & 0.006 & 0.019 & 0.031                      &  & 0.012 & 0.012 & 0.012 & 0.012 & 0.012 & 0.059                      \\
                                & Moderate                                           & 0.001 & 0.001 & 0.003 & 0.006 & 0.019 & 0.030                      &  & 0.008 & 0.008 & 0.008 & 0.008 & 0.008 & 0.042                      \\ \hline
\end{tabular}

\end{threeparttable}
\end{sidewaystable}

\begin{sidewaystable}
\setlength\extrarowheight{1pt} 
\centering
\captionsetup{size=small}
\tiny

\begin{threeparttable}
\caption[]{Definition 3: The observed prediction error and expected prediction error for different definition of death rates.}
\label{Tab:Table-Peh-Def-3}

\begin{tabular}{ccccccccccccccc}
\hline
                                & \multicolumn{1}{l}{}                               & \multicolumn{6}{c}{Observed Prediction Error}                      &  & \multicolumn{6}{c}{Expected Prediction Error}                      \\ \cline{1-8} \cline{10-15} 
Method                          & \multicolumn{1}{l}{$\sigma_e^2$ (or $\sigma_u^2$)} & Day 1 & Day 2 & Day 3 & Day 4 & Day 5 & $\sum_{h=1}^H{\rm OPE}(h)$ &  & Day 1 & Day 2 & Day 3 & Day 4 & Day 5 & $\sum_{h=1}^H{\rm EPE}(h)$ \\ \hline
                                & \multicolumn{14}{c}{Definition 3}                                                                                                                                                               \\ \cline{3-15} 
                                & \multicolumn{1}{l}{}                               & \multicolumn{13}{l}{British Columbia}                                                                                                      \\
Naive                           & -                                                  & 0.000 & 0.003 & 0.020 & 0.057 & 0.090 & 0.170                      &  & 0.030 & 0.031 & 0.031 & 0.031 & 0.031 & 0.155                      \\
\multirow{2}{*}{Additive}       & Mild                                               & 0.001 & 0.001 & 0.000 & 0.005 & 0.009 & 0.016                      &  & 0.029 & 0.030 & 0.030 & 0.030 & 0.030 & 0.151                      \\
                                & Moderate                                           & 0.001 & 0.001 & 0.000 & 0.005 & 0.009 & 0.016                      &  & 0.026 & 0.028 & 0.028 & 0.028 & 0.028 & 0.137                      \\
\multirow{2}{*}{Multiplicative} & Mild                                               & 0.001 & 0.001 & 0.000 & 0.005 & 0.009 & 0.016                      &  & 0.007 & 0.008 & 0.008 & 0.008 & 0.008 & 0.038                      \\
                                & Moderate                                           & 0.001 & 0.001 & 0.000 & 0.006 & 0.010 & 0.017                      &  & 0.005 & 0.005 & 0.005 & 0.005 & 0.005 & 0.023                      \\ \cline{3-15} 
                                &                                                    & \multicolumn{13}{l}{Ontario}                                                                                                               \\
Naive                           & -                                                  & 0.048 & 0.132 & 0.243 & 0.333 & 0.464 & 1.219                      &  & 0.039 & 0.039 & 0.039 & 0.042 & 0.042 & 0.202                      \\
\multirow{2}{*}{Additive}       & Mild                                               & 0.002 & 0.004 & 0.011 & 0.017 & 0.024 & 0.058                      &  & 0.039 & 0.039 & 0.039 & 0.042 & 0.042 & 0.200                      \\
                                & Moderate                                           & 0.002 & 0.004 & 0.011 & 0.016 & 0.023 & 0.057                      &  & 0.036 & 0.036 & 0.036 & 0.039 & 0.039 & 0.187                      \\
\multirow{2}{*}{Multiplicative} & Mild                                               & 0.002 & 0.004 & 0.011 & 0.016 & 0.023 & 0.057                      &  & 0.011 & 0.011 & 0.011 & 0.012 & 0.012 & 0.056                      \\
                                & Moderate                                           & 0.002 & 0.004 & 0.011 & 0.015 & 0.023 & 0.055                      &  & 0.009 & 0.009 & 0.009 & 0.010 & 0.010 & 0.048                      \\ \hline
\end{tabular}

\end{threeparttable}
\end{sidewaystable}


\clearpage\pagebreak\newpage
\pagenumbering{arabic}


\end{document}